\documentclass[sigconf]{acmart}

\AtBeginDocument{%
  \providecommand\BibTeX{{%
    \normalfont B\kern-0.5em{\scshape i\kern-0.25em b}\kern-0.8em\TeX}}}

\copyrightyear{2024}
\acmYear{2024}
\setcopyright{rightsretained}
\acmConference[CHI '24]{Proceedings of the CHI Conference on Human Factors in Computing Systems}{May 11--16, 2024}{Honolulu, HI, USA}
\acmBooktitle{Proceedings of the CHI Conference on Human Factors in Computing Systems (CHI '24), May 11--16, 2024, Honolulu, HI, USA}
\acmDOI{10.1145/3613904.3642823}
\acmISBN{979-8-4007-0330-0/24/05}

\acmSubmissionID{9829}

\usepackage{acmart-taps} 
\usepackage{subcaption} 
\usepackage{colortbl} 
\usepackage[inline]{enumitem} 

\usepackage{stfloats}
\usepackage{framed}
\definecolor{shadecolor}{gray}{0.9}

\ifdefined\toprule\else\def\toprule{\hline}\fi
\ifdefined\midrule\else\def\midrule{\hline}\fi
\ifdefined\bottomrule\else\def\bottomrule{\hline}\fi

\definecolor{okabe-green}{HTML}{018571}
\definecolor{okabe-red}{HTML}{FF6666}
\definecolor{pinegreen-rba}{HTML}{008B72}
\definecolor{royalblue-rba}{HTML}{0071BC}
\definecolor{somewhatgray}{HTML}{e6e6e6}
\definecolor{was_me}{HTML}{33a02c} 
\definecolor{probably_was_me}{HTML}{a6cee3} 
\definecolor{fatigue}{HTML}{fdbf6f} 
\definecolor{unsure}{HTML}{cab2d6} 
\definecolor{someone_else}{HTML}{e31a1c}
\definecolor{other}{HTML}{dddddd} 
\definecolor{dont_remember}{HTML}{eeeeee} 
\definecolor{spontaneous}{HTML}{a6cee3} 
\definecolor{feel_protected}{HTML}{1f78b4} 
\definecolor{low_value}{HTML}{b2df8a}  
\definecolor{was_me}{HTML}{33a02c} 
\definecolor{didnt_understand}{HTML}{fb9a99} 
\definecolor{someone_else}{HTML}{e31a1c} 
\definecolor{fatigue}{HTML}{fdbf6f} 
\definecolor{suspicious}{HTML}{ff7f00} 
\definecolor{unsure}{HTML}{cab2d6}  

\aptLtoX[graphic=no, type=html]{

}{

}

\newcommand{\nnnewmoon}{\includegraphics[height=0.01\textwidth]{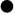}}
\newcommand{\nnfullmoon}{\includegraphics[height=0.01\textwidth]{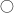}}

\aptLtoX[graphic=no, type=html]{

}{

}

\newenvironment{myquote}{
  \begin{itemize}[topsep=0pt,itemsep=3pt,partopsep=0pt,parsep=0pt,leftmargin=0pt,rightmargin=0pt,itemindent=1em]
  \item[]}{\end{itemize}}

\makeatletter
\newcommand{\mylabel}[2]{%
  \renewcommand{\@currentlabel}{\textbf{#2}}
  \label{#1}
}
\makeatother

\newcommand{\legendboxBlack}[2]{\colorbox{#1}{\parbox{0.75em}{\centering #2}}}  

\begin{document}

\title{Understanding Users' Interaction with Login Notifications}

\author{Philipp Markert}
\orcid{0000-0002-9232-4496}
\affiliation{%
  \institution{Ruhr University Bochum}
  \country{}}
\email{philipp.markert@rub.de}

\author{Leona Lassak}
\orcid{0000-0001-8309-3211}
\affiliation{%
  \institution{Ruhr University Bochum}
  \country{}}
\email{leona.lassak@rub.de}

\author{Maximilian Golla}
\orcid{0000-0003-2204-2132}
\affiliation{%
  \institution{CISPA Helmholtz Center for Information Security}
  \country{}}
\email{golla@cispa.de}

\author{Markus D\"urmuth}
\orcid{0000-0001-5048-3723}
\affiliation{%
  \institution{Leibniz University Hannover}
  \country{}}
\email{markus.duermuth@itsec.uni-hannover.de}

\renewcommand{\shortauthors}{Markert et al.}

\begin{abstract}
Login notifications intend to inform users about sign-ins and help them protect their accounts from unauthorized access.
Notifications are usually sent if a login deviates from previous ones, potentially indicating malicious activity. 
They contain information like the location, date, time, and device used to sign in.
Users are challenged to verify whether they recognize the login (because it was them or someone they know) or to protect their account from unwanted access. 
In a user study, we explore users' comprehension, reactions, and expectations of login notifications. 
We utilize two treatments to measure users' behavior in response to notifications sent for a login they initiated or based on a malicious actor relying on statistical sign-in information. 
We find that users identify legitimate logins but need more support to halt malicious sign-ins. 
We discuss the identified problems and give recommendations for service providers to ensure usable and secure logins for everyone.
\end{abstract}

\ccsdesc[500]{Security and privacy~Authentication}
\ccsdesc[500]{Security and privacy~Usability in security and privacy}

\keywords{notification, email, authentication, risk-based authentication, password change}

\begin{teaserfigure}
    \centering
    \includegraphics[width=.78\textwidth]{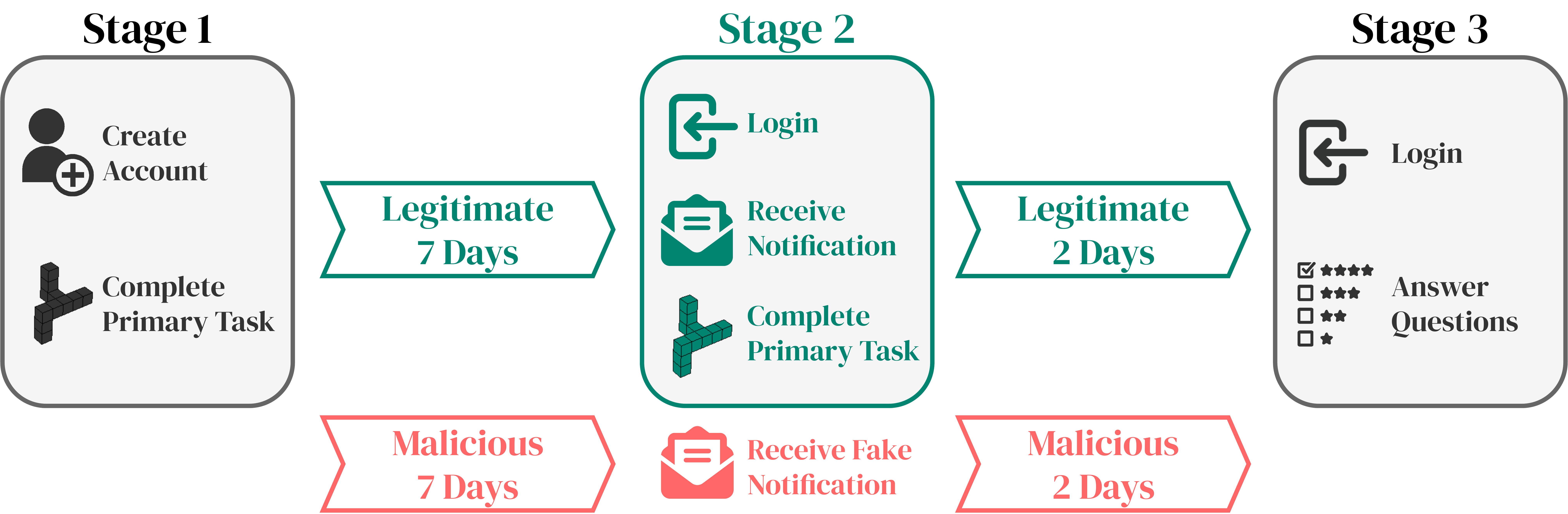}
    \caption{Structure of the user study ($n=229$). After 7 days, participants in the {\color{okabe-green}legitimate} treatment were invited to return to Stage 2 and received a notification after logging in. The {\color{okabe-red}malicious} group received a notification without actually signing in to mimic the scenario when an unexpected login occurred. Before the final stage, we gave participants 2~days to react.}
    \label{fig:teaser}
    \Description[Study flow with 3 stages for legitimate and malicious groups]{Structure of the conducted user study and the flow for the two treatments, legitimate and malicious. The prior group received a notification for a login they initiated. The latter group received a login notification without actually signing in to mimic the scenario when an unexpected login occurred.}
\end{teaserfigure}

\maketitle

\section{Introduction}\label{sec:intro}
Login notifications intend to inform users about recent sign-ins, to protect accounts from unauthorized access.
Depending on the service, notifications are sent if the login occurred from an \textit{unknown location} or \textit{new device}, which may indicate malicious activity.

Notifications are often delivered via email and include details about the device (browser and OS), approximate location, date, and time of the sign-in.
Users need to decide whether the reported login is legitimate or malicious and are recommended to change the password in case the login is unfamiliar.
Logins can be confused to be malicious when users \textit{share accounts}, and friends or family log in unknowingly.
%
While the notification is intended to protect users and provide a feeling of security, it can also be perceived as burdening and overwhelming by requiring a decision based on technical jargon and highlighting negative consequences.

Previous work~\cite{redmiles-19-should-worry} focused on challenge-based notifications and studied incident-response information-seeking and mental models about attackers.
In contrast, we focus on \emph{granted access} notifications informing users about a recent sign-in and analyze users' comprehension, expectations, and reaction to the notification.

In this work, we collected and analyzed 72~login notifications sent by real-world services and developed a \emph{baseline} notification that we employed in a user study.
The structure of the study is shown in Figure~\ref{fig:teaser}.
We disguised the study ($n = 229$)  as a psychological test, and let users create an account they had to sign into during different stages of the study.
Participants then either received a legitimate notification to their email upon signing in themselves (\emph{Legitimate}) or unexpectedly received a notification prefilled with sign-in information a non-targeted statistical attacker would use after around one week (\emph{Malicious}).\\

\noindent We sought to answer the following questions:
\aptLtoX[graphic=no, type=html]{
\begin{enumerate}
    \item[]{\textbf{RQ1}} \emph{[Reaction \& Comprehension]
    Which actions do users take in response to receiving notifications, and is resolving the situation a priority?
    Do users understand why they received the notification and which factors may have caused receiving it?}
        
    We found that participants correctly understood that ``a login'' caused receiving the notification.
    However, they are unaware of or misinterpret the trigger and are thus unsure how to react appropriately, especially in the malicious case.
    
    \item[]{\textbf{RQ2}} \emph{[Decision-Making \& Execution] 
    Do the state of the art notifications help users distinguish malicious and legitimate logins?
    Which information helps account owners with their decision, and do current notifications appropriately guide users in resolving the situation?}

    Based on device and location, participants can correctly attribute notifications caused by their own logins, but they are confused when the notification is unexpected (Malicious) and struggle to identify the correct reaction even if (as it was the case in our study) all necessary information is provided by the notification.
    
    \item[]{\textbf{RQ3}} \emph{[Perception \& Expectation]
    How do login notifications make users feel?
    When do they expect notifications to be sent, and how does prior experience affect their decision?}
    
    Notifications about malicious logins evoke (more) negative emotions, but participants who changed their password also felt empowered by taking action to protect their account.
    Interestingly, more than 90\% of the participants expect services to send login notifications because it makes them feel protected.
\end{enumerate}
}{
\begin{enumerate}[wide,format=\bfseries,topsep=5pt,itemsep=5pt,labelindent=0pt,partopsep=0pt,parsep=0pt,labelwidth=0pt,label=\textbf{RQ\arabic*}, ref=\textbf{RQ\arabic*}]
    \item\label{RQ:reaction-comprehension} \emph{[Reaction \& Comprehension]
    Which actions do users take in response to receiving notifications, and is resolving the situation a priority?
    Do users understand why they received the notification and which factors may have caused receiving it?}
        
    We found that participants correctly understood that ``a login'' caused receiving the notification.
    However, they are unaware of or misinterpret the trigger and are thus unsure how to react appropriately, especially in the malicious case.
    
    \item\label{RQ:decision-making} \emph{[Decision-Making \& Execution] 
    Do the state of the art notifications help users distinguish malicious and legitimate logins?
    Which information helps account owners with their decision, and do current notifications appropriately guide users in resolving the situation?}

    Based on device and location, participants can correctly attribute notifications caused by their own logins, but they are confused when the notification is unexpected (Malicious) and struggle to identify the correct reaction even if (as it was the case in our study) all necessary information is provided by the notification.
    
    \item\label{RQ:perception-expectation} \emph{[Perception \& Expectation]
    How do login notifications make users feel?
    When do they expect notifications to be sent, and how does prior experience affect their decision?}
    
    Notifications about malicious logins evoke (more) negative emotions, but participants who changed their password also felt empowered by taking action to protect their account.
    Interestingly, more than 90\% of the participants expect services to send login notifications because it makes them feel protected.
\end{enumerate}
}

Analyzing 72 real-world notifications revealed malformed login notifications and problematic anti-phishing advice.
Our user study shows that login notifications contribute to account security, yet our results suggest room for improvement.
We find that only 22\% of the participants who should have changed their password to protect their account did.
While participants appreciate when companies decide to monitor their accounts for incidents, services that send notifications for every, or almost every login in a ``better safe than sorry'' manner contribute to warning fatigue.
We give clear recommendations for service providers to improve their notifications.
While login notifications can help reinforce account security, protecting their accounts by identifying malicious logins should not be solely the user's responsibility.
\begin{figure*}[!ht]
    \centering
    \begin{subfigure}[b]{0.315\textwidth}
        \centering
        \includegraphics[width=\textwidth]{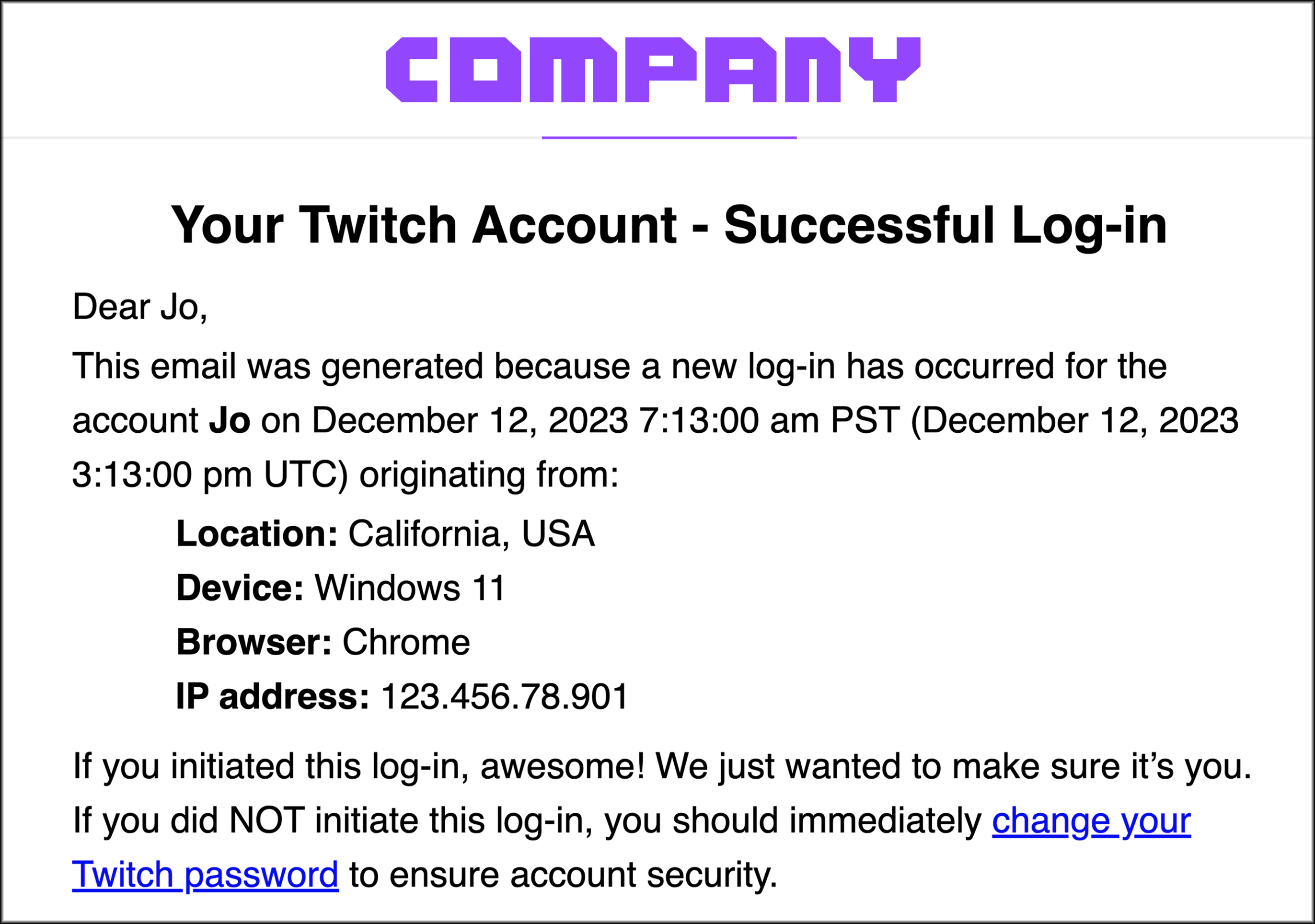}
        \caption{\textbf{Granted Access} from Twitch.}
        \label{fig:example-granted-notification}
    \end{subfigure}
    \hspace{.25em}
    \begin{subfigure}[b]{0.315\textwidth}
        \centering
        \includegraphics[width=\textwidth]{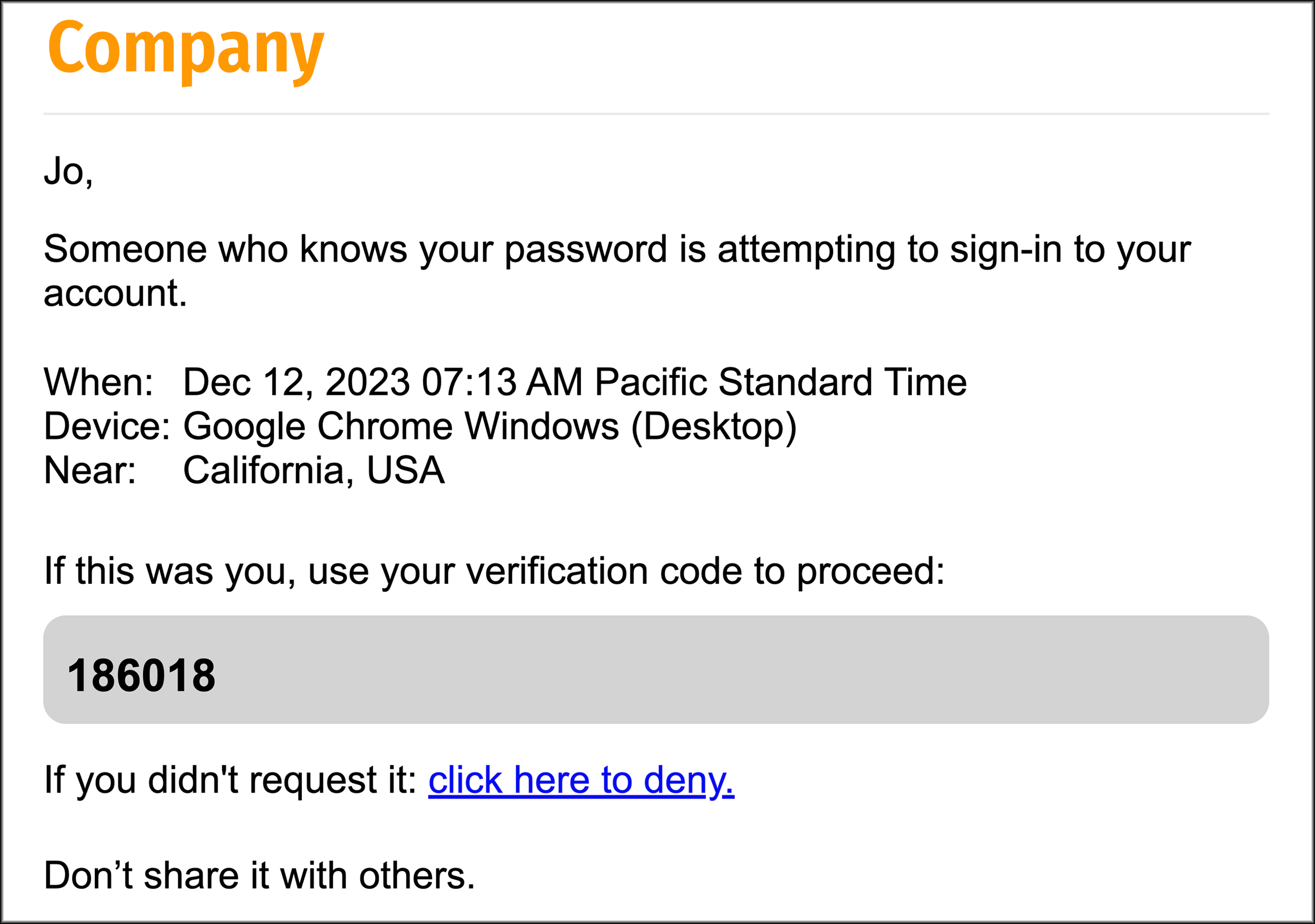}
        \caption{\textbf{Additional Challenge} from Amazon.}
        \label{fig:example-challenge-notification}
    \end{subfigure}
    \hspace{.25em}
    \begin{subfigure}[b]{0.315\textwidth}
        \centering
        \includegraphics[width=\textwidth]{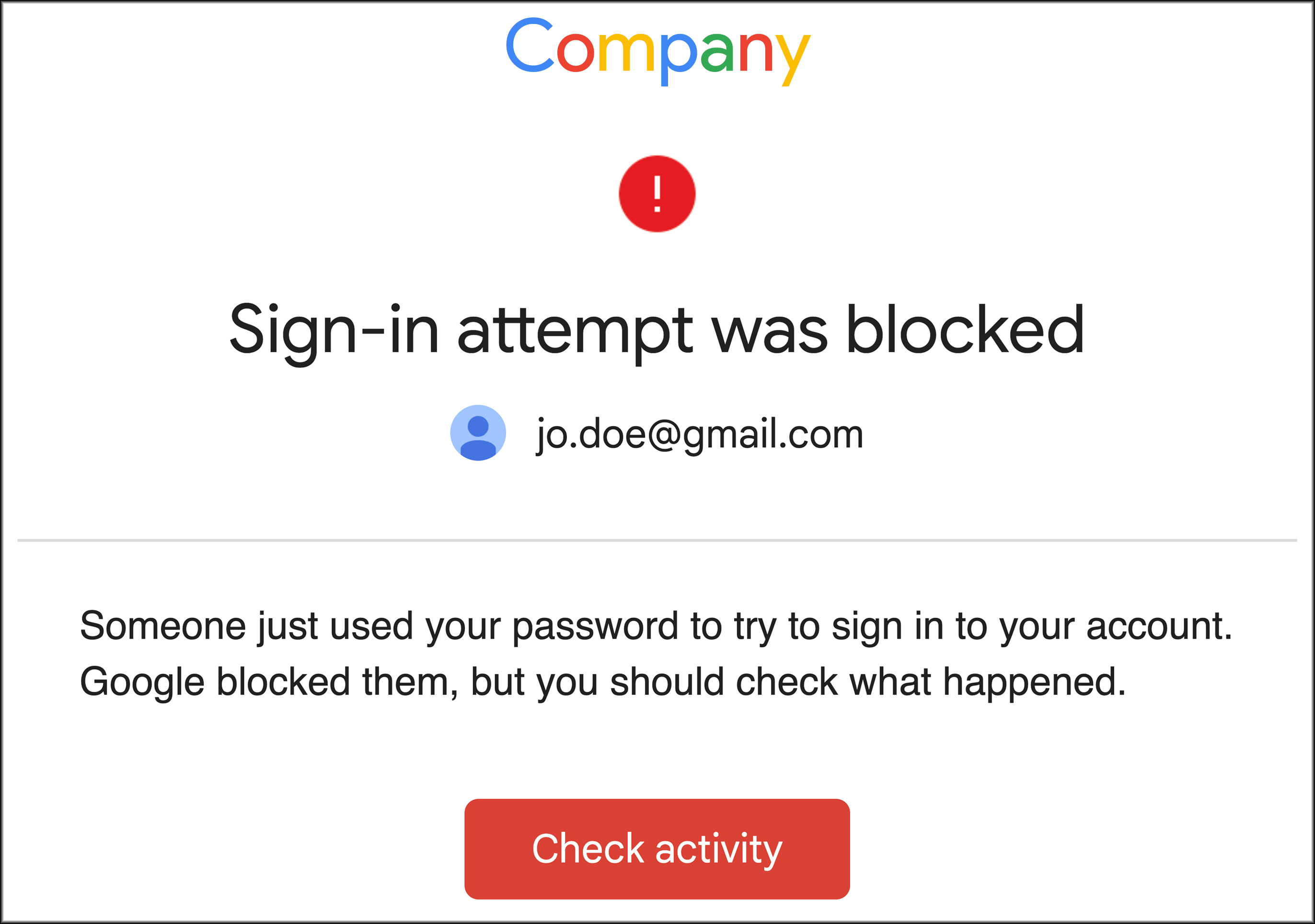}
        \caption{\textbf{Blocked Access} from Google.}
        \label{fig:example-blocked-notification}
    \end{subfigure}
    \caption{Real-world examples of the three sign-in notification types (logos removed due to copyright, cropped, as of Dec. 2023).}
    \label{fig:examples-notifications}
    \Description[Example notifications from real-world services]{The figure contains three screenshots from real-world services. Figure a) displays an example of a granted access notification from Twitch containing login information such as timestamp, location, device, browser, and IP address as well as common email contents such as greetings. Figure b) is an example of a challenge notification from Amazon which in addition to the information displayed in Figure a) also provides a 6-digit code which the user is asked to enter on the webpage in order to complete the login. Figure c) shows a blocked access notification from Google displaying none of the information from the previous notifications but only informing users that a ``Sign-in attempt was blocked'' and providing a button to ``check activity.''}
\end{figure*}

\section{Related Work}
Next, we outline how our research extends related work.

\subsection{Risk-Based Authentication, Login~Notifications, and Account Sharing}\label{sec:RW:RBA}
Only few have studied login notifications in the context of risk-based authentication so far. A qualitative interview study ($n=67$) by Redmiles~\cite{redmiles-19-should-worry} explores the account security incident response at Facebook by interviewing users who experienced a login incident.
Unlike our work, Redmiles focused on ``secondary authentication'' notifications that prompt users to enter a code to regain access to their accounts after the account has been temporarily disabled due to suspicious account activity.
Redmiles interviewed participants from 5~countries and reported on incident-response information-seeking and mental models about attackers.
Regarding the notifications' effectiveness, Redmiles identified a lack of key information as problematic, e.g., the likelihood that the notification is about a legitimate threat.
In contrast, our work studies a different type of login notification (see Section~\ref{sec:background}).
It focuses on users' comprehension, expectations, and reaction to the notification, not on regaining access or mental models about attackers.
Markert et al.~\cite{markert-22-rba-admin} studied administrators' risk-based authentication~(RBA) configuration. Administrators are responsible for the content of the login notifications users receive.
The researchers found that the predefined notifications were often barely customized, and only a few administrators opted to disable them entirely. 
Also, participants lacked consensus about which information to include, indicating a knowledge gap.
The administrators also wished for more context and explanation to prevent phishing attacks and pointed out the inaccuracy of IP-based location estimation.
Our research helps identify key features of notifications that yield correct user comprehension.
These results can help administrators align RBA configurations with users' expectations.

A study by Doerfler et al.~\cite{doerfler-19-login-challenges} evaluated the efficacy of login challenges in preventing account takeovers, finding that up to 94\% of phishing-rooted hijacking attempts and 100\% of automated hijacking attempts can be prevented. This highlights the efficacy of login notifications in account protection and motivates the design of usable and understandably designed notifications.
Still, Gavazzi et al.~\cite{gavazzi-23-rba-availability} found that only about 20\% of popular websites employ risk-based measures.
Wiefling et al.~\cite{wiefling-19-rba-in-the-wild} showed that verification codes sent via email are the de~facto standard for login challenges enforced by RBA.
In a subsequent study, they demonstrated that providing this code in the subject of the notification can reduce the login time~\cite{wiefling-20-rba-evaluation}.
Using account login notifications, Wardle~\cite{wardle-19-get-owned} measured the time it takes for leaked credentials to be abused by creating accounts on web services and intentionally leaking the credentials online. 
Adding to this literature, our research contributes insights on concrete user behavior, identifying key success features to deepen the understanding of the notifications' efficacy.

Shared passwords and accounts are of particular concern when it comes to login notifications.
When multiple individuals access the same account, the intended account owner might find it challenging to maintain control and recognize logins.
In this context, Obada-Obieh et al.~\cite{obada-obieh-20-account-sharing} investigated online account sharing and found that users struggle to remember which accounts they share and with whom.
Similarly, Song et al.~\cite{song-19-sharing-workplace} studied account-sharing practices in the workplace and observed conflicts over simultaneous access and difficulties controlling access.
While account sharing is out of scope in our research, it can have influence on users' understanding of notifications which we also address in our discussion.

\subsection{Security Warning \& Notification Design}
There is a large body of literature on security warning design~\cite{bauer-13-warning-guidelines, reeder-18-warning-reaction, walkington-19-better-warnings}.
The most prominent applications are notifications in the context of TLS~\cite{akhawe-13-alice, felt-15-ssl-warnings}, phishing~\cite{petelka-19-phishing-warnings}, malware~\cite{almuhimedi-14-malware-warning}, and cookie banners under GDPR~\cite{biselli-24-cookies, utz-19-uninformed-consent, krisam-21-dark-german500, nouwens-20-dark-patterns}, as well as warnings for developers~\cite{gorski-20-pd-for-crypto-apis} or for countering misinformation~\cite{kaiser-20-warning-disinformation}.
For user authentication, there is work on breach notifications~\cite{huh-17-linkedin, zou-19-might-be-affected}, 
password-reuse notifications~\cite{golla-18-reuse-notification, thomas-19-pw-checkup}, notifications to promote the use of 2FA~\cite{redmiles-17-2fa-msg-design, golla-21-2fa-adoption}, or FIDO2~\cite{lassak-21-webauthn-misconceptions}, or protect users from using common PINs~\cite{markert-21-pin-unlock}.

While considering the best practices for notification design in other domains is important, this is not the main focus of our study.
However, in our notification analysis (see Section~\ref{sec:background}), we try to identify common design patterns.
\section{Login Notifications in the Wild}\label{sec:background}
Login notifications intend to inform users about recent sign-ins and often include technical details such as the login time, used device, or approximate sign-in location.
However, depending on the service, they are not sent for every login.
While theoretically significant location or device changes trigger notifications, the probabilistic nature involving factors like sign-in history and user behavior makes it difficult to predict when notifications are sent.
Some services sent notifications for every login; others only sent notifications in case of significant location and device changes, causing a higher risk level.
For example, we noticed receiving fewer sign-in notifications if the affected account had two-factor authentication enabled.
Interestingly, the cause for receiving a login notification is not always transparent to the user.
We encountered multiple instances where notifications were not triggered by the account owner logging into their account.
Most commonly, the phenomenon of unexpectedly receiving a login notification is related to shared accounts (i.e., Netflix or Amazon)~\cite{angelini-23-netflix-who-logged-in} but is also known from third-party apps or services automatically signing into an account on behalf of the user~\cite{1password-22-new-login-email, redfox-22-netflix-new-device}.

\subsection{Notification Types}\label{sec:background:notification-types}
Based on the type of information they convey, notifications can be divided into three different types~\cite{markert-22-rba-admin}. Examples of each of them are shown in Figure~\ref{fig:examples-notifications}.
\aptLtoX[graphic=no, type=html]{
\begin{enumerate}
    \item[(1)]{} \textbf{Granted Access:} The notification informs about \textit{granted access}. Some services send such notifications for every sign-in, while others follow a risk-based approach.

    \item[(2)]{} \textbf{Additional Challenge:} These notifications inform about a new sign-in attempt for which an \textit{additional challenge} needs to be solved (i.e., insert a code or click a link).

    \item[(3)]{} \textbf{Blocked Access:} The notification informs users about \textit{blocked access}, which can happen because the risk-based authentication system ranks the sign-in as too risky.
\end{enumerate}
}{
\renewcommand{\labelenumi}{(\theenumi)}
\begin{enumerate}[nolistsep,leftmargin=1.7em]
    \item \textbf{Granted Access:} The notification informs about \textit{granted access}. Some services send such notifications for every sign-in, while others follow a risk-based approach.

    \item \textbf{Additional Challenge:} These notifications inform about a new sign-in attempt for which an \textit{additional challenge} needs to be solved (i.e., insert a code or click a link).

    \item \textbf{Blocked Access:} The notification informs users about \textit{blocked access}, which can happen because the risk-based authentication system ranks the sign-in as too risky.
\end{enumerate}
}

For the remainder of this work, we focus on the first type, i.e., notifications informing the user about \textit{granted access}.
This popular notification type is deployed by well-known companies such as Alphabet~\cite{alphabet-24-login-notification}, Amazon~\cite{amazon-24-login-notification}, Apple~\cite{apple-24-login-notification}, Meta~\cite{meta-24-login-notification}, and Microsoft~\cite{microsoft-24-login-notification}.
Every organization can send this type, as it does not require an advanced risk assessment (i.e., basic logic and the ability to display login details are enough).
Moreover, we limited our dataset to email-based notifications.
While notifications can also be sent via other channels, e.g., SMS or push notifications, establishing them requires additional effort.

\subsection{Analysis Method}\label{sec:background:analysis}
To familiarize ourselves with the state of the art of granted access notifications, we collected over $90$ login-related emails from real-world services by enumerating over $500$ existing accounts.
To trigger the notification, we signed in using the Tor browser, which is often classified as suspicious activity, and monitored our inbox.
We also searched through account remediation pages~\cite{netflix-23-faq} and community support forums~\cite{apple-22-forum} and learned about them via friends and colleagues.
In both cases, we created an account on the service to obtain a notification.
Our collection is limited to the top Tranco list~\cite{lepochat-19-tranco} websites (as of June~2023), with about \(\frac{1}{3}\) being in the top~$100$/$1,000$/$50,000$ respectively.
The dataset includes popular websites from social media, streaming, shopping, finance, travel, email, and gaming services.
Most of them are US-based (44) and the rest are from Europe (18) and Asia (10).
The dataset is biased towards English notifications; few non-English notifications have been translated.
The full list can be found in Appendix~\ref{app:real-world:emailmetadata} and \ref{app:real-world:features}.

For the analysis, two authors categorized $72$ emails as \emph{granted access} notifications.
The authors then independently analyzed the notifications based on a set of features derived following an iterative coding approach until no new codes and themes emerged~\cite{birks-22-grounded-theory,urquhart-13-grounded-theory}.
In particular, the authors checked which sign-in information the notification includes (i.e., login time, location, device), what the main components are (i.e., headline, malicious instructions), salient design and wording decisions (i.e., logo, highlighting of sign-in details, neutral language), and metadata such as sender and subject.
Conflicts were resolved when they emerged by consensus discussion with a third member of the team (resulting in a hypothetical final agreement of 100\%).

\begin{figure}[!t]
    \centering
    \includegraphics[width=0.89\columnwidth]{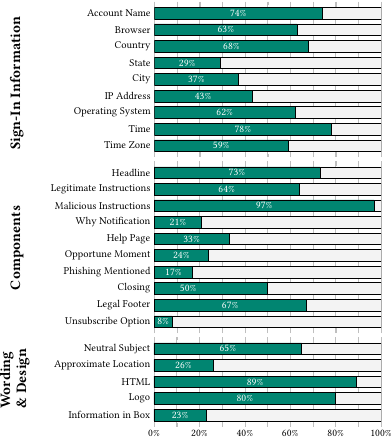}
    \caption{The information included in login notifications for granted access notifications ($n=72$) sent by real-world services.}
    \label{fig:notification-statistics}
    \Description[Barplot with percentages of elements of login notifications]{The figure displays a barplot representing all identified components of the 72 analyzed login notifications. Results correspond to the text in Section ``3.3 Findings of Notification Analysis.'' All percentages displayed in this graphic are mentioned in the text.}
\end{figure}
\vspace{5pt}
\begin{figure}[!t]
    \centering
    \includegraphics[width=0.89\columnwidth]{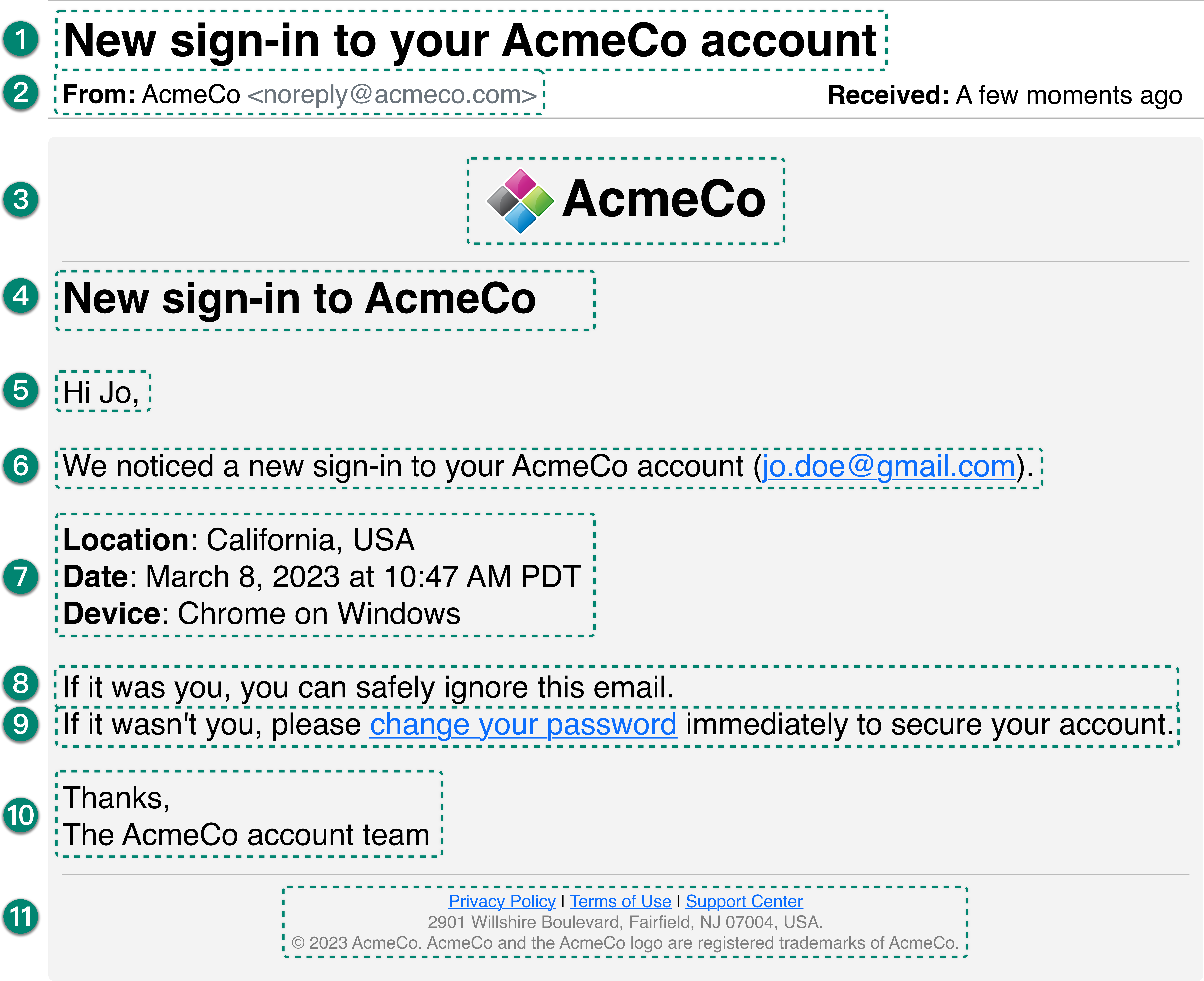}
    \caption{The \emph{baseline} login notification, which we derived from ($n=72$) real-world notifications. For our user study, we rebranded the text and the look to match the study website.}
    \label{fig:baselineemail}
    \Description[Baseline notification]{This figure displays the baseline notification. All components are marked with numbered boxes (1-11) that correspond to numbers in the text of Section ``3.3 Findings of Notification Analysis.''}
\end{figure}

\subsection{Findings of Notification Analysis}\label{sec:background:findings}
We summarize our findings in Figure~\ref{fig:notification-statistics}.
Please refer to Appendix~\ref{app:real-world:emailmetadata} and \ref{app:real-world:features} for the full details.

\textbf{Sign-In Information}
As depicted in Figure~\ref{fig:notification-statistics}, the majority of notifications included the login \includegraphics[height=0.015\textwidth]{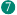}~\emph{Time}~($78\%$), \includegraphics[height=0.015\textwidth]{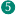}~\emph{Account Name}~($74\%$), \includegraphics[height=0.015\textwidth]{7.pdf}~\emph{Country}~($68\%$), \includegraphics[height=0.015\textwidth]{7.pdf}~\emph{Browser}~($63\%$), and  \includegraphics[height=0.015\textwidth]{7.pdf}~\emph{Operating System}~($62\%$).
Less frequently, the notifications included the \includegraphics[height=0.015\textwidth]{7.pdf}~\emph{Time~Zone}~($59\%$) or a login \emph{IP~Address}~($43\%$).
The small number of notifications, including the login \emph{City}~($37\%$) or \emph{State}~($29\%$), is explained by geographical differences between the U.S. and Europe.
For our dataset, we collected notifications from different sources and observed that notifications for logins in the U.S. mostly reported the state. Notifications for logins from Europe often also included a city.

\textbf{Components} 
Most notifications made use of a \includegraphics[height=0.015\textwidth]{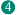}~\emph{Headline}~($73\%$) that was often~($76\%$) different from the email subject.
Another critical component were the instructions describing how users should respond to the notification.
While only $64\%$ provided instructions in the \includegraphics[height=0.015\textwidth]{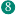}~\emph{Legitimate} case, more than $97\%$ explained how to react in the \includegraphics[height=0.015\textwidth]{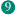}~\emph{Malicious} case if the user does not recognize the login.
The large majority ($66\%$) recommended changing the password. 
Fewer (high-ranked) web services included a button to report the login as malicious or legitimate on a separate web page ($9\%$) displaying account remediation steps.
Similarly, a small number ($9\%$) suggested to visit the account activity page.
Prominent among financial services was the option to contact support ($4\%$).
A dedicated \emph{Why Notification} component was included in $21\%$ of the notifications.
It primarily creates context and explains to users why they received the notification.
It often gives examples of legitimate (i.e., new device) and malicious causes (``someone unauthorized gained access'') that might have triggered the notification.
$33\%$ included a link to a dedicated \emph{Help Page} (note: \emph{regular} support links in the email footer were not counted).

About $24\%$ of the emails tried to use the \emph{Opportune Moment} to tell the user about other options to secure their account (i.e., enabling 2FA).
The dangers of \emph{Phishing} and methods to double-check the legitimacy of the notification were mentioned in $17\%$ of the emails, with the most prominent suggestion being not to click the ``change password'' link and instead sign in to the website by manually pasting or typing in the URL.
About half of the notifications included a \includegraphics[height=0.015\textwidth]{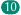}~\emph{Closing}~($50\%$) text that often thanked the user and included the name of a ``\{service\} account team.''
A footer with \includegraphics[height=0.015\textwidth]{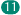}~\emph{Legal} information was included in $68\%$ of the emails, and an \emph{Unsubscribe} link was present in $8\%$ of the notifications.

\textbf{Wording \& Design}
Using affinity diagramming, we identified the wording of most email subjects as \emph{Neutral}~($65\%$), with a strong focus on ``\emph{New login to \{service\}}.'' 
In some cases, it is alarming~($23\%$), like ``\emph{Security alert}'' or a prompt~($9\%$), like ``\emph{Please review this sign in!}.'' 
In two cases, it was a question~($3\%$), e.g., ``\emph{Did you recently sign into \{service\}?}.''
Almost all emails ($92\%$) referred to \includegraphics[height=0.015\textwidth]{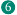}~``your account'' to emphasize the importance of the notification.
A few services tried to address the inaccuracies of IP-based location estimation by describing it as \emph{Approximate}~($26\%$).
Most notifications~($89\%$) were sent as \emph{HTML} emails; the rest were sent in plaintext.
For a ``corporate look-and-feel,'' $80\%$ of all notifications included a \includegraphics[height=0.015\textwidth]{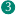}~\emph{Logo}, with an even split between a centered or left alignment.
Interestingly, $23\%$ of the emails displayed the sign-in information in a visually detached box, most likely to draw the user's attention to the login details.

\textbf{Subject \& Sender}
We identified three different types of email subjects:
\aptLtoX[graphic=no, type=html]{
(a) The majority of email subjects~($79\%$) did not include specific details and were relatively generic, e.g., ``Your account has been logged into'' (Tumblr). (b) A small fraction~($15\%$) made use of login metadata, e.g., ``New login to Twitter from \{browser\} on \{OS\}'' (X, formerly Twitter). (c) Only a few~($6\%$) included the account name, e.g., ``\{Name\}, did you recently sign into Etsy?'' (Etsy). Five services did not use an email sender name (display name).
}{
\begin{enumerate*}[label=(\alph*)]
    \item The majority of email subjects~($79\%$) did not include specific details and were relatively generic, e.g., ``Your account has been logged into'' (Tumblr).
    \item A small fraction~($15\%$) made use of login metadata, e.g., ``New login to Twitter from \{browser\} on \{OS\}'' (X, formerly Twitter).
    \item Only a few~($6\%$) included the account name, e.g., ``\{Name\}, did you recently sign into Etsy?'' (Etsy). Five services did not use an email sender name (display name).
\end{enumerate*}
}

\textbf{Technical Details}
Throughout our analysis, we found numerous areas for improvement regarding the parsing and displaying of technical details such as location data, browser, and OS.
We found incorrectly escaped HTML ``\&Icirc;le-de-France'' instead of ``Île-de-France,'' empty placeholders, e.g., ``Browser: N/A,'' and cryptic smartphone model numbers ``SM-S908B/DS'' instead of accessible names like ``Samsung Galaxy S22.'' 
Operating systems were often reported with their full version number, e.g., ``iOS 17.1.2.''
We even found four~notifications that included the raw User-Agent string.

\textbf{Questionable Advice}
Some notifications included information about phishing, which does not always align with state of the art recommendations on those topics.
Questionable advice is given by X (and three other major services), which suggests that the presence of a padlock icon will ``let you know a site is secure'' and that users should check for the presence of ``https://'' and ``\{domain\}'' in the hyperlink.
Similarly, Amazon suggests better copying and pasting the ``It wasn't me''-link into a browser ``just to be safe.''
Spotify advises users to verify that the email was sent from ``@spotify.com,'' which is only expedient if the email server and DNS are configured correctly.
In line with the latest research, PayPal's advice~\cite{paypal-22-phishing} explicitly mentions to ``not rely on the padlock symbol and the `s' in HTTPS''.
Interestingly, LinkedIn added a security footer message~\cite{linkedin-22-security-footer} to their login notification that includes the affected account name and corresponding profession to authenticate official emails.

\subsection{Selecting a Representative Notification}
For our user study (see Section~\ref{sec:measurment-method}), we aimed to use a notification that closely resembles the state of the art of real-world login notifications.
Our data-driven \emph{baseline} (see Figure~\ref{fig:baselineemail}) includes all components used by at least $50\%$ of the analyzed notifications, leading to 11 components comprising the notification:
It uses a neutral subject and a slightly modified headline.
We adjusted the email sender, opted for an HTML email, and included a logo.
We also mentioned the affected account name and referred to ``your account.''
We listed the most popular sign-in details and legitimate and malicious instructions with actions for users to take after receiving the notification.
Our study sample was U.S.-based, so we included the \includegraphics[height=0.015\textwidth]{7.pdf}~\emph{State} in the sign-in details.
The email also had a closing and footer with fictional legal information.

By deriving a representative notification, we could test users' general understanding , their reaction, and their perceptions.
While the majority of the 72 collected notifications included slightly fewer components (57\% include at least 9 components) we still decided to include all 11 components in the \textit{baseline} to be able to make a statement about each components' usefulness in an idealized scenario. 
Components omitted in real-world notifications most often were the \includegraphics[height=0.015\textwidth]{5.pdf}~\emph{Account Name}, e.g., ``Hi \{Account Name\},'' and the \includegraphics[height=0.015\textwidth]{10.pdf}~\emph{Closing}, e.g., ``Thanks, The AcmeCo account team.''
\section{Method}\label{sec:measurment-method}

The following section outlines the protocol, treatments, recruitment, ethics, and limitations of our user study.

\subsection{Study Protocol}
\aptLtoX[graphic=no, type=html]{
Participants in this study should receive a notification for a concrete account. 
To resemble a real-world setting, the protocol had to fulfill four criteria: (1) A \emph{real account} gets created, (2) participants are \emph{unaware} that the study is about login notifications, (3) participants receive the notification in their \emph{personal email account}, (4) and reactions to login notifications are \emph{measurable}.
}{
Participants in this study should receive a notification for a concrete account. 
To resemble a real-world setting, the protocol had to fulfill four criteria: \begin{enumerate*}
    \item A \emph{real account} gets created,
    \item participants are \emph{unaware} that the study is about login notifications,
    \item participants receive the notification in their \emph{personal email account},
    \item and reactions to login notifications are \emph{measurable}.
\end{enumerate*}
}

For this, we invited participants to take part in a multi-stage study about changes in the cognitive ability of mental rotation over time~\cite{shepard-71-mental-rotation, vandenberg-78-mental-rotation}.
This framing allowed us to inform people about the length of the commitment without revealing our interest and justified the account creation on our website.
The task was also a strong cognitive distractor that prevented participants from paying too much attention to the notification and authentication task.

We used two treatments, and the baseline notification (see Figure~\ref{fig:baselineemail}) was adapted to the branding of our study's website:
The legitimate group ($n=110$) received a notification only after they logged in.
The location, date, and device in the notification were derived from the metadata of their login. 
The malicious group ($n=119$) received a notification unexpectedly at a time when they had not interacted with the account for multiple days.
This resembled a login attempt by a malicious actor.
The location (``California, USA'') and device (``Chrome on Windows'') were selected to have the highest statistical chance of matching any user in our U.S.-based sample~\cite{us-22-population, statcounter-22-browser-market, statcounter-22-os-market}.
Trawling (untargeted) attackers would likely use a similar configuration when signing into the account, to minimize the risk of being detected by RBA systems~\cite{doerfler-19-login-challenges}.
Because the details are intentionally chosen to closely align with common user configurations -- just like in real life -- some participants in our study (10 of 119) also matched these details, making it more challenging for them to recognize that the notification is a result of a malicious login.
We did not allow mobile devices and recorded if the email was opened via a small self-hosted image, which itself was invisible in the email.

\noindent \textbf{Stage~1:}\hspace{.5em} The first page on the study's website explained the mental rotation test.
To ensure participants regularly check their email and understand the value of the account, they saw a privacy notice after giving their consent, which highlighted the importance of the account as it would be used to store the study data and email address.
It also explained that the email would be used to send invitations to subsequent stages, and the compensation would be in the form of Amazon gift cards.
After the account creation, participants solved 5~mental rotation tests and provided demographic information (\aptLtoX[graphic=no, type=html]{\textbf{MD1}--\textbf{MD4}}{\ref{app:part2:d1}--\ref{app:part2:d4}}).
At the end, participants in the legitimate treatment were informed that invitations to Stage~2 would be sent in approx. 7~days; in the malicious group, the note said 14~days. 

\noindent \textbf{Stage~2:}\hspace{.5em} After 7~days, participants in the legitimate group received an email inviting them to return to our website for another mental rotation test. 
To do so, they had to log into their account, which triggered a notification.
Participants in the malicious group expected their next email after 14~days. 
However, to imitate a malicious login, we sent them an (unexpected) login notification filled with our statistical sign-in data 7~days after they completed the first stage. 

\noindent \textbf{Stage~3:}\hspace{.5em} For the legitimate group, invitations to the final Stage~3 were sent 48~hours after they completed Stage~2; in the malicious group, 48~hours after they received a notification for a login they did not initiate. 
We chose this time frame to give participants enough time to react. 
After logging into Stage~3, participants were debriefed and told about the purpose of the study. 
This was followed by our questionnaire (see Appendix~\ref{app:part2}).
From then on, the notification we sent was shown on the left side of their screen for reference.\vspace{-1.8pt}
\aptLtoX[graphic=no, type=html]{
\begin{enumerate}
\item[(1)]{} \textit{Email:} First, we asked if participants remember receiving the notification (\textbf{MQ0}); if not, they were forwarded to a different section (see Appendix~\ref{app:part2}). Participants for whom we received a read receipt or who changed their password skipped this question.

\item[(2)]{} \textit{I-PANAS-SF:} To learn about the feelings and emotions in reaction to the notification, in \textbf{MP} we utilized the Positive and Negative Affect Schedule (I-PANAS-SF)~\cite{thompson-07-panas}.

\item[(3)]{} \textit{Reaction:} Next, we asked how thoroughly participants read the notification (\textbf{MQ1}) and how and why they chose to react to it (\textbf{MQ2a}--\textbf{MQ3a}). 
Participants who changed their password were specifically asked about any other actions (\textbf{MQ2b}--\textbf{MQ3b}). 

\item[(4)]{} \textit{Content \& Design:} To better understand the reactions, \textbf{MQ4} asked about influencing factors like metadata, content, and design.
\textbf{MQ5} specifically asked about the helpfulness of the account name, location, date, and device.

\item[(5)]{} \textit{Time \& Location:} \textbf{MQ6}--\textbf{MQ10} investigated the time when and location where the notification was read. 
With \textbf{MQ7}, we verified if the location, which had been derived automatically, was actually accurate or could have led to confusion, and \textbf{MAC2} was an attention check.

\item[(6)]{} \textit{Comprehension \& Expectation:} With \textbf{MQ11}, we captured if participants understood why they received the notification. 
\textbf{MQ12} and \textbf{MQ13} asked participants when they expect real companies to send notifications. 

\item[(7)]{} \textit{Prior Experience:} We concluded with three questions covering negative experiences with security incidents (\textbf{MQ14}), as well as their opinion on regular (\textbf{MQ15}) and event-driven password changes (\textbf{MQ16}).
\end{enumerate}
}{
\begin{enumerate}[nolistsep,leftmargin=1.2em]
\item \textit{Email:} First, we asked if participants remember receiving the notification (\ref{app:part2:s1}); if not, they were forwarded to a different section (see Appendix~\ref{app:part2}). Participants for whom we received a read receipt or who changed their password skipped this question.

\item \textit{I-PANAS-SF:} To learn about the feelings and emotions in reaction to the notification, in \ref{app:part2:panas} we utilized the Positive and Negative Affect Schedule (I-PANAS-SF)~\cite{thompson-07-panas}.

\item \textit{Reaction:} Next, we asked how thoroughly participants read the notification (\ref{app:part2:q1}) and how and why they chose to react to it (\ref{app:part2:q2a}--\ref{app:part2:q3a}). 
Participants who changed their password were specifically asked about any other actions (\ref{app:part2:q2b}--\ref{app:part2:q3b}). 

\item \textit{Content \& Design:} To better understand the reactions, \ref{app:part2:q4} asked about influencing factors like metadata, content, and design.
\ref{app:part2:q5} specifically asked about the helpfulness of the account name, location, date, and device.

\item \textit{Time \& Location:} \ref{app:part2:q6}--\ref{app:part2:q10} investigated the time when and location where the notification was read. 
With \ref{app:part2:q7}, we verified if the location, which had been derived automatically, was actually accurate or could have led to confusion, and \ref{app:part2:ac2} was an attention check.

\item \textit{Comprehension \& Expectation:} With \ref{app:part2:q11}, we captured if participants understood why they received the notification. 
\ref{app:part2:q12} and \ref{app:part2:q13} asked participants when they expect real companies to send notifications. 

\item \textit{Prior Experience:} We concluded with three questions covering negative experiences with security incidents (\ref{app:part2:q14}), as well as their opinion on regular (\ref{app:part2:q15}) and event-driven password changes (\ref{app:part2:q16}).
\end{enumerate}
}

\begin{table}[!tb]
    \centering
    \caption{Participants' demographics.}
    \label{tab:demoTwo}
    \Description[Demographics table]{Columns separate demographics by male, female, other, and total. Each column has two sub-columns, the left displaying the number of participants, the right displaying the percentages within the group. Rows first show the total numbers for age and then separated by age groups. Secondly they show total education numbers as well as separated by education level. Lastly the technical background, separated by total, yes, no, and prefer not to answer is presented.}
    {\def\arraystretch{1}
    \setlength\tabcolsep{3pt}
    \begin{tabular}{r|rrrrrr|rr}
    \toprule
    & \multicolumn{2}{c}{\textbf{Male}} & \multicolumn{2}{c}{\textbf{Female}}
    & \multicolumn{2}{c|}{\textbf{Other}} & \multicolumn{2}{c}{\textbf{Total}} \\
    & \textbf{No.} & \textbf{\%} & \textbf{No.} & \textbf{\%}
    & \textbf{No.} & \textbf{\%} & \textbf{No.} & \textbf{\%}\\
    \midrule
    \textbf{Age} & 149 & 65 & 79 & 34 & 1 & 0 & 229 & 100 \\ 
    \midrule
    18--24 & 4 & 2 & 6 & 3 & 0 & 0 & 10 & 4 \\ 
    25--34 & 17 & 7 & 15 & 7 & 0 & 0 & 32 & 14 \\ 
    35--44 & 27 & 12 & 17 & 7 & 0 & 0 & 44 & 19 \\ 
    45--54 & 24 & 10 & 15 & 7 & 0 & 0 & 39 & 17 \\ 
    55--64 & 35 & 15 & 13 & 6 & 0 & 0 & 48 & 21 \\ 
    65--74 & 31 & 14 & 11 & 5 & 0 & 0 & 42 & 18 \\ 
    75+ & 11 & 5 & 2 & 1 & 1 & 0 & 14 & 6 \\ 
    \midrule
    \textbf{Education} & 149 & 65 & 79 & 34 & 1 & 0 & 229 & 100 \\ 
    \midrule
    High School & 47 & 21 & 31 & 14 & 0 & 0 & 78 & 34 \\ 
    Trade & 39 & 17 & 12 & 5 & 1 & 0 & 52 & 23 \\ 
    Bachelor's & 34 & 15 & 24 & 10 & 0 & 0 & 58 & 25 \\ 
    Master's & 23 & 10 & 10 & 4 & 0 & 0 & 33 & 14 \\ 
    Doctorate & 4 & 2 & 2 & 1 & 0 & 0 & 6 & 3 \\ 
    Prefer not to say & 2 & 1 & 0 & 0 & 0 & 0 & 2 & 1 \\ 
    \midrule
    \textbf{Background} & 149 & 65 & 79 & 34 & 1 & 0 & 229 & 100 \\ 
    \midrule
    Technical & 10 & 4 & 25 & 11 & 0 & 0 & 35 & 15 \\ 
    Non-Technical & 134 & 59 & 53 & 23 & 1 & 0 & 188 & 82 \\ 
    Prefer not to say & 5 & 2 & 1 & 0 & 0 & 0 & 6 & 3 \\ 
    \bottomrule
    \end{tabular}
    }
\end{table}

\subsection{Recruitment \& Demographics}
We used the panel provider Cint for the recruitment of the study. They are a comparable platform to larger providers such as Respondi and Kantar, operating numerous sub-panels for different locations across the globe.
Criteria for participation were being 18 or older, being willing to participate in deception studies, and being US-based.
For Stage~1, we recruited 625~participants, about 3 times more than the desired number of completions as recommended by the panel provider.
Our a priori power analysis determined the minimum sample size to be $N = 100$ per group - in order to achieve 80\% power for detecting a medium effect (Cohen's $d = .4$), at a significance criterion of $\alpha = .05$.
After filtering~12~participants who failed the attention check (\aptLtoX[graphic=no, type=html]{\textbf{MAC1}}{\ref{app:part2:ac1}}),~613~participants remained.
At the end of Stage~3, we had 252~completions.
This high number of dropouts is almost exclusively attributed to participants who did not return after the first stage of the study.
Lastly, we removed 23 participants who provided unrelated answers or failed the second attention check (\aptLtoX[graphic=no, type=html]{\textbf{MAC2}}{\ref{app:part2:ac2}}) for a final number of $n=229$.
Stage~1 took, on average, 2.5~minutes and was compensated with \$3.00~USD.
Stage~3 took, on average, 6~minutes and was compensated with \$4.00~USD.
Participants in the legitimate group received an additional \$1.00~USD for the completion of Stage~2, which took 2~minutes on average.
Table~\ref{tab:demoTwo} shows the participants' demographics. 
Regarding the demographics, we observe a shift towards male-identifying participants (65\%).
The age distribution is diverse, ranging from 14\% to 21\% for all age groups between 25 and 74.
Most participants had a high school (33\%), Bachelor's (26\%), or trade degree (23\%) and did not have a technical background (82\%). 

\subsection{Ethical Considerations}\label{sec:measurment-method:ethics}
At the time we conducted the study, none of the authors worked at an institution with an IRB.
However, we carefully followed the guidelines provided in the Menlo~Report, including a risk-benefit evaluation, developing the protocol with peers familiar with conducting user studies and following the legal requirements.
The study included deception and sent a login notification to participants' personal email accounts, which could have caused more anxiety than just imagining to have received a login notification.

To protect participants from unnecessary risks, we implemented several safeguards:
\aptLtoX[graphic=no, type=html]{
i) Our panel provider offered the study only to participants who agreed to studies that might involve deception.
ii) The affected spatial reasoning account had no subjective value to the participants and only allowed to access the email address. Participants might have been concerned about their answers to the questionnaire, which, in their impression, were also tied to the same account.
However, at the time the malicious login notification was sent, participants had not been asked any sensitive or personal questions yet.
iii) All participants (including those who decided to withdraw early or drop out) have been debriefed. 
In particular, we told them about the true purpose of the study, and in case they belonged to the \emph{malicious} treatment that ``This sign-in did not take place; at no time was your account at risk,'' and asked them whether they prefer to leave the study early (while being fully compensated), which nobody did.
iv) We provided an optional contact address and feedback form that we closely monitored (we have not received any complaints).
v) We shared a website (also accessible from outside the study) that participants could visit and share to learn more about login notifications and related account security measures.
vi) We created a distinct email account for sending notifications that applied all state of the art email security features, which can prevent email spoofing attacks. We also allowed participants to reply to the notification and ask for assistance. Finally, all email addresses were only stored encrypted, separated from the study responses, and were deleted after the study in accordance with progressive data protection laws like GDPR and CCPA.
}{
\begin{enumerate*}[label=\roman*)]
\item Our panel provider offered the study only to participants who agreed to studies that might involve deception.
\item The affected spatial reasoning account had no subjective value to the participants and only allowed to access the email address. Participants might have been concerned about their answers to the questionnaire, which, in their impression, were also tied to the same account.
However, at the time the malicious login notification was sent, participants had not been asked any sensitive or personal questions yet.
\item All participants (including those who decided to withdraw early or drop out) have been debriefed. 
In particular, we told them about the true purpose of the study, and in case they belonged to the \emph{malicious} treatment that ``This sign-in did not take place; at no time was your account at risk,'' and asked them whether they prefer to leave the study early (while being fully compensated), which nobody did.
\item We provided an optional contact address and feedback form that we closely monitored (we have not received any complaints).
\item We shared a website (also accessible from outside the study) that participants could visit and share to learn more about login notifications and related account security measures.
\item We created a distinct email account for sending notifications that applied all state of the art email security features, which can prevent email spoofing attacks. We also allowed participants to reply to the notification and ask for assistance. Finally, all email addresses were only stored encrypted, separated from the study responses, and were deleted after the study in accordance with progressive data protection laws like GDPR and CCPA.
\end{enumerate*}
}
\vspace{-7pt}
\subsection{Limitations}\label{sec:measurment-method:limitations}
For this study, we relied on a controllable artificial account setting, which might lack ecological validity. 
However, only 7~participants mentioned the non-real-world setting as a reason for not reacting to the notification.
We expect more participants to change their password if the notification was sent for an account with a higher subjective value.
We did not control for VPN usage, which might also slightly influence results in real-world settings.
Like many human-subject studies, there is the potential for a bias in question-wording.
To circumvent this, we piloted the study and tried to keep the questions short and clear.
The full survey instrument can be found in Appendix~\ref{app:part2}.
Lastly, we only recruited US-based participants, which can have culture-based influences on the results.
\section{Results}\label{sec:measurement-results}
\begin{figure}[!ht]
    \centering
    \includegraphics[width=\columnwidth]{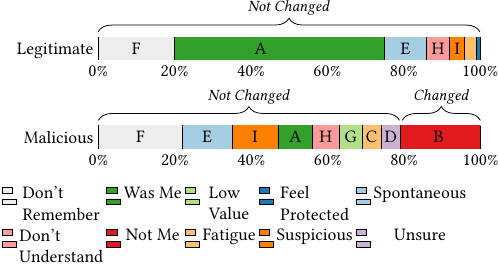}
    \caption{Breakdown of treatments into participants who \textit{have} or \textit{have not} changed their password and their reasoning.}
    \label{fig:StudyTwoBreakdown}
    \Description[Stacked barplot of reasoning for reaction]{Two barplots are displayed. The one at the top represents the legitimate group, showing which reasoning they provided for their reaction to receiving the login notification. The plot at the bottom represents the malicious groups' reasonings for their reactions. Reasons of those who did not change their password are represented on the left, reasons of those who changed the password are represented on the right.}
\end{figure}

Next, we present the results of the study.
The qualitative coding was done by two of the researchers, who started by separately coding 10\% of the answers.
Afterward, they agreed on a joint codebook (see Tables~\ref{tab:codebook:study2-rq1}--\ref{tab:codebook:study2-rq3} in Appendix~\ref{sec:codebook}) and used it to code the remaining 90\%.
The agreement between the two coders was high ($\kappa = 0.77$).
When quoting individual participants, e.g., {\color{pinegreen-rba}L}61-{\color{royalblue-rba}N}, one can derive their treatment (\textit{\textbf{\color{pinegreen-rba}L}egitimate} or \textit{\textbf{\color{pinegreen-rba}M}alicious}) and password change behavior (\textit{\textbf{\color{royalblue-rba}N}o Change} or \textit{\textbf{\color{royalblue-rba}C}hange}).
Similarly, we use color codes like \legendboxBlack{was_me}{A}~\textit{Was~Me} corresponding to Figure~\ref{fig:StudyTwoBreakdown} within the text when referring to participants' explanations.

\subsection{RQ1: Reaction \& Comprehension} 
\label{sec:study2-comprehension-and-reaction}

\textbf{Reaction General}
Out of the total 229 participants, 48 participants, 23 in the legitimate and 25 in the malicious treatment, \legendboxBlack{dont_remember}{F}~cannot remember the notification. Still, for 26 of them, we received a read receipt, so they must have at least opened the notification. 
Among the large majority of participants who saw the notification (181; 79\%), it was very rare that they completely ignored its content.
In response to \aptLtoX[graphic=no, type=html]{\textbf{MQ1}}{\ref{app:part2:q1}}, just 6\% said that they only read the subject. 
About 90\% read the notification completely or at least skimmed the body. 

\textbf{Reaction Legitimate}
No participant in the legitimate treatment changed their password.
As shown in Figure~\ref{fig:StudyTwoBreakdown}, the majority of participants (60; 55\%) explained their reaction in response to~\aptLtoX[graphic=no, type=html]{\textbf{MQ3b}}{\ref{app:part2:q3b}} by saying \legendboxBlack{was_me}{A} it was their own login. Another 12, i.e., 11\%, described it as a \legendboxBlack{spontaneous}{E} \textit{spontaneous} reaction, e.g., M42-N: ``I just didn't think much of it.''
We also see that some participants do not understand what the notification is saying, which was the driving reason for \legendboxBlack{didnt_understand}{H}~6\% (6) to ignore it.
Finally, we recorded themes of participants who were \legendboxBlack{suspicious}{I} \textit{suspicious} about the legitimacy of the notification (4; 4\%), felt \legendboxBlack{fatigue}{C}~\textit{fatigued}~(3;~3\%), or \legendboxBlack{feel_protected}{J} \textit{protected} (3; 3\%).

\textbf{Reaction Malicious}
In the malicious group, only 26 of the 119 participants, i.e., 22\%, changed their password; all of them correctly saying \legendboxBlack{someone_else}{B} it was not them logging in. 
The reasons given by the other 78\% (93) in response to~\aptLtoX[graphic=no, type=html]{\textbf{MQ3b}}{\ref{app:part2:q3b}} for not changing their password mostly overlapped with responses given by participants in the legitimate treatment: \legendboxBlack{spontaneous}{E} \textit{spontaneous} reaction (15; 13\%), notification looked \legendboxBlack{suspicious}{I} \textit{suspicious} (14; 12\%), or was \legendboxBlack{didnt_understand}{H} \textit{not understood} (8; 7\%), \legendboxBlack{fatigue}{C} being \textit{fatigued} (6; 5\%) or \legendboxBlack{unsure}{D} \textit{unsure} how to react (6; 5\%).
Finally, there are two justifications that are owed to the study design: participants describing they \legendboxBlack{was_me}{A} logged in themselves although they did not (11; 9\%), likely an example of social desirability, and those who assigned a \legendboxBlack{low_value}{G} low value to the account (7; 6\%):
\begin{myquote}
``\emph{This account has no value, it was not a streaming or banking account or amazon account}''~(M74-N)
\end{myquote}
This justification can be reasonable, but users need to keep in mind that an attacker can also target other accounts that verbatim or partially reused the compromised password~\cite{pal-19-personal-psm}.

\textbf{Comprehension}
When asked why they have received the notification (\aptLtoX[graphic=no, type=html]{\textbf{MQ11}}{\ref{app:part2:q11}}), 85\% (93) in the legitimate and 79\% (94) of the participants in the malicious treatment realized that a new login happened. Very few who gave a different explanation believed it was a phishing attempt (3; 1\%), most simply did not understand what has happened at all (39;~17\%):
\begin{myquote}
``\emph{I had no idea, which is why I deleted it.}''~(M93-N)
\end{myquote}
Those in the legitimate treatment who mapped the notification to a new login usually perceived it as a simple info email (42; 38\%), followed by those who saw it as a prompt to review the login (28; 26\%). Fewer responses (15; 13\%) explicitly mention that the login must have been abnormal. In the malicious treatment, most participants who understood that a new login happened described that they were (potentially) compromised (46; 36\%). Another 19\% (22) perceived it as an informative but non-critical email. The remainder (13; 11\% each) either mentioned that the system rated the login as unusual or wants them to review the login.

We observed a low comprehension of what might have caused the notification, especially in the malicious group. One explanation might be the temporal connection between logging in and receiving the notification. From \aptLtoX[graphic=no, type=html]{\textbf{MQ6}}{\ref{app:part2:q6}}, we know that about two-thirds read it immediately, and most of the others within a few hours. Hence, participants in the legitimate treatment had indeed a connection, and their understanding was substantially better. 
This influence of contextual factors was already observed by prior work on warning design~\cite{reeder-18-warning-reaction, gorski-20-pd-for-crypto-apis} and could be achieved by including a \emph{Why Notification} section. Some websites already do (see Section~\ref{sec:background}), and we will elaborate on this in the discussion.\\

\begin{shaded}
\noindent{\textbf{Summary}} {About 80\% saw the notification.
Participants in the legitimate treatment who triggered it themselves understood what it was telling them and reacted accordingly.
In the malicious treatment where participants did not have this context, only 22\% changed their password, and they had more difficulties explaining the circumstances.
Hence, the number of password changes in the malicious treatment is substantially lower than expected.}\end{shaded}

\subsection{RQ2: Decision-Making \& Execution} 
\label{sec:study2-decision-and-execution}

We now focus on the decision-making process to understand if participants struggle with determining whether it was them or not, especially for malicious logins.

\begin{figure}[!ht]
    \centering
    \includegraphics[width=1.0\columnwidth]{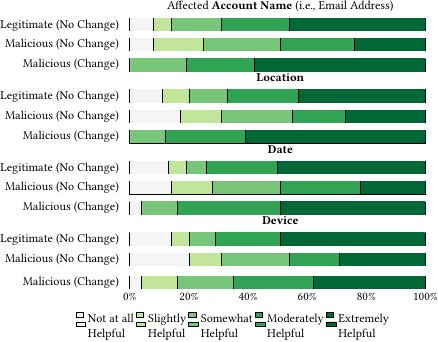}
    \caption{Helpfulness of the details for deciding (\ref{app:part2:q5}).}
    \label{fig:helpfulnessInformation}
    \Description[Stacked barplots for helpfulness of information]{Stacked barplots representing the helpfulness of specific information in the login notification. There are 12 bars in total, separated into 4 sections. Section 1 represents the helpfulness of the affected account name, section 2 represents location, section 3 the date, and section 4 the device. All sections show three separate bars. The first for the legitimate group, the second for malicious (no change), and the third for malicious (change). Answer choices are represented with different colors ranging from ``not at all helpful'' on the left to ``extremely helpful'' on the right.}
\end{figure}

\textbf{Helpfulness of Login Information}
Foremost, we wanted to get insights into the helpfulness of the displayed login information~(\aptLtoX[graphic=no, type=html]{\textbf{MQ5}}{\ref{app:part2:q5}}).
In Figure~\ref{fig:helpfulnessInformation}, we can see that for those in the \textit{Legitimate} and \textit{Malicious (Change)}
group, all information is about equally helpful: 22--35\% find the different types \textit{moderately} and 38--62\% even \textit{extremely} helpful. 
Participants in the \textit{Malicious (No Change)} group, in contrast, appear to have a less distinct opinion as ratings are more equally distributed, ranging from 8--30\%.
A Kruskal-Wallis test also showed significant differences for all types of information when comparing \textit{Malicious (No Change)} to \textit{Legitimate} and \textit{Malicious (Change)}, respectively.
This uncertainty of participants in the \textit{Malicious (No Change)} group regarding the displayed information aligns with the previous section, where we found that those participants misattributed or did not understand the cause of the notification.

\begin{figure}[!ht]
    \centering
    \includegraphics[width=0.85\columnwidth]{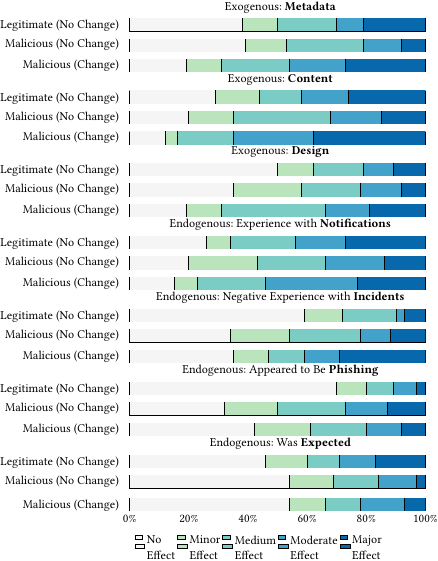}
    \caption{Influence of factors on participants reaction (\ref{app:part2:q4}).}
    \label{fig:effectReaction}
    \Description[Stacked barplots for influence of factors on reaction]{Stacked barplots representing the influence of different factors on participants' reactions. There are 21 bars in total, separated into 7 sections. Section 1 represents the influence of metadata, section 2 represents content, and section 3 the design of the notification. Section 4 represents prior experience with such notifications, section 5 represents negative experience with security incidents. Section 6 shows the influence of whether participants thought the notification was phishing and section 7 whether the notification was expected. All sections show three separate bars. The first for the legitimate group, the second for malicious (no change), and the third for malicious (change). Answer choices are represented with different colors ranging from ``no effect'' on the left to ``major effect'' on the right.}
\end{figure}

\textbf{Effect of Other Factors}
In addition to the already-known influence of the login information, we were also interested in the effect of other exogenous and endogenous factors~(\aptLtoX[graphic=no, type=html]{\textbf{MQ4}}{\ref{app:part2:q4}}). Figure~\ref{fig:effectReaction} gives an overview. 
Generally speaking, the content (e.g., provided information, instructions, wording) and prior experience in dealing with such notifications had the highest effect on participants' reactions, with 42\% expressing a \textit{moderate} or \textit{major effect} on average. 
Followed by that is the metadata (e.g., sender, subject, time of arrival) with 29\%.
All other factors seemed to have a weaker influence, with only 18\% (appeared to be phishing) to 23\% (was expected) of the participants reporting a \textit{moderate} or \textit{major effect}.

When comparing groups, \textit{Legitimate} is the one where most participants reported a factor having no effect. The \textit{Malicious (Change)} group, on the other hand, is the one where participants describe the strongest influences of the factors. Using a Kruskal-Wallis test with Bonferroni-correction for pairwise comparisons, we found that metadata had a significantly higher effect for \textit{Malicious (Change)} participants compared to \textit{Malicious (No Change)} participants ($\chi^{2}(2) = 6.65, p < 0.05$). 
The same is true for the email content ($\chi^{2}(2) = 7.73, p < 0.05$). 
Thus, to nudge more users to change their password upon receiving potentially malicious login notifications, focusing on designing the content and metadata is vital.

\textbf{Influence of Negative Experiences}
Overall, 30\% of participants described falling victim to a security breach within the last two years (\aptLtoX[graphic=no, type=html]{\textbf{MQ14}}{\ref{app:part2:q14}}). 
In the malicious treatment, 42\% of those who changed their password reported prior negative experiences. Only 32\% of those who did not change their password said so. 
The difference is not statistically significant, $\chi^{2}(2) = 2.61, p = 0.271$, but suggests that prior breach experience increases the likelihood of users changing their password upon receiving a notification.\\

\begin{shaded}
\noindent{\textbf{Summary}} {When comparing login information side-by-side, we can conclude that all factors are equally essential. We also observed that the helpfulness of the information for the \textit{Malicious (No Change)} participants is significantly lower, which further explains the issues of this group when determining what happened. 
Regarding other factors, the content of the notification, its metadata, and prior experience in dealing with it had the highest effect across all treatments. Negative experience tends to influence the reaction as well; other aspects appeared to be less crucial.}\end{shaded}

\subsection{RQ3: Perception \& Expectation} 
\label{sec:study2-perception-and-expectation}

\textbf{Perception}
The PANAS (\ref{app:part2:panas}) reveals that participants who changed their password feel more positive but also more negative. The average positive \emph{affect} of the \textit{Malicious (Change)} group is 15.0 (SD: 4.5) but only 11.6 (SD: 5.0) and 12.6 (SD: 5.6) for the \textit{Malicious (No Change)} 
and \textit{Legitimate}, respectively.
Using a Kruskal-Wallis test (Bonferroni corrected), we were also able to confirm the significance between the two malicious groups, $\chi^{2}(2) = 8.29, p < 0.05$.
For the negative \emph{affect}, \textit{Malicious (Change)} averages 9.8 (SD: 4.1), \textit{Malicious (No Change)} 8.1 (SD: 4.3), and \textit{Legitimate} 5.7 (SD: 1.6). 
Again, Kruskal-Wallis was used yielding significance between both malicious groups and \textit{Legitimate} ($p < 0.01$); the comparison between the two malicious groups nearly did, $\chi^{2}(2) = 5.26, p = 0.0654$. 

\textbf{Expectation}
So far, the study showed that there is a substantial number of participants who have not changed their password although they should, some of them mentioning that it was a spontaneous reaction, which this fatigue may also explain.
Hence, we used \aptLtoX[graphic=no, type=html]{\textbf{MQ12}}{\ref{app:part2:q12}} to learn when users expect to receive notifications.
A majority of participants (151; 66\%) expressed they want to receive notifications after suspicious account activity. 
On average, 60\% want to be notified if a login takes place from a new device, 47\% for logins from a new location, 31\% if they have not logged in for a while, and 22\% for logins that take place at an unusual time of the day.
Only 9\% want to receive a login for \textit{every} login, and even fewer (9; 3\%) do not want to receive login notifications at all.
\vspace{2pt}

\begin{shaded}
\noindent{\textbf{Summary}} {We can conclude that participants who changed their password felt both more positively and negatively, probably because they assumed some form of compromise but also had a sense of achievement after preventing it by changing the password.
The other groups had lower scores, aligning with them not expecting any harm. 
We showed that participants expect services to send login notifications and can further specify this by saying that participants want to be notified after suspicious logins, logins from new devices, and logins from new locations.
Fewer participants expect to receive notifications based on temporal deviations.}
\end{shaded}

\section{Discussion}\label{sec:discussion}
Next, we discuss the takeaways and give recommendations.

\subsection{Effectiveness of Login Notifications}
We wondered if login notifications that many services use daily, achieve their goal of increasing the security of online accounts.

\textbf{Effectiveness Depends on Trigger} According to our findings, the notifications achieve what they intend---at least to a certain extent.
In the malicious login case, we saw that only about 20\% of the users in the study reacted to our \emph{baseline} notification and changed their password.
Hence, we conclude that login notifications can improve account security partially, as every password change can help to stop a malicious actor.
However, at the same time, 80\% of the participants in the malicious group who should have changed their password did not.
While some participants might have decided not to change their password due to the study accounts' low value, it is still a high number, questioning the overall cost-benefit trade-off of the notifications.

\textbf{Arguments against Notifications} On the cost side of things, we found several participants being \textit{annoyed}, which is in line with research on fatigue in the context of security warnings and notifications~\cite{akhawe-13-alice, amran-18-warning-fatigue}.
Another argument against the notifications is that they shift the responsibility for account security away from the service provider onto the user.
In a sense, such notifications can be perceived as \textit{burdening and blaming}~\cite{sasse-15-bullying, herley-09-rejecting-advice}.
If service providers which hold exhaustive records about a user's login history are uncertain, why should the user be able to determine the legitimacy of a login?
It is fair to say that in some cases users may know better whether i.e., their location has changed. 
However, in real life with shared accounts~\cite{angelini-23-netflix-who-logged-in}, third-party apps and services that automatically sign in to an account~\cite{1password-22-new-login-email, redfox-22-netflix-new-device}, or simply on busy days only few of us can realistically remember which accounts they used (note that we did not explicitly test these factors in our study). 
Thus expecting users to determine the legitimacy of a login better than a service provider is unfair. 
From a service provider's perspective, allowing logins and hedging them with a notification rather than blocking them makes sense; for them, it is the easiest ``solution'' to the problem. 
On a conceptual level, it all boils down to whether users should be made responsible for damages or if it should rather be the service provider's duty to implement robust security measures~\cite{lin-22-phish-in-sheeps}.

\textbf{Arguments for Notifications} Contrary to concerns about \mbox{burdening} users, it can be argued that users took the appropriate action in the legitimate case---namely, ignoring the notification.
Our study showed that most \textit{participants correctly followed the instructions} when prompted by a legitimate login notification and our qualitative results proved that they even correctly understand its meaning and cause.
Additionally, some users \textit{felt more protected} and satisfied when receiving such notifications.
According to their qualitative feedback, such notifications' reassurance contributes to a positive user experience and reinforces trust in the service.

\subsection{Refining Notifications}
As indicated before, research and development should focus on refining notification systems to ensure their maximum effectiveness and usability.
Past examples in the warning design space, i.e., TLS warnings, have demonstrated how improved warning designs can increase comprehension and adherence and decrease click-through rates~\cite{felt-15-ssl-warnings}.
One approach to facilitate appropriate reactions may be to align notifications' implementations with users' understanding.
As shown in Section~\ref{sec:measurement-results}, participants appear to expect and need contextual factors to determine what caused a notification.
Especially, we saw significant differences in the helpfulness ratings of information between those in the malicious group, who changed their password, and those who did not.

While future research needs to investigate the exact root cause for this difference, we can certainly say that the information provided and users' ability to understand it correlates with their behavior in terms of password change. 
Malicious (No Change) participants, in particular, often misattributed or did not understand the cause of the notification---indicating that this information needs to be refined.
Services could address this and adhere to users' expectations by including a distinct \emph{Why Notification} component, e.g., by explicitly saying that a login happened from a previously unseen device. 
Our initial analysis of real-world notifications only found this contextualizing section in about $20\%$ of the real-world notifications.
Moreover, from a security standpoint, services need to provide more help than just suggesting to change the password.
While password change is a first line of defense upon account compromise, it is by far not sufficient to ensure that the compromised account and other accounts are safe.
Service providers should thus initiate a thorough remediation process, including expiring all sessions, reviewing third-party access, enabling 2FA, suggesting using a password manager, and checking related accounts~\cite{neil-21-acc-remediation-adv, walsh-21-intercultural-analysis}.

\subsection{Expectation vs. Fatigue}
Besides the correct reaction and understanding of notifications---user satisfaction with notifications is equally important.
Fortunately, we found that more than 95\% of the participants expect services to send login notifications, and only 3\% do not want notifications to be sent at all, underlining an overall positive assessment.
However, it is crucial to find a balance between sending them too often and too rarely. 
For service providers, sending notifications following a ``better safe than sorry''-mentality may be tempting. 
Yet, for users, this leads to security warning fatigue~\cite{egelman-08-phishing-warnings, sunshine-09-crying-wolf, akhawe-13-alice}.
This fatigue is most likely caused by unnecessary login notifications, i.e., those that do not convey a real risk teach users that all notifications are unimportant.
The situation is aggravated by services like Etsy, GitLab, Mozilla, Tumblr, and others that send notifications for each and every login.
We saw that over 90\% did not want to receive a notification on every login, and 15\% even explicitly expressed ``fatigue.''
Thus, we strongly dissuade sending notifications on every login.
Instead, they should only be sent if the service suspects malicious account activity.
Concretely, the majority of the participants wants to be notified when a login takes place from a new device or location, and especially if a login appears ``suspicious.''
Service providers can accomplish this with advanced logic provided by risk-based algorithms~\cite{wiefling-19-rba-in-the-wild, wiefling-20-rba-evaluation}.
Time-related notifications (i.e., a login after a long or at an unusual time) are less demanded.
In favor of sparsity, time-related notifications should be omitted unless there is a concrete reason for suspecting malicious account activity.

\subsection{Good Intentions \& Questionable Advice}
In the study, about 8\% of the participants questioned the legitimacy of the notification or referred to it as phishing.
Prior work explains how to best advise users on this topic~\cite{secuso-22-detect-phishing},
yet most of the 10~real-world notifications that include information about phishing do not follow the recommendations.
While contradicting security advice, as well as no consensus among security experts about its prioritization, is nothing new in the community~\cite{reeder-17-152-simple-steps, redmiles-20-security-advice}, for us, it was surprising how potentially dangerous and obsolete some of the given advice is (see Section~\ref{sec:background:findings}).
For example, many large services like X (Twitter), Spotify, and Amazon portrayed the padlock icon of the browser as a type of trust and legitimacy indicator.
While this is not only false, in early September 2023, Google removed the padlock icon with Chrome~117, as HTTPS should be considered the default state~\cite{adrian-23-remove-padlock}.

\subsection{Recommendations}
Based on our findings, we give some recommendations for service providers below.

\textbf{Notify about Devices, Locations, and Suspicious Logins}
We advise against sending notifications after every login, mainly because some of our participants reported being annoyed by the frequency of real-world services sending notifications (see Section~\ref{sec:study2-decision-and-execution}).
Instead, we recommend that services send login notifications when a login takes place from a new device or location, and especially if a login appears ``suspicious.''

\textbf{Describe What Happened}
What triggers a notification, e.g., an ``unusual login,'' is often unclear to participants (see Section~\ref{sec:study2-comprehension-and-reaction}).
Services could easily address this issue by explaining what triggered the notification, yet only 21\% of the evaluated emails currently provide examples of common triggers (see Section~\ref{sec:background:findings}).
Explaining the circumstances would also help to create context, which is especially important when users receive unexpected notifications and struggle to assess the situation correctly.

\textbf{Include Information in Metadata}
We found that the metadata is an influential factor, and 75\% of the participants paid attention to the email subject (see Section~\ref{sec:study2-decision-and-execution}).
Hence, in addition to the most important information, the email subject should already provide context for deciding how to react.
Currently, only $15\%$ of our analyzed notifications make use of subjects like ``New login to Instagram from \{browser\} on \{OS\}'' (see Section~\ref{sec:background:findings}).
Similarly, websites should make use of the email sender's name so that recipients can quickly parse the information about the sender.

\textbf{Specify Instructions}
Based on the findings of our email analysis (see Section~\ref{sec:background:findings}), notifications should include instructions for both outcomes, i.e., legitimate and malicious logins.
For the legitimate case, most services suggest to ignore the message. 
For malicious logins, the recommendation needs to prompt users to visit the website and change or reset the password or, even better, initiate a thorough remediation process.
Most services facilitate this by including a link, which is a controversial practice.
However, 8\% of the participants were suspicious, some of them due to the presence of a link, and did not change their password.

\textbf{Provide Comprehensible Details}
We found that all types of information (account name, location, time, and device) have a positive influence (see Section~\ref{sec:study2-decision-and-execution}).
Still, services need to be careful when it comes to parsing and displaying technical details such as location data, browser, and OS (see Section~\ref{sec:background:findings}).
Here, special care and testing are required, as a badly parsed or displayed detail could impact the overall perceived legitimacy of the notification.

By addressing the identified areas, service providers can continue to strengthen account security and foster user trust.
\section{Conclusion}\label{sec:conclusion}

We explored users' comprehension, reactions, and expectations of login notifications that are sent by services to help users protect their accounts from unauthorized access.

In a three-stages user study ($n=229$), we evaluated a \emph{baseline} notification that was created by collecting and analyzing 72~notifications sent by real-world services.
To prevent participants from spending most of their attention on the notification and authentication task, we introduced a strong cognitive distractor by implementing a mental rotation test.
We split participants into two treatments:
a) The legitimate group received a notification only after they logged in, using metadata derived from their login information.
b) The malicious group received a notification unexpectedly at a time when they had not interacted with the study website for multiple days using a generic location (``California, USA'') and device (``Chrome on Windows'') with the highest statistical chance of matching any user in our U.S.-based sample.

Overall, we find that login notifications achieve their goal of increasing account security.
However, the tested notification failed to convince the majority of participants to change their password.
Participants expressed the need for more contextual factors to help determine what caused the notification.
Thus, instead of talking about an ``unusual login,'' services need to explain what triggered the notification to assist users who received the notification unexpectedly.
Even though some participants expressed feeling more protected and satisfied after receiving a notification, we argue that the service and not the user must be held accountable for implementing robust account protection measures.
Interestingly, we find that more than 90\% of the participants expect services to send login notifications when a login takes place from a new device or location, and especially if a login appears ``suspicious.''
However, participants also expressed that they do not want to be notified for every login, highlighting the importance of finding the right balance between sending notifications too often or too rarely.

\begin{acks}
This research was supported by the research training group ``Human Centered Systems Security'' sponsored by the state of North Rhine-Westphalia and funded by the Deutsche Forschungsgemeinschaft (DFG, German Research Foundation) under Germany's Excellence Strategy -- EXC 2092 CASA -- 390781972.
\end{acks}

\bibliographystyle{ACM-Reference-Format}
\bibliography{bibliography}


\begin{thebibliography}{64}


\ifx \showCODEN    \undefined \def \showCODEN     #1{\unskip}     \fi
\ifx \showDOI      \undefined \def \showDOI       #1{#1}\fi
\ifx \showISBNx    \undefined \def \showISBNx     #1{\unskip}     \fi
\ifx \showISBNxiii \undefined \def \showISBNxiii  #1{\unskip}     \fi
\ifx \showISSN     \undefined \def \showISSN      #1{\unskip}     \fi
\ifx \showLCCN     \undefined \def \showLCCN      #1{\unskip}     \fi
\ifx \shownote     \undefined \def \shownote      #1{#1}          \fi
\ifx \showarticletitle \undefined \def \showarticletitle #1{#1}   \fi
\ifx \showURL      \undefined \def \showURL       {\relax}        \fi
\providecommand\bibfield[2]{#2}
\providecommand\bibinfo[2]{#2}
\providecommand\natexlab[1]{#1}
\providecommand\showeprint[2][]{arXiv:#2}

\bibitem[{1Password Community Member}(2022)]%
        {1password-22-new-login-email}
\bibfield{author}{\bibinfo{person}{{1Password Community Member}}.}
  \bibinfo{year}{2022}\natexlab{}.
\newblock \bibinfo{title}{{1Password ``New Login'' E-Mail Notification}}.
\newblock
\newblock
\newblock
\shownote{\url{https://1password.community/discussion/comment/643758/}, as of
  \today}.


\bibitem[Adrian et~al\mbox{.}(2023)]%
        {adrian-23-remove-padlock}
\bibfield{author}{\bibinfo{person}{David Adrian}, \bibinfo{person}{Serena
  Chen}, \bibinfo{person}{Joe DeBlasio}, \bibinfo{person}{Emily Stark}, {and}
  \bibinfo{person}{Emanuel {von Zezschwitz}}.} \bibinfo{year}{2023}\natexlab{}.
\newblock \bibinfo{title}{{Chromium Blog: An Update on the Lock Icon}}.
\newblock
\newblock
\newblock
\shownote{\url{https://blog.chromium.org/2023/05/an-update-on-lock-icon.html},
  as of \today}.


\bibitem[Akhawe and Felt(2013)]%
        {akhawe-13-alice}
\bibfield{author}{\bibinfo{person}{Devdatta Akhawe} {and}
  \bibinfo{person}{Adrienne~Porter Felt}.} \bibinfo{year}{2013}\natexlab{}.
\newblock \showarticletitle{{Alice in Warningland: A Large-Scale Field Study of
  Browser Security Warning Effectiveness}}. In \bibinfo{booktitle}{\emph{USENIX
  Security Symposium}} \emph{(\bibinfo{series}{SSYM~'13})}.
  \bibinfo{publisher}{USENIX}, \bibinfo{address}{Washington, District of
  Columbia, USA}, \bibinfo{pages}{257--272}.
\newblock


\bibitem[Almuhimedi et~al\mbox{.}(2014)]%
        {almuhimedi-14-malware-warning}
\bibfield{author}{\bibinfo{person}{Hazim Almuhimedi},
  \bibinfo{person}{Adrienne~Porter Felt}, \bibinfo{person}{Robert~W. Reeder},
  {and} \bibinfo{person}{Sunny Consolvo}.} \bibinfo{year}{2014}\natexlab{}.
\newblock \showarticletitle{{Your Reputation Precedes You: History, Reputation,
  and the Chrome Malware Warning}}. In \bibinfo{booktitle}{\emph{Symposium on
  Usable Privacy and Security}} \emph{(\bibinfo{series}{SOUPS~'14})}.
  \bibinfo{publisher}{USENIX}, \bibinfo{address}{Menlo Park, California, USA},
  \bibinfo{pages}{113--128}.
\newblock


\bibitem[{Alphabet, Inc.}(2024)]%
        {alphabet-24-login-notification}
\bibfield{author}{\bibinfo{person}{{Alphabet, Inc.}}}
  \bibinfo{year}{2024}\natexlab{}.
\newblock \bibinfo{title}{{Respond to Security Alerts: When You'll Get an
  Alert}}.
\newblock
\newblock
\newblock
\shownote{\url{https://support.google.com/accounts/answer/2590353}, as of
  \today}.


\bibitem[{Amazon.com, Inc.}(2024)]%
        {amazon-24-login-notification}
\bibfield{author}{\bibinfo{person}{{Amazon.com, Inc.}}}
  \bibinfo{year}{2024}\natexlab{}.
\newblock \bibinfo{title}{{If You Get a Security Alert about Activity You Don't
  Recognize}}.
\newblock
\newblock
\newblock
\shownote{\url{https://www.amazon.com/gp/help/customer/display.html?nodeId=GLXNK37D6R3WGXKW},
  as of \today}.


\bibitem[Amran et~al\mbox{.}(2018)]%
        {amran-18-warning-fatigue}
\bibfield{author}{\bibinfo{person}{Ammar Amran}, \bibinfo{person}{Zarul~Fitri
  Zaaba}, {and} \bibinfo{person}{Manmeet Kaur~Mahinderjit Singh}.}
  \bibinfo{year}{2018}\natexlab{}.
\newblock \showarticletitle{{Habituation Effects in Computer Security
  Warning}}.
\newblock \bibinfo{journal}{\emph{Information Security Journal: A Global
  Perspective}} \bibinfo{volume}{27}, \bibinfo{number}{4} (\bibinfo{date}{Oct.}
  \bibinfo{year}{2018}), \bibinfo{pages}{192--204}.
\newblock


\bibitem[Angelini(2023)]%
        {angelini-23-netflix-who-logged-in}
\bibfield{author}{\bibinfo{person}{Francesca Angelini}.}
  \bibinfo{year}{2023}\natexlab{}.
\newblock \bibinfo{title}{{Do You Know Who's Logging into Your Netflix
  Account?}}
\newblock
\newblock
\newblock
\shownote{\url{https://www.thetimes.co.uk/article/do-you-know-whos-logging-into-your-netflix-account-m728z3678},
  as of \today}.


\bibitem[{Apple, Inc.}(2022)]%
        {apple-22-forum}
\bibfield{author}{\bibinfo{person}{{Apple, Inc.}}}
  \bibinfo{year}{2022}\natexlab{}.
\newblock \bibinfo{title}{{Apple Support Community: Notification of Sign In
  with Apple ID}}.
\newblock
\newblock
\newblock
\shownote{\url{https://discussions.apple.com/thread/253581666}, as of \today}.


\bibitem[{Apple, Inc.}(2024)]%
        {apple-24-login-notification}
\bibfield{author}{\bibinfo{person}{{Apple, Inc.}}}
  \bibinfo{year}{2024}\natexlab{}.
\newblock \bibinfo{title}{{Signs That Your Apple ID Has Been Compromised}}.
\newblock
\newblock
\newblock
\shownote{\url{https://support.apple.com/en-us/102560}, as of \today}.


\bibitem[Bauer et~al\mbox{.}(2013)]%
        {bauer-13-warning-guidelines}
\bibfield{author}{\bibinfo{person}{Lujo Bauer}, \bibinfo{person}{Cristian
  Bravo-Lillo}, \bibinfo{person}{Lorrie Cranor}, {and} \bibinfo{person}{Elli
  Fragkaki}.} \bibinfo{year}{2013}\natexlab{}.
\newblock \bibinfo{booktitle}{\emph{{Warning Design Guidelines}}}.
\newblock \bibinfo{type}{Technical Report} CMU-CyLab-13-002.
  \bibinfo{institution}{Carnegie Mellon University}.
\newblock


\bibitem[Birks and Mills(2022)]%
        {birks-22-grounded-theory}
\bibfield{author}{\bibinfo{person}{Melanie Birks} {and} \bibinfo{person}{Jane
  Mills}.} \bibinfo{year}{2022}\natexlab{}.
\newblock \bibinfo{booktitle}{\emph{{Grounded Theory: A Practical Guide}}
  (\bibinfo{edition}{3} ed.)}.
\newblock \bibinfo{publisher}{{SAGE Publications, Ltd.}},
  \bibinfo{address}{Thousand Oaks, California, USA}.
\newblock


\bibitem[Biselli et~al\mbox{.}(2024)]%
        {biselli-24-cookies}
\bibfield{author}{\bibinfo{person}{Tom Biselli}, \bibinfo{person}{Laura Utz},
  {and} \bibinfo{person}{Christian Reuter}.} \bibinfo{year}{2024}\natexlab{}.
\newblock \showarticletitle{{Supporting Informed Choices about Browser Cookies:
  The Impact of Personalised Cookie Banners}}. In
  \bibinfo{booktitle}{\emph{Privacy Enhancing Technologies Symposium}}
  \emph{(\bibinfo{series}{PETS~'24})}. \bibinfo{publisher}{PoPETS},
  \bibinfo{address}{Bristol, United Kingdom}, \bibinfo{pages}{171--191}.
\newblock


\bibitem[Doerfler et~al\mbox{.}(2019)]%
        {doerfler-19-login-challenges}
\bibfield{author}{\bibinfo{person}{Periwinkle Doerfler}, \bibinfo{person}{Kurt
  Thomas}, \bibinfo{person}{Maija Marincenko}, \bibinfo{person}{Juri Ranieri},
  \bibinfo{person}{Yu Jiang}, \bibinfo{person}{Angelika Moscicki}, {and}
  \bibinfo{person}{Damon McCoy}.} \bibinfo{year}{2019}\natexlab{}.
\newblock \showarticletitle{{Evaluating Login Challenges as a Defense Against
  Account Takeover}}. In \bibinfo{booktitle}{\emph{The World Wide Web
  Conference}} \emph{(\bibinfo{series}{WWW~'19})}. \bibinfo{publisher}{ACM},
  \bibinfo{address}{San Francisco, California, USA}, \bibinfo{pages}{372--382}.
\newblock


\bibitem[Egelman et~al\mbox{.}(2008)]%
        {egelman-08-phishing-warnings}
\bibfield{author}{\bibinfo{person}{Serge Egelman},
  \bibinfo{person}{Lorrie~Faith Cranor}, {and} \bibinfo{person}{Jason Hong}.}
  \bibinfo{year}{2008}\natexlab{}.
\newblock \showarticletitle{{You've Been Warned: An Empirical Study of the
  Effectiveness of Web Browser Phishing Warnings}}. In
  \bibinfo{booktitle}{\emph{ACM Conference on Human Factors in Computing
  Systems}} \emph{(\bibinfo{series}{CHI~'08})}. \bibinfo{publisher}{ACM},
  \bibinfo{address}{Florence, Italy}, \bibinfo{pages}{1065--1074}.
\newblock


\bibitem[Felt et~al\mbox{.}(2015)]%
        {felt-15-ssl-warnings}
\bibfield{author}{\bibinfo{person}{Adrienne~Porter Felt}, \bibinfo{person}{Alex
  Ainslie}, \bibinfo{person}{Robert~W. Reeder}, \bibinfo{person}{Sunny
  Consolvo}, \bibinfo{person}{Somas Thyagaraja}, \bibinfo{person}{Alan Bettes},
  \bibinfo{person}{Helen Harris}, {and} \bibinfo{person}{Jeff Grimes}.}
  \bibinfo{year}{2015}\natexlab{}.
\newblock \showarticletitle{{Improving SSL Warnings: Comprehension and
  Adherence}}. In \bibinfo{booktitle}{\emph{ACM Conference on Human Factors in
  Computing Systems}} \emph{(\bibinfo{series}{CHI~'15})}.
  \bibinfo{publisher}{ACM}, \bibinfo{address}{Seoul, Republic of Korea},
  \bibinfo{pages}{2893--2902}.
\newblock


\bibitem[Gavazzi et~al\mbox{.}(2023)]%
        {gavazzi-23-rba-availability}
\bibfield{author}{\bibinfo{person}{Anthony Gavazzi}, \bibinfo{person}{Ryan
  Williams}, \bibinfo{person}{Engin Kirda}, \bibinfo{person}{Long Lu},
  \bibinfo{person}{Andre King}, \bibinfo{person}{Andy Davis}, {and}
  \bibinfo{person}{Tim Leek}.} \bibinfo{year}{2023}\natexlab{}.
\newblock \showarticletitle{{A Study of Multi-Factor and Risk-Based
  Authentication Availability}}. In \bibinfo{booktitle}{\emph{USENIX Security
  Symposium}} \emph{(\bibinfo{series}{SSYM~'23})}. \bibinfo{publisher}{USENIX},
  \bibinfo{address}{Anaheim, California, USA}, \bibinfo{pages}{2043--2060}.
\newblock


\bibitem[Golla et~al\mbox{.}(2021)]%
        {golla-21-2fa-adoption}
\bibfield{author}{\bibinfo{person}{Maximilian Golla}, \bibinfo{person}{Grant
  Ho}, \bibinfo{person}{Marika Lohmus}, \bibinfo{person}{Monica Pulluri}, {and}
  \bibinfo{person}{Elissa~M. Redmiles}.} \bibinfo{year}{2021}\natexlab{}.
\newblock \showarticletitle{{Driving 2FA Adoption at Scale: Optimizing
  Two-Factor Authentication Notification Design Patterns}}. In
  \bibinfo{booktitle}{\emph{USENIX Security Symposium}}
  \emph{(\bibinfo{series}{SSYM~'21})}. \bibinfo{publisher}{USENIX},
  \bibinfo{address}{Virtual Conference}, \bibinfo{pages}{109--126}.
\newblock


\bibitem[Golla et~al\mbox{.}(2018)]%
        {golla-18-reuse-notification}
\bibfield{author}{\bibinfo{person}{Maximilian Golla}, \bibinfo{person}{Miranda
  Wei}, \bibinfo{person}{Juliette Hainline}, \bibinfo{person}{Lydia Filipe},
  \bibinfo{person}{Markus D\"{u}rmuth}, \bibinfo{person}{Elissa Redmiles},
  {and} \bibinfo{person}{Blase Ur}.} \bibinfo{year}{2018}\natexlab{}.
\newblock \showarticletitle{{``What was that site doing with my Facebook
  password?'' Designing Password-Reuse Notifications}}. In
  \bibinfo{booktitle}{\emph{ACM Conference on Computer and Communications
  Security}} \emph{(\bibinfo{series}{CCS~'18})}. \bibinfo{publisher}{ACM},
  \bibinfo{address}{Toronto, Ontario, Canada}, \bibinfo{pages}{1549--1566}.
\newblock


\bibitem[Gorski et~al\mbox{.}(2020)]%
        {gorski-20-pd-for-crypto-apis}
\bibfield{author}{\bibinfo{person}{Peter~Leo Gorski}, \bibinfo{person}{Yasemin
  Acar}, \bibinfo{person}{Luigi Lo~Iacono}, {and} \bibinfo{person}{Sascha
  Fahl}.} \bibinfo{year}{2020}\natexlab{}.
\newblock \showarticletitle{{Listen to Developers! A Participatory Design Study
  on Security Warnings for Cryptographic APIs}}. In
  \bibinfo{booktitle}{\emph{ACM Conference on Human Factors in Computing
  Systems}} \emph{(\bibinfo{series}{CHI~'20})}. \bibinfo{publisher}{ACM},
  \bibinfo{address}{Honolulu, Hawaii, USA}, \bibinfo{pages}{1--13}.
\newblock


\bibitem[Herley(2009)]%
        {herley-09-rejecting-advice}
\bibfield{author}{\bibinfo{person}{Cormac Herley}.}
  \bibinfo{year}{2009}\natexlab{}.
\newblock \showarticletitle{{So Long, and No Thanks for the Externalities: The
  Rational Rejection of Security Advice by Users}}. In
  \bibinfo{booktitle}{\emph{New Security Paradigms Workshop}}
  \emph{(\bibinfo{series}{NSPW~'09})}. \bibinfo{publisher}{ACM},
  \bibinfo{address}{Oxford, United Kingdom}, \bibinfo{pages}{133--144}.
\newblock


\bibitem[Huh et~al\mbox{.}(2017)]%
        {huh-17-linkedin}
\bibfield{author}{\bibinfo{person}{Jun~Ho Huh}, \bibinfo{person}{Hyoungshick
  Kim}, \bibinfo{person}{Swathi~S.V.P. Rayala}, \bibinfo{person}{Rakesh~B.
  Bobba}, {and} \bibinfo{person}{Konstantin Beznosov}.}
  \bibinfo{year}{2017}\natexlab{}.
\newblock \showarticletitle{{I'm Too Busy to Reset My LinkedIn Password: On the
  Effectiveness of Password Reset Emails}}. In \bibinfo{booktitle}{\emph{ACM
  Conference on Human Factors in Computing Systems}}
  \emph{(\bibinfo{series}{CHI~'17})}. \bibinfo{publisher}{ACM},
  \bibinfo{address}{Denver, Colorado, USA}, \bibinfo{pages}{387--391}.
\newblock


\bibitem[Kaiser et~al\mbox{.}(2021)]%
        {kaiser-20-warning-disinformation}
\bibfield{author}{\bibinfo{person}{Ben Kaiser}, \bibinfo{person}{Jerry Wei},
  \bibinfo{person}{Elena Lucherini}, \bibinfo{person}{Kevin Lee},
  \bibinfo{person}{J.~Nathan Matias}, {and} \bibinfo{person}{Jonathan Mayer}.}
  \bibinfo{year}{2021}\natexlab{}.
\newblock \showarticletitle{{Adapting Security Warnings to Counter Online
  Disinformation}}. In \bibinfo{booktitle}{\emph{USENIX Security Symposium}}
  \emph{(\bibinfo{series}{SSYM~'21})}. \bibinfo{publisher}{USENIX},
  \bibinfo{address}{Virtual Conference}, \bibinfo{pages}{1163--1180}.
\newblock


\bibitem[Krisam et~al\mbox{.}(2021)]%
        {krisam-21-dark-german500}
\bibfield{author}{\bibinfo{person}{Chiara Krisam}, \bibinfo{person}{Heike
  Dietmann}, \bibinfo{person}{Melanie Volkamer}, {and} \bibinfo{person}{Oksana
  Kulyk}.} \bibinfo{year}{2021}\natexlab{}.
\newblock \showarticletitle{{Dark Patterns in the Wild: Review of Cookie
  Disclaimer Designs on Top 500 German Websites}}. In
  \bibinfo{booktitle}{\emph{European Workshop on Usable Security}}
  \emph{(\bibinfo{series}{EuroUSEC~'21})}. \bibinfo{publisher}{ACM},
  \bibinfo{address}{Virtual Conference}, \bibinfo{pages}{1--8}.
\newblock


\bibitem[Lassak et~al\mbox{.}(2021)]%
        {lassak-21-webauthn-misconceptions}
\bibfield{author}{\bibinfo{person}{Leona Lassak}, \bibinfo{person}{Annika
  Hildebrandt}, \bibinfo{person}{Maximilian Golla}, {and}
  \bibinfo{person}{Blase Ur}.} \bibinfo{year}{2021}\natexlab{}.
\newblock \showarticletitle{{``It's Stored, Hopefully, on an Encrypted
  Server'': Mitigating Users' Misconceptions About FIDO2 Biometric WebAuthn}}.
  In \bibinfo{booktitle}{\emph{USENIX Security Symposium}}
  \emph{(\bibinfo{series}{SSYM~'21})}. \bibinfo{publisher}{USENIX},
  \bibinfo{address}{Virtual Conference}, \bibinfo{pages}{91--108}.
\newblock


\bibitem[{Le Pochat} et~al\mbox{.}(2019)]%
        {lepochat-19-tranco}
\bibfield{author}{\bibinfo{person}{Victor {Le Pochat}}, \bibinfo{person}{Tom
  {Van Goethem}}, \bibinfo{person}{Samaneh Tajalizadehkhoob},
  \bibinfo{person}{Maciej Korczy\'{n}ski}, {and} \bibinfo{person}{Wouter
  Joosen}.} \bibinfo{year}{2019}\natexlab{}.
\newblock \showarticletitle{{Tranco: A Research-Oriented Top Sites Ranking
  Hardened Against Manipulation}}. In \bibinfo{booktitle}{\emph{Symposium on
  Network and Distributed System Security}}
  \emph{(\bibinfo{series}{NDSS~'19})}. \bibinfo{publisher}{ISOC},
  \bibinfo{address}{San Diego, California, USA}.
\newblock


\bibitem[Lin et~al\mbox{.}(2022)]%
        {lin-22-phish-in-sheeps}
\bibfield{author}{\bibinfo{person}{Xu Lin}, \bibinfo{person}{Panagiotis Ilia},
  \bibinfo{person}{Saumya Solanki}, {and} \bibinfo{person}{Jason Polakis}.}
  \bibinfo{year}{2022}\natexlab{}.
\newblock \showarticletitle{{Phish in Sheep's Clothing: Exploring the
  Authentication Pitfalls of Browser Fingerprinting}}. In
  \bibinfo{booktitle}{\emph{USENIX Security Symposium}}
  \emph{(\bibinfo{series}{SSYM~'22})}. \bibinfo{publisher}{USENIX},
  \bibinfo{address}{Boston, Massachusetts, USA}, \bibinfo{pages}{1651--1668}.
\newblock


\bibitem[{LinkedIn, Inc.}(2023)]%
        {linkedin-22-security-footer}
\bibfield{author}{\bibinfo{person}{{LinkedIn, Inc.}}}
  \bibinfo{year}{2023}\natexlab{}.
\newblock \bibinfo{title}{{Security Footer Message in LinkedIn Emails}}.
\newblock
\newblock
\newblock
\shownote{\url{https://www.linkedin.com/help/linkedin/answer/a1339250}, as of
  \today}.


\bibitem[Markert et~al\mbox{.}(2021)]%
        {markert-21-pin-unlock}
\bibfield{author}{\bibinfo{person}{Philipp Markert}, \bibinfo{person}{Daniel~V.
  Bailey}, \bibinfo{person}{Maximilian Golla}, \bibinfo{person}{Markus
  D\"{u}rmuth}, {and} \bibinfo{person}{Adam~J. Aviv}.}
  \bibinfo{year}{2021}\natexlab{}.
\newblock \showarticletitle{{On the Security of Smartphone Unlock PINs}}.
\newblock \bibinfo{journal}{\emph{ACM Transactions on Privacy and Security}}
  \bibinfo{volume}{24}, \bibinfo{number}{4} (\bibinfo{date}{Sept.}
  \bibinfo{year}{2021}), \bibinfo{pages}{30:1--30:36}.
\newblock


\bibitem[Markert et~al\mbox{.}(2022)]%
        {markert-22-rba-admin}
\bibfield{author}{\bibinfo{person}{Philipp Markert}, \bibinfo{person}{Theodor
  Schnitzler}, \bibinfo{person}{Maximilian Golla}, {and}
  \bibinfo{person}{Markus D\"{u}rmuth}.} \bibinfo{year}{2022}\natexlab{}.
\newblock \showarticletitle{{``As soon as it's a risk, I want to require MFA'':
  How Administrators Configure Risk-based Authentication}}. In
  \bibinfo{booktitle}{\emph{Symposium on Usable Privacy and Security}}
  \emph{(\bibinfo{series}{SOUPS~'22})}. \bibinfo{publisher}{USENIX},
  \bibinfo{address}{Boston, Massachusetts, USA}, \bibinfo{pages}{483--501}.
\newblock


\bibitem[{Meta Platforms, Inc.}(2024)]%
        {meta-24-login-notification}
\bibfield{author}{\bibinfo{person}{{Meta Platforms, Inc.}}}
  \bibinfo{year}{2024}\natexlab{}.
\newblock \bibinfo{title}{{Get Alerts about Unrecognized Logins to Facebook}}.
\newblock
\newblock
\newblock
\shownote{\url{https://www.facebook.com/help/162968940433354}, as of \today}.


\bibitem[{Microsoft, Corporation}(2024)]%
        {microsoft-24-login-notification}
\bibfield{author}{\bibinfo{person}{{Microsoft, Corporation}}.}
  \bibinfo{year}{2024}\natexlab{}.
\newblock \bibinfo{title}{{What Happens If There's an Unusual Sign-in to Your
  Account}}.
\newblock
\newblock
\newblock
\shownote{\url{https://support.microsoft.com/en-us/account-billing/what-happens-if-there-s-an-unusual-sign-in-to-your-account-eba43e04-d348-b914-1e95-fb5052d3d8f0},
  as of \today}.


\bibitem[Neil et~al\mbox{.}(2021)]%
        {neil-21-acc-remediation-adv}
\bibfield{author}{\bibinfo{person}{Lorenzo Neil}, \bibinfo{person}{Elijah
  Bouma-Sims}, \bibinfo{person}{Evan Lafontaine}, \bibinfo{person}{Yasemin
  Acar}, {and} \bibinfo{person}{Bradley Reaves}.}
  \bibinfo{year}{2021}\natexlab{}.
\newblock \showarticletitle{{Investigating Web Service Account Remediation
  Advice}}. In \bibinfo{booktitle}{\emph{Symposium on Usable Privacy and
  Security}} \emph{(\bibinfo{series}{SOUPS~'21})}. \bibinfo{publisher}{USENIX},
  \bibinfo{address}{Virtual Conference}, \bibinfo{pages}{359--376}.
\newblock


\bibitem[{Netflix, Inc.}(2023)]%
        {netflix-23-faq}
\bibfield{author}{\bibinfo{person}{{Netflix, Inc.}}}
  \bibinfo{year}{2023}\natexlab{}.
\newblock \bibinfo{title}{{I Received an Email Stating There Was a New Sign-in
  to My Account}}.
\newblock
\newblock
\newblock
\shownote{\url{https://help.netflix.com/en/node/100775}, as of \today}.


\bibitem[Nouwens et~al\mbox{.}(2020)]%
        {nouwens-20-dark-patterns}
\bibfield{author}{\bibinfo{person}{Midas Nouwens}, \bibinfo{person}{Ilaria
  Liccardi}, \bibinfo{person}{Michael Veale}, \bibinfo{person}{David Karger},
  {and} \bibinfo{person}{Lalana Kagal}.} \bibinfo{year}{2020}\natexlab{}.
\newblock \showarticletitle{{Dark Patterns after the GDPR: Scraping Consent
  Pop-ups and Demonstrating their Influence}}. In \bibinfo{booktitle}{\emph{ACM
  Conference on Human Factors in Computing Systems}}
  \emph{(\bibinfo{series}{CHI~'20})}. \bibinfo{publisher}{ACM},
  \bibinfo{address}{Honolulu, Hawaii, USA}, \bibinfo{pages}{1--13}.
\newblock


\bibitem[Obada-Obieh et~al\mbox{.}(2020)]%
        {obada-obieh-20-account-sharing}
\bibfield{author}{\bibinfo{person}{Borke Obada-Obieh}, \bibinfo{person}{Yue
  Huang}, {and} \bibinfo{person}{Konstantin Beznosov}.}
  \bibinfo{year}{2020}\natexlab{}.
\newblock \showarticletitle{{The Burden of Ending Online Account Sharing}}. In
  \bibinfo{booktitle}{\emph{ACM Conference on Human Factors in Computing
  Systems}} \emph{(\bibinfo{series}{CHI~'20})}. \bibinfo{publisher}{ACM},
  \bibinfo{address}{Honolulu, Hawaii, USA}, \bibinfo{pages}{503:1--503:13}.
\newblock


\bibitem[Pal et~al\mbox{.}(2019)]%
        {pal-19-personal-psm}
\bibfield{author}{\bibinfo{person}{Bijeeta Pal}, \bibinfo{person}{Tal Daniel},
  \bibinfo{person}{Rahul Chatterjee}, {and} \bibinfo{person}{Thomas
  Ristenpart}.} \bibinfo{year}{2019}\natexlab{}.
\newblock \showarticletitle{{Beyond Credential Stuffing: Password Similarity
  Models using Neural Networks}}. In \bibinfo{booktitle}{\emph{IEEE Symposium
  on Security and Privacy}} \emph{(\bibinfo{series}{SP~'19})}.
  \bibinfo{publisher}{IEEE}, \bibinfo{address}{San Francisco, California, USA},
  \bibinfo{pages}{866--883}.
\newblock


\bibitem[{PayPal, Inc.}(2023)]%
        {paypal-22-phishing}
\bibfield{author}{\bibinfo{person}{{PayPal, Inc.}}}
  \bibinfo{year}{2023}\natexlab{}.
\newblock \bibinfo{title}{{PayPal Security Center: How to Identify Fake
  Messages}}.
\newblock
\newblock
\newblock
\shownote{\url{https://www.paypal.com/us/security/learn-about-fake-messages},
  as of \today}.


\bibitem[Petelka et~al\mbox{.}(2019)]%
        {petelka-19-phishing-warnings}
\bibfield{author}{\bibinfo{person}{Justin Petelka}, \bibinfo{person}{Yixin
  Zou}, {and} \bibinfo{person}{Florian Schaub}.}
  \bibinfo{year}{2019}\natexlab{}.
\newblock \showarticletitle{{Put Your Warning Where Your Link Is: Improving and
  Evaluating Email Phishing Warnings}}. In \bibinfo{booktitle}{\emph{ACM
  Conference on Human Factors in Computing Systems}}
  \emph{(\bibinfo{series}{CHI~'19})}. \bibinfo{publisher}{ACM},
  \bibinfo{address}{Glasgow, Scotland, United Kingdom},
  \bibinfo{pages}{518:1--518:15}.
\newblock


\bibitem[{RedFox Community Member}(2022)]%
        {redfox-22-netflix-new-device}
\bibfield{author}{\bibinfo{person}{{RedFox Community Member}}.}
  \bibinfo{year}{2022}\natexlab{}.
\newblock \bibinfo{title}{{Netflix Email Message: ``A new device is using your
  account''}}.
\newblock
\newblock
\newblock
\shownote{\url{https://forum.redfox.bz/threads/netflix-email-message-a-new-device-is-using-your-account.86205/},
  as of \today}.


\bibitem[Redmiles(2019)]%
        {redmiles-19-should-worry}
\bibfield{author}{\bibinfo{person}{Elissa~M. Redmiles}.}
  \bibinfo{year}{2019}\natexlab{}.
\newblock \showarticletitle{{``Should I Worry?'' A Cross-Cultural Examination
  of Account Security Incident Response}}. In \bibinfo{booktitle}{\emph{IEEE
  Symposium on Security and Privacy}} \emph{(\bibinfo{series}{SP~'19})}.
  \bibinfo{publisher}{IEEE}, \bibinfo{address}{San Francisco, California, USA},
  \bibinfo{pages}{920--934}.
\newblock


\bibitem[Redmiles et~al\mbox{.}(2017)]%
        {redmiles-17-2fa-msg-design}
\bibfield{author}{\bibinfo{person}{Elissa~M. Redmiles},
  \bibinfo{person}{Everest Liu}, {and} \bibinfo{person}{Michelle~L. Mazurek}.}
  \bibinfo{year}{2017}\natexlab{}.
\newblock \showarticletitle{{You Want Me To Do What? A Design Study of
  Two-Factor Authentication Messages}}. In \bibinfo{booktitle}{\emph{Who Are
  You?! Adventures in Authentication Workshop}}
  \emph{(\bibinfo{series}{WAY~'17})}. \bibinfo{publisher}{USENIX},
  \bibinfo{address}{Santa Clara, California, USA}, \bibinfo{pages}{1--5}.
\newblock


\bibitem[Redmiles et~al\mbox{.}(2020)]%
        {redmiles-20-security-advice}
\bibfield{author}{\bibinfo{person}{Elissa~M. Redmiles}, \bibinfo{person}{Noel
  Warford}, \bibinfo{person}{Amritha Jayanti}, \bibinfo{person}{Aravind
  Koneru}, \bibinfo{person}{Sean Kross}, \bibinfo{person}{Miraida Morales},
  \bibinfo{person}{Rock Stevens}, {and} \bibinfo{person}{Michelle~L. Mazurek}.}
  \bibinfo{year}{2020}\natexlab{}.
\newblock \showarticletitle{{A Comprehensive Quality Evaluation of Security and
  Privacy Advice on the Web}}. In \bibinfo{booktitle}{\emph{USENIX Security
  Symposium}} \emph{(\bibinfo{series}{SSYM~'20})}. \bibinfo{publisher}{USENIX},
  \bibinfo{address}{Virtual Conference}, \bibinfo{pages}{89--108}.
\newblock


\bibitem[Reeder et~al\mbox{.}(2018)]%
        {reeder-18-warning-reaction}
\bibfield{author}{\bibinfo{person}{Robert~W. Reeder},
  \bibinfo{person}{Adrienne~Porter Felt}, \bibinfo{person}{Sunny Consolvo},
  \bibinfo{person}{Nathan Malkin}, \bibinfo{person}{Christopher Thompson},
  {and} \bibinfo{person}{Serge Egelman}.} \bibinfo{year}{2018}\natexlab{}.
\newblock \showarticletitle{{An Experience Sampling Study of User Reactions to
  Browser Warnings in the Field}}. In \bibinfo{booktitle}{\emph{ACM Conference
  on Human Factors in Computing Systems}} \emph{(\bibinfo{series}{CHI~'18})}.
  \bibinfo{publisher}{ACM}, \bibinfo{address}{Montreal, Quebec, Canada},
  \bibinfo{pages}{512:1--512:13}.
\newblock


\bibitem[Reeder et~al\mbox{.}(2017)]%
        {reeder-17-152-simple-steps}
\bibfield{author}{\bibinfo{person}{Robert~W. Reeder}, \bibinfo{person}{Iulia
  Ion}, {and} \bibinfo{person}{Sunny Consolvo}.}
  \bibinfo{year}{2017}\natexlab{}.
\newblock \showarticletitle{{152 Simple Steps to Stay Safe Online: Security
  Advice for Non-Tech-Savvy Users}}.
\newblock \bibinfo{journal}{\emph{IEEE Security \& Privacy}}
  \bibinfo{volume}{15}, \bibinfo{number}{5} (\bibinfo{date}{Oct.}
  \bibinfo{year}{2017}), \bibinfo{pages}{55--64}.
\newblock


\bibitem[Sasse(2015)]%
        {sasse-15-bullying}
\bibfield{author}{\bibinfo{person}{Angela Sasse}.}
  \bibinfo{year}{2015}\natexlab{}.
\newblock \showarticletitle{{Scaring and Bullying People into Security Won't
  Work}}.
\newblock \bibinfo{journal}{\emph{IEEE Security \& Privacy}}
  \bibinfo{volume}{13}, \bibinfo{number}{3} (\bibinfo{date}{May}
  \bibinfo{year}{2015}), \bibinfo{pages}{80--83}.
\newblock


\bibitem[{SECUSO Research Group, KIT}(2022)]%
        {secuso-22-detect-phishing}
\bibfield{author}{\bibinfo{person}{{SECUSO Research Group, KIT}}.}
  \bibinfo{year}{2022}\natexlab{}.
\newblock \bibinfo{title}{{How to Detect Fraudulent and Phishing Messages}}.
\newblock
\newblock
\newblock
\shownote{\url{https://secuso.aifb.kit.edu/betr-nachrichten-flyer2EN}, as of
  \today}.


\bibitem[Shepard and Metzler(1971)]%
        {shepard-71-mental-rotation}
\bibfield{author}{\bibinfo{person}{Roger~N. Shepard} {and}
  \bibinfo{person}{Jacqueline Metzler}.} \bibinfo{year}{1971}\natexlab{}.
\newblock \showarticletitle{{Mental Rotation of Three-Dimensional Objects}}.
\newblock \bibinfo{journal}{\emph{Science}} \bibinfo{volume}{171},
  \bibinfo{number}{3972} (\bibinfo{date}{Feb.} \bibinfo{year}{1971}),
  \bibinfo{pages}{701--703}.
\newblock


\bibitem[Song et~al\mbox{.}(2019)]%
        {song-19-sharing-workplace}
\bibfield{author}{\bibinfo{person}{Yunpeng Song}, \bibinfo{person}{Cori
  Faklaris}, \bibinfo{person}{Zhongmin Cai}, \bibinfo{person}{Jason~I. Hong},
  {and} \bibinfo{person}{Laura Dabbish}.} \bibinfo{year}{2019}\natexlab{}.
\newblock \showarticletitle{{Normal and Easy: Account Sharing Practices in the
  Workplace}}. In \bibinfo{booktitle}{\emph{ACM Conference on
  Computer-Supported Cooperative Work and Social Computing}}
  \emph{(\bibinfo{series}{CSCW~'19})}. \bibinfo{publisher}{ACM},
  \bibinfo{address}{Austin, Texas, USA}, \bibinfo{pages}{83:1--83:25}.
\newblock


\bibitem[StatCounter(2023a)]%
        {statcounter-22-browser-market}
\bibfield{author}{\bibinfo{person}{StatCounter}.}
  \bibinfo{year}{2023}\natexlab{a}.
\newblock \bibinfo{title}{{Desktop Browser Market Share Worldwide -- June
  2023}}.
\newblock
\newblock
\newblock
\shownote{\url{https://gs.statcounter.com/browser-market-share/desktop/worldwide},
  as of \today}.


\bibitem[StatCounter(2023b)]%
        {statcounter-22-os-market}
\bibfield{author}{\bibinfo{person}{StatCounter}.}
  \bibinfo{year}{2023}\natexlab{b}.
\newblock \bibinfo{title}{{Desktop Operating System Market Share Worldwide --
  June 2023}}.
\newblock
\newblock
\newblock
\shownote{\url{https://gs.statcounter.com/os-market-share/desktop/worldwide},
  as of \today}.


\bibitem[Sunshine et~al\mbox{.}(2009)]%
        {sunshine-09-crying-wolf}
\bibfield{author}{\bibinfo{person}{Joshua Sunshine}, \bibinfo{person}{Serge
  Egelman}, \bibinfo{person}{Hazim Almuhimedi}, \bibinfo{person}{Neha Atri},
  {and} \bibinfo{person}{Lorrie~Faith Cranor}.}
  \bibinfo{year}{2009}\natexlab{}.
\newblock \showarticletitle{{Crying Wolf: An Empirical Study of SSL Warning
  Effectiveness}}. In \bibinfo{booktitle}{\emph{USENIX Security Symposium}}
  \emph{(\bibinfo{series}{SSYM~'09})}. \bibinfo{publisher}{USENIX},
  \bibinfo{address}{San Diego, California, USA}, \bibinfo{pages}{399--416}.
\newblock


\bibitem[Thomas et~al\mbox{.}(2019)]%
        {thomas-19-pw-checkup}
\bibfield{author}{\bibinfo{person}{Kurt Thomas}, \bibinfo{person}{Jennifer
  Pullman}, \bibinfo{person}{Kevin Yeo}, \bibinfo{person}{Ananth Raghunathan},
  \bibinfo{person}{Patrick~Gage Kelley}, \bibinfo{person}{Luca Invernizzi},
  \bibinfo{person}{Borbala Benko}, \bibinfo{person}{Tadek Pietraszek},
  \bibinfo{person}{Sarvar Patel}, \bibinfo{person}{Dan Boneh}, {and}
  \bibinfo{person}{Elie Bursztein}.} \bibinfo{year}{2019}\natexlab{}.
\newblock \showarticletitle{{Protecting Accounts from Credential Stuffing with
  Password Breach Alerting}}. In \bibinfo{booktitle}{\emph{USENIX Security
  Symposium}} \emph{(\bibinfo{series}{SSYM~'19})}. \bibinfo{publisher}{USENIX},
  \bibinfo{address}{Santa Clara, California, USA}, \bibinfo{pages}{1556--1571}.
\newblock


\bibitem[Thompson(2007)]%
        {thompson-07-panas}
\bibfield{author}{\bibinfo{person}{Edmund~R. Thompson}.}
  \bibinfo{year}{2007}\natexlab{}.
\newblock \showarticletitle{{Development and Validation of an Internationally
  Reliable Short-Form of the Positive and Negative Affect Schedule}}.
\newblock \bibinfo{journal}{\emph{Journal of Cross-Cultural Psychology}}
  \bibinfo{volume}{38}, \bibinfo{number}{2} (\bibinfo{date}{March}
  \bibinfo{year}{2007}), \bibinfo{pages}{227--242}.
\newblock


\bibitem[Urquhart(2013)]%
        {urquhart-13-grounded-theory}
\bibfield{author}{\bibinfo{person}{Cathy Urquhart}.}
  \bibinfo{year}{2013}\natexlab{}.
\newblock \bibinfo{booktitle}{\emph{{Grounded Theory for Qualitative Research:
  A Practical Guide}} (\bibinfo{edition}{1} ed.)}.
\newblock \bibinfo{publisher}{{SAGE Publications, Ltd.}},
  \bibinfo{address}{Thousand Oaks, California, USA}.
\newblock


\bibitem[{U.S. Census Bureau}(2023)]%
        {us-22-population}
\bibfield{author}{\bibinfo{person}{{U.S. Census Bureau}}.}
  \bibinfo{year}{2023}\natexlab{}.
\newblock \bibinfo{title}{{U.S. and World Population Clock}}.
\newblock
\newblock
\newblock
\shownote{\url{https://www.census.gov/popclock/}, as of \today}.


\bibitem[Utz et~al\mbox{.}(2019)]%
        {utz-19-uninformed-consent}
\bibfield{author}{\bibinfo{person}{Christine Utz}, \bibinfo{person}{Martin
  Degeling}, \bibinfo{person}{Sascha Fahl}, \bibinfo{person}{Florian Schaub},
  {and} \bibinfo{person}{Thorsten Holz}.} \bibinfo{year}{2019}\natexlab{}.
\newblock \showarticletitle{{(Un)informed Consent: Studying GDPR Consent
  Notices in the Field}}. In \bibinfo{booktitle}{\emph{ACM Conference on
  Computer and Communications Security}} \emph{(\bibinfo{series}{CCS~'19})}.
  \bibinfo{publisher}{ACM}, \bibinfo{address}{London, United Kingdom},
  \bibinfo{pages}{973--990}.
\newblock


\bibitem[Vandenberg and Kuse(1978)]%
        {vandenberg-78-mental-rotation}
\bibfield{author}{\bibinfo{person}{Steven~G. Vandenberg} {and}
  \bibinfo{person}{Allan~R. Kuse}.} \bibinfo{year}{1978}\natexlab{}.
\newblock \showarticletitle{{Mental Rotations, a Group Test of
  Three-Dimensional Spatial Visualization}}.
\newblock \bibinfo{journal}{\emph{Perceptual and Motor Skills}}
  \bibinfo{volume}{47}, \bibinfo{number}{2} (\bibinfo{date}{Oct.}
  \bibinfo{year}{1978}), \bibinfo{pages}{599--604}.
\newblock


\bibitem[Walkington(2019)]%
        {walkington-19-better-warnings}
\bibfield{author}{\bibinfo{person}{Meridel Walkington}.}
  \bibinfo{year}{2019}\natexlab{}.
\newblock \bibinfo{title}{{Designing Better Security Warnings}}.
\newblock
\newblock
\newblock
\shownote{\url{https://blog.mozilla.org/ux/2019/03/designing-better-security-warnings/},
  as of \today}.


\bibitem[Walsh et~al\mbox{.}(2021)]%
        {walsh-21-intercultural-analysis}
\bibfield{author}{\bibinfo{person}{Kathryn Walsh}, \bibinfo{person}{Faiza
  Tazi}, \bibinfo{person}{Philipp Markert}, {and} \bibinfo{person}{Sanchari
  Das}.} \bibinfo{year}{2021}\natexlab{}.
\newblock \showarticletitle{{My Account Is Compromised -- What Do I Do? Towards
  an Intercultural Analysis of Account Remediation for Websites}}. In
  \bibinfo{booktitle}{\emph{Workshop on Inclusive Privacy and Security}}
  \emph{(\bibinfo{series}{WIPS~'21})}. \bibinfo{publisher}{SSRN Electronic
  Journal}, \bibinfo{address}{Virtual Conference}, \bibinfo{pages}{1--6}.
\newblock


\bibitem[Wardle(2019)]%
        {wardle-19-get-owned}
\bibfield{author}{\bibinfo{person}{David Wardle}.}
  \bibinfo{year}{2019}\natexlab{}.
\newblock \bibinfo{booktitle}{\emph{{How Long Does It Take To Get Owned?}}}
\newblock \bibinfo{type}{Technical Report} RHUL–ISG–2019–4.
  \bibinfo{institution}{Royal Holloway University of London}.
\newblock


\bibitem[Wiefling et~al\mbox{.}(2019)]%
        {wiefling-19-rba-in-the-wild}
\bibfield{author}{\bibinfo{person}{Stephan Wiefling}, \bibinfo{person}{Luigi~Lo
  Iacono}, {and} \bibinfo{person}{Markus D\"{u}rmuth}.}
  \bibinfo{year}{2019}\natexlab{}.
\newblock \showarticletitle{{Is This Really You? An Empirical Study on
  Risk-Based Authentication Applied in the Wild}}. In
  \bibinfo{booktitle}{\emph{International Conference on ICT Systems Security
  and Privacy Protection}} \emph{(\bibinfo{series}{IFIP~SEC~'19})}.
  \bibinfo{publisher}{IFIP}, \bibinfo{address}{Lisbon, Portugal},
  \bibinfo{pages}{134--148}.
\newblock


\bibitem[Wiefling et~al\mbox{.}(2020)]%
        {wiefling-20-rba-evaluation}
\bibfield{author}{\bibinfo{person}{Stephan Wiefling}, \bibinfo{person}{Tanvi
  Patil}, \bibinfo{person}{Markus D\"{u}rmuth}, {and} \bibinfo{person}{Luigi
  Lo~Iacono}.} \bibinfo{year}{2020}\natexlab{}.
\newblock \showarticletitle{{Evaluation of Risk-based Re-Authentication
  Methods}}. In \bibinfo{booktitle}{\emph{International Conference on ICT
  Systems Security and Privacy Protection}}
  \emph{(\bibinfo{series}{IFIP~SEC~'20})}. \bibinfo{publisher}{IFIP},
  \bibinfo{address}{Virtual Conference}, \bibinfo{pages}{280--294}.
\newblock


\bibitem[Zou et~al\mbox{.}(2019)]%
        {zou-19-might-be-affected}
\bibfield{author}{\bibinfo{person}{Yixin Zou}, \bibinfo{person}{Shawn Danino},
  \bibinfo{person}{Kaiwen Sun}, {and} \bibinfo{person}{Florian Schaub}.}
  \bibinfo{year}{2019}\natexlab{}.
\newblock \showarticletitle{{You `Might' Be Affected: An Empirical Analysis of
  Readability and Usability Issues in Data Breach Notifications}}. In
  \bibinfo{booktitle}{\emph{ACM Conference on Human Factors in Computing
  Systems}} \emph{(\bibinfo{series}{CHI~'19})}. \bibinfo{publisher}{ACM},
  \bibinfo{address}{Glasgow, Scotland, United Kingdom},
  \bibinfo{pages}{194:1--194:14}.
\newblock


\end{thebibliography}
\clearpage

\appendix

\section{Survey Instrument}\label{app:part2}
\newcommand{\mychoice}[1]{{$\circ$}~#1 \, }
\newcommand{\mymultchoice}[1]{{$\square$}~#1 \, }
\newcommand{\myspecial}[1]{{$\bowtie$}~#1 \, }

\footnotesize
\noindent\small{\textbf{Stage 1: Enrollment}}

\footnotesize
\noindent\textbf{Landing Page}\\
This study is used to measure the spatial reasoning ability by letting participants decide whether two displayed objects have the same shape and size. Since there is no detailed information about changes in this ability over time, you can help us to close this gap by participating in this multi-stage study.
What do you have to do?
\begin{itemize}
    \item Create an account. This allows us to observe changes over time.
    \item Assess your spatial reasoning ability by completing the five rounds.
    \item Participate in the additionally payed recall stages to enable us to analyze how your abilities change over time.
\end{itemize}
If you want to learn more about spatial reasoning or the study itself, visit the About page. In case you have any questions, please do not hesitate to contact us via our email address. A typical reply is within 24 hours, or sooner.

\noindent\textbf{Consent Form}\\
Please indicate, in the boxes below, that you are at least 18 years old, have read and understood this consent form and agree to participate in this study.
\begin{itemize}
    \item[$\square$] I am at least 18 years old.
    \item[$\square$] I have read and understood this consent form.
    \item[$\square$] I voluntarily agree to participate in this study.
\end{itemize}
\textit{The full consent form is left out for space reasons.}

\noindent\textbf{Privacy Note}\\
On the next page, you are asked to create an account. Make sure to use an email address you frequently check, as we will use it to send you the invitations to the subsequent stages. At the end of the study, we will delete your email address. Please use a secure and unique password to prevent others from accessing your personal information during the study. We recommend treating this account like other important accounts you have, e.g., your email account.
\begin{itemize}
    \item[$\square$] I understand this is an important account, and I am responsible for it.
\end{itemize}

\noindent\textbf{Account Creation}\\
Please create an account by providing the data in the fields below.\\
Email: \_\_\_\_\_\_\_\_\_ \\
Password: \_\_\_\_\_\_\_\_\_ \\
Confirm Password: \_\_\_\_\_\_\_\_\_

\noindent\textbf{Email Address Confirmation}\\
Confirm your email address by providing the code or clicking the link we just sent you via email to: \textit{\{participant's email address\}} \\
If you need to change the email address, you can go back to the previous step.

\noindent\textbf{Explanation}\\
On the following five pages you will see pairs of perspective line drawings. Please decide for each pair whether the two drawings portray objects with the \textbf{same} shape and size, i.e., are congruent with respect to three-dimensional shape, or depict objects of \textbf{different} three-dimensional shapes.

\noindent\textit{{\color{teal}The following page was shown 5 times}}\\
\noindent\textbf{\textit{\{No.\}} Perspective Line Pair} \\
Please decide for each pair whether the two drawings portray objects with the \textbf{same} shape and size, i.e., are congruent with respect to three-dimensional shape, or depict objects of \textbf{different} three-dimensional shapes.

\noindent\textbf{Demography}\\
To improve the quality of our research, we kindly ask you to provide some demographic information in the form below.
\aptLtoX[graphic=no, type=html]{
\begin{enumerate}

\item[MD1]{} What is your age range?\\
\mychoice{18--24}
\mychoice{25--34}
\mychoice{35--44} 
\mychoice{45--54}
\mychoice{55--64}
\mychoice{65--74}
\mychoice{75+}\\
\mychoice{Prefer not to answer}
\label{app:part2:d1}

\item[MD2]{} Which of these best describes your current gender identity? \\
\mychoice{Woman}
\mychoice{Man}
\mychoice{Non-binary}
\mychoice{Prefer to self-describe: \_\_\_\_\_\_\_\_\_}
\mychoice{Prefer not to answer}
\label{app:part2:d2}

\item[MD3]{} What is the highest degree or level of school you have completed? \\
\mychoice{No schooling completed}
\mychoice{Some high school, no diploma}\\
\mychoice{High school graduate, diploma, or equivalent}
\mychoice{Some college}\\
\mychoice{Trade, technical, or vocational training}
\mychoice{Associate's degree}\\
\mychoice{Bachelor's degree} 
\mychoice{Master's degree}
\mychoice{Professional degree}
\mychoice{Doctorate}\\
\mychoice{Prefer not to answer}
\label{app:part2:d3}
\end{enumerate}

\begin{enumerate}
\item[MAC1]{} Please select `Agree' as the answer to this question. \\
\mychoice{Strongly disagree}
\mychoice{Disagree}
\mychoice{Neither agree or disagree} 
\mychoice{Agree} 
\mychoice{Strongly agree}
\label{app:part2:ac1}
\end{enumerate}

\begin{enumerate}
\item[MD4]{} Which of the following best describes your educational background or job field?  \\
\mychoice{I have an education in, or work in, the field of computer science, computer engineering or IT.} \\
\mychoice{I do not have an education in, nor do I work in, the field of computer science, computer engineering or IT.} \\
\mychoice{Prefer not to answer}
\label{app:part2:d4}
\end{enumerate}
}{
\begin{enumerate}[leftmargin=3em, label=\textbf{MD\arabic*},nolistsep]

\item What is your age range?\\
\mychoice{18--24}
\mychoice{25--34}
\mychoice{35--44} 
\mychoice{45--54}
\mychoice{55--64}
\mychoice{65--74}
\mychoice{75+}\\
\mychoice{Prefer not to answer}
\label{app:part2:d1}

\item Which of these best describes your current gender identity? \\
\mychoice{Woman}
\mychoice{Man}
\mychoice{Non-binary}
\mychoice{Prefer to self-describe: \rule{1.5cm}{.1pt}}
\mychoice{Prefer not to answer}
\label{app:part2:d2}

\item What is the highest degree or level of school you have completed? \\
\mychoice{No schooling completed}
\mychoice{Some high school, no diploma}\\
\mychoice{High school graduate, diploma, or equivalent}
\mychoice{Some college}\\
\mychoice{Trade, technical, or vocational training}
\mychoice{Associate's degree}\\
\mychoice{Bachelor's degree} 
\mychoice{Master's degree}
\mychoice{Professional degree}
\mychoice{Doctorate}\\
\mychoice{Prefer not to answer}
\label{app:part2:d3}
\end{enumerate}

\begin{enumerate}[leftmargin=3em, label=\textbf{MAC1},nolistsep]
\item Please select `Agree' as the answer to this question. \\
\mychoice{Strongly disagree}
\mychoice{Disagree}
\mychoice{Neither agree or disagree} 
\mychoice{Agree} 
\mychoice{Strongly agree}
\label{app:part2:ac1}
\end{enumerate}

\begin{enumerate}[leftmargin=3em, label=\textbf{MD\arabic*},nolistsep]
\setcounter{enumi}{3}

\item  Which of the following best describes your educational background or job field?  \\
\mychoice{I have an education in, or work in, the field of computer science, computer engineering or IT.} \\
\mychoice{I do not have an education in, nor do I work in, the field of computer science, computer engineering or IT.} \\
\mychoice{Prefer not to answer}
\label{app:part2:d4}
\end{enumerate}
}

\noindent\textbf{Thank you for taking the survey!} \\
The invitation for the second stage will be sent in \textbf{2 weeks}. You can now close this window.

\noindent\textit{{\color{teal}Stage 2 was only completed by participants in the legitimate group}}\\
\noindent\small{\textbf{Stage 2: Recall} }

\footnotesize
\noindent\textbf{Landing Page}\\
This study is used to measure the spatial reasoning ability by letting participants decide whether two displayed objects have the same shape and size. Since there is no detailed information about changes in this ability over time, you can help us to close this gap by participating in this multi-stage study.
What do you have to do?
\begin{itemize}
    \item Log in with your account that you created at the beginning of stage one.
    \item Assess your spatial reasoning ability by completing the five rounds which enables us to analyze how your ability has changed over time.
\end{itemize}
If you want to learn more about spatial reasoning or the study itself, visit the About page. In case you have any questions, please do not hesitate to contact us via our email address. A typical reply is within 24 hours, or sooner.

\noindent\textbf{Sign In}\\
Please sign in with the account you created for this study.\\
Email: \_\_\_\_\_\_\_\_\_ \\
Password: \_\_\_\_\_\_\_\_\_

\noindent\textbf{Explanation}\\
On the following five pages you will see pairs of perspective line drawings. Please decide for each pair whether the two drawings portray objects with the \textbf{same} shape and size, i.e., are congruent with respect to three-dimensional shape, or depict objects of \textbf{different} three-dimensional shapes.

\noindent\textit{{\color{teal}The following page was shown 5 times}} \\
\noindent\textbf{\textit{\{No.\}} Perspective Line Pair}\\
Please decide for each pair whether the two drawings portray objects with the \textbf{same} shape and size, i.e., are congruent with respect to three-dimensional shape, or depict objects of \textbf{different} three-dimensional shapes.

\noindent\textbf{Thank you for taking the survey!} \\
We will send you the compensation for completing the second stage of this study shortly. In \textbf{2 days}, we will send the invitation for the third and final stage, for which an \textbf{additional \$4.00} is paid. You can now close this window.
\vspace{1em}

\noindent\small{\textbf{Stage 3: Questionnaire}}
\footnotesize

\noindent\textbf{Landing Page}\\
This study is used to measure the spatial reasoning ability by letting participants decide whether two displayed objects have the same shape and size. Since there is no detailed information about changes in this ability over time, you can help us to close this gap by participating in this multi-stage study.
What do you have to do?
\begin{itemize}
    \item Log in with your account that you created at the beginning of stage one.
    \item Assess your spatial reasoning ability by completing the five rounds which enables us to analyze how your ability has changed over time.
\end{itemize}
If you want to learn more about spatial reasoning or the study itself, visit the About page. In case you have any questions, please do not hesitate to contact us via our email address. A typical reply is within 24 hours, or sooner.

\noindent\textbf{Sign In}\\
Please sign in with the account you created for this study.\\
Email: \_\_\_\_\_\_\_\_\_ \\
Password: \_\_\_\_\_\_\_\_\_

\noindent\textbf{Debriefing} \\
\noindent\textit{The main part of the research that is relevant for us starts on the next page.\\Please do not close the browser window yet.} \\
What are we trying to learn in this research? \\
Unlike initially explained, the goal of this study is to better understand the effectiveness of \textbf{sign-in emails}. Our only interest surrounds your interaction with the \textbf{sign-in email} you received for your account during this study. The spatial reasoning task's only purpose was giving the study a meaningful primary task that does not hint at the true purpose of our study. Your answers in the spatial reasoning task have been stored, and may be analyzed, but are not of primary interest for our research. \\
What data was collected? \\
As part of this study, we collected usage data about the sign-in emails, including whether users changed their password and how long it took them to react to the sign-in email. All your responses are only stored anonymously \textbf{using a random identifier}. Moreover, we separated all your survey responses from your email address, to\textbf{ prevent any chance of re-identification}. All collected data was \textbf{encrypted}, and all identifiable data (such as your email addresses) \textbf{will be deleted} at the end of the study. \\
Why is this important to scientists or the general public? \\
Our work is concerned with designing systems to help users keep their accounts secure. Part of designing good security systems is usability: if people cannot use a system, they will not be able to keep their accounts secure. By better understanding the usability and effectiveness of sign-in emails, we will be able to create systems that are usable and secure. \\
What if I have question later? \\
If you have any remaining concerns, questions, or comments about the experiment, please feel free to contact us.
To continue in the study, please continue to the next page. If you do not want to participate anymore you can click here.

\noindent\textbf{Email} \\
\textit{{\color{teal} The \underline{individual} login notification we sent to the participant is displayed for later reference (see Figure~\ref{fig:baselineemail}, but re-branded to match the SRS~study).}}

\noindent \textit{{\color{teal}If participant has not changed their password.}}

\aptLtoX[graphic=no, type=html]{
\begin{enumerate}
\item[]{\textbf{MQ0}} Do you \textbf{remember} receiving \textbf{this email}?\\
\mychoice{Yes}
\mychoice{No}
\label{app:part2:s1}
\end{enumerate}
}{
\begin{enumerate}[leftmargin=3em, label=\textbf{MQ\arabic*}]
\setcounter{enumi}{-1}
\item Do you \textbf{remember} receiving \textbf{this email}?\\
\mychoice{Yes}
\mychoice{No}
\label{app:part2:s1}
\end{enumerate}
}

\noindent \textit{{\color{teal}Participants who selected `No' in \aptLtoX[graphic=no, type=html]{\textbf{MQ0}}{\ref{app:part2:s1}} were forwarded to \aptLtoX[graphic=no, type=html]{\textbf{MQ10}}{\ref{app:part2:q10}}.}}

\mylabel{app:part2:panas}{MP}
\noindent\textbf{~MP\hspace{1.1em}I-PANAS-SF} \\
\noindent Now we would like to know how you felt in reaction to the email. The list below consists of a number of words that describe different feelings and emotions. Read each item and then mark the appropriate answer on the list. \textbf{Indicate to what extent you felt this way when you noticed the email}.
\vspace{-1em}
\begin{table}[htbp]
\centering
\footnotesize
\setlength{\tabcolsep}{.4\tabcolsep}
\begin{tabular}{lccccc}
\toprule
& Very slightly  \\
& or not at all  & A little & Moderately & Quite a bit & Extremely \\
& (1) & (2) & (3) & (4) & (5) \\
\midrule
Upset       & $\bigcirc$ & $\bigcirc$ & $\bigcirc$ & $\bigcirc$ & $\bigcirc$ \\
Hostile     & $\bigcirc$ & $\bigcirc$ & $\bigcirc$ & $\bigcirc$ & $\bigcirc$ \\
Alert       & $\bigcirc$ & $\bigcirc$ & $\bigcirc$ & $\bigcirc$ & $\bigcirc$ \\
Ashamed     & $\bigcirc$ & $\bigcirc$ & $\bigcirc$ & $\bigcirc$ & $\bigcirc$ \\
Inspired    & $\bigcirc$ & $\bigcirc$ & $\bigcirc$ & $\bigcirc$ & $\bigcirc$ \\
Nervous     & $\bigcirc$ & $\bigcirc$ & $\bigcirc$ & $\bigcirc$ & $\bigcirc$ \\
Determined  & $\bigcirc$ & $\bigcirc$ & $\bigcirc$ & $\bigcirc$ & $\bigcirc$ \\
Attentive   & $\bigcirc$ & $\bigcirc$ & $\bigcirc$ & $\bigcirc$ & $\bigcirc$ \\
Afraid      & $\bigcirc$ & $\bigcirc$ & $\bigcirc$ & $\bigcirc$ & $\bigcirc$ \\
Active      & $\bigcirc$ & $\bigcirc$ & $\bigcirc$ & $\bigcirc$ & $\bigcirc$ \\
\bottomrule
\end{tabular}
\end{table}
\vspace{-1em}

\noindent\textbf{Reaction}
\aptLtoX[graphic=no, type=html]{
\begin{enumerate}
\item[MQ1]{} Did you \textbf{read} this \textbf{email} when you received it? (email as shown on the left)\\
\mychoice{I did not read it at all}
\mychoice{I only read the subject but not the body}\\
\mychoice{I read the subject and skimmed the body}
\mychoice{I fully read it}
\label{app:part2:q1}
\end{enumerate}

\begin{enumerate}
\item[MQ2a]{} In reaction to this email, you decided to change your password.\\Please \textbf{describe} any \textbf{other actions} you took.\\
Answer: \_\_\_\_\_\_\_\_\_\_\_\_\_\_\_\_\_\_\_\_\_\_\_\_\_\_\_
\label{app:part2:q2a}
\end{enumerate}

\begin{enumerate}
\item[MQ3a]{} \textbf{Why} did you react this way, i.e., change your password and take the other actions you described.\\
Answer: \_\_\_\_\_\_\_\_\_\_\_\_\_\_\_\_\_\_\_\_\_\_\_\_\_\_\_
\label{app:part2:q3a}
\end{enumerate}

\noindent \textit{{\color{teal}If participant has not changed their password.}}

\begin{enumerate}
\item[MQ2b]{} \textbf{What} did you do \textbf{in reaction} to it?\\
Answer: \_\_\_\_\_\_\_\_\_\_\_\_\_\_\_\_\_\_\_\_\_\_\_\_\_\_\_
\label{app:part2:q2b}
\end{enumerate}

\begin{enumerate}
\item[MQ3b]{} \textbf{Why} did you react this way?\\
Answer: \_\_\_\_\_\_\_\_\_\_\_\_\_\_\_\_\_\_\_\_\_\_\_\_\_\_\_
\label{app:part2:q3b}
\end{enumerate}

\noindent\textbf{Content \& Design}
\begin{enumerate}
\item[MQ4]{} How much did the following \textbf{factors influence} your \textbf{reaction}?\\
\textit{{\color{teal}Answer choice per item: No effect (1) -- Major effect (5).}}\\
\myspecial{Email metadata (e.g., sender, subject, time of arrival)}\\
\myspecial{Email content (e.g., information, instructions, wording)}\\
\myspecial{Email design (e.g., structure, color, font size)}\\
\myspecial{Experience in dealing with such emails}\\
\myspecial{Negative experience with security and privacy incidents (e.g., data breach, identity theft)} \\
\myspecial{Email appeared to be phishing}\\
\myspecial{Expected to receive such an email}\\
\myspecial{Other: \_\_\_\_\_\_\_\_\_}\\
\textit{{\color{teal}Answer choices were randomly ordered.}}
\label{app:part2:q4}

\item[MQ5]{} Please rate how \textbf{helpful} the following \textbf{information} was for \textbf{deciding how to react} to this email?\\
\textit{{\color{teal}Answer choice per item: Not at all helpful (1) -- Extremely helpful (5).}}\\
\myspecial{Affected account name (i.e., email address)}
\myspecial{Location}
\myspecial{Date}
\myspecial{Device}
\label{app:part2:q5}
\end{enumerate}

\noindent \textbf{Time \& Location}
\begin{enumerate}
\item[MQ6]{} \textbf{When} did you read the email?\\
\mychoice{I never read it}
\mychoice{Immediately after I noticed it}\\
\mychoice{Less than 1 hour after I noticed it}
\mychoice{A few hours after I noticed it}\\
\mychoice{One day after I noticed it}
\mychoice{More than one day after I noticed it}\\
\mychoice{I do not remember}
\label{app:part2:q6}
\end{enumerate}

\noindent \textit{{\color{teal}If participant has not selected ``Never'' in \textbf{MQ6}: }}

\begin{enumerate}
\item[MQ7]{} \textbf{In which US state} have you been when you \textbf{read} the \textbf{email}?\\
\textit{Dropdown with all 50 US states + District of Columbia}.\\If somewhere outside the USA: \_\_\_\_\_\_\_\_\_
\label{app:part2:q7}

\item[MQ8]{} \textbf{Where} did you read the email?\\
\mychoice{At home}
\mychoice{At work}
\mychoice{On the go}
\mychoice{Somewhere else: \_\_\_\_\_\_\_\_\_}\\
\mychoice{I do not remember}
\label{app:part2:q8}
\end{enumerate}

\begin{enumerate}
\item[MAC2]{} Please select `Agree' as the \textbf{answer} to this question. \\
\mychoice{Strongly disagree}
\mychoice{Disagree}
\mychoice{Neither agree or disagree}
\mychoice{Agree} 
\mychoice{Strongly agree}
\label{app:part2:ac2}
\end{enumerate}

\begin{enumerate}
\item[MQ9]{} In case you received the email at a \textbf{different location or different time}, would your \textbf{reaction} to it been any \textbf{different}?\\
\mychoice{Yes}
\mychoice{No}
\mychoice{Do not know}
\label{app:part2:q9}
\end{enumerate}

\noindent \textit{{\color{teal}If participant selected ``Yes'' in \textbf{MQ8}: }}

\begin{enumerate}
\item[MQ10]{} \textbf{What} would you have \textbf{done differently}, if you had received the email at a \textbf{different location or different time}?\\
Answer: \_\_\_\_\_\_\_\_\_\_\_\_\_\_\_\_\_\_\_\_\_\_\_\_\_\_\_
\label{app:part2:q10}
\end{enumerate}

\noindent\textbf{Comprehension}
\begin{enumerate}
\item[MQ11]{} In your opinion, \textbf{why} have you \textbf{received} this email?\\
Answer: \_\_\_\_\_\_\_\_\_\_\_\_\_\_\_\_\_\_\_\_\_\_\_\_\_\_\_
\label{app:part2:q11}
\end{enumerate}

\noindent \textbf{Expectation}
\begin{enumerate}
\item[MQ12]{} In your opinion, \textbf{when} should real companies \textbf{send emails} like this one? (Select all that apply)\\
\mymultchoice{Never}\\
\mymultchoice{After every detected sign-in which suggests that something is suspicious or wrong} \\
\mymultchoice{After every detected sign-in when I have not signed in for a while}\\
\mymultchoice{After every detected sign-in from a new device}\\
\mymultchoice{After every detected sign-in at an unusual time of the day (e.g., in the middle of the night)}\\
\mymultchoice{After every detected sign-in from a new location}\\
\mymultchoice{After every detected sign-in}\\
\mymultchoice{Other: \_\_\_\_\_\_\_\_\_}
\label{app:part2:q12}
\end{enumerate}

\noindent \textit{{\color{teal}If participant selected ``Never'' in \textbf{MQ12}: }}

\begin{enumerate}
\item[MQ13]{} In your opinion, \textbf{why} do you think real companies should \textbf{never send emails} like this one?\\
Answer: \_\_\_\_\_\_\_\_\_\_\_\_\_\_\_\_\_\_\_\_\_\_\_\_\_\_\_
\label{app:part2:q13}
\end{enumerate}

\noindent \textbf{Prior Experience}
\begin{enumerate}
\item[MQ14]{} Have you had any \textbf{negative experiences} with a \textbf{security or privacy} incident within the \textbf{last two years} (e.g., data breach, identity theft)?\\
\mychoice{Yes}
\mychoice{No}
\label{app:part2:q14}

\item[MQ15]{} \textbf{Regularly changing} my \textbf{password} (e.g., every 90 days) \textbf{increases} the \textbf{security} of my account.\\
\mychoice{Strongly disagree}
\mychoice{Disagree}
\mychoice{Neither agree or disagree}
\mychoice{Agree} 
\mychoice{Strongly agree}
\label{app:part2:q15}

\item[MQ16]{} \textbf{Changing} my \textbf{password} after it \textbf{has been breached} \textbf{increases} the \textbf{security} of my account.\\
\mychoice{Strongly disagree}
\mychoice{Disagree}
\mychoice{Neither agree or disagree}
\mychoice{Agree} 
\mychoice{Strongly agree}
\label{app:part2:q16}

\end{enumerate}
}{
\begin{enumerate}[leftmargin=3em, label=\textbf{MQ\arabic*}]
\item Did you \textbf{read} this \textbf{email} when you received it? (email as shown on the left)\\
\mychoice{I did not read it at all}
\mychoice{I only read the subject but not the body}\\
\mychoice{I read the subject and skimmed the body}
\mychoice{I fully read it}
\label{app:part2:q1}
\end{enumerate}

\begin{enumerate}[leftmargin=3em, label=\textbf{MQ2a}]
\item In reaction to this email, you decided to change your password.\\Please \textbf{describe} any \textbf{other actions} you took.\\
Answer: \rule{3.5cm}{.1pt}
\label{app:part2:q2a}
\end{enumerate}

\begin{enumerate}[leftmargin=3em, label=\textbf{MQ3a}]
\item \textbf{Why} did you react this way, i.e., change your password and take the other actions you described.\\
Answer: \rule{3.5cm}{.1pt}
\label{app:part2:q3a}
\end{enumerate}

\noindent \textit{{\color{teal}If participant has not changed their password.}}

\begin{enumerate}[leftmargin=3em, label=\textbf{MQ2b}]
\item \textbf{What} did you do \textbf{in reaction} to it?\\
Answer: \rule{3.5cm}{.1pt}
\label{app:part2:q2b}
\end{enumerate}

\begin{enumerate}[leftmargin=3em, label=\textbf{MQ3b}]
\item \textbf{Why} did you react this way?\\
Answer: \rule{3.5cm}{.1pt}
\label{app:part2:q3b}
\end{enumerate}

\noindent\textbf{Content \& Design}
\begin{enumerate}[leftmargin=3em, label=\textbf{MQ\arabic*}]
\setcounter{enumi}{3}
\item How much did the following \textbf{factors influence} your \textbf{reaction}?\\
\textit{{\color{teal}Answer choice per item: No effect (1) -- Major effect (5).}}\\
\myspecial{Email metadata (e.g., sender, subject, time of arrival)}\\
\myspecial{Email content (e.g., information, instructions, wording)}\\
\myspecial{Email design (e.g., structure, color, font size)}\\
\myspecial{Experience in dealing with such emails}\\
\myspecial{Negative experience with security and privacy incidents (e.g., data breach, identity theft)} \\
\myspecial{Email appeared to be phishing}\\
\myspecial{Expected to receive such an email}\\
\myspecial{Other: \rule{1.5cm}{.1pt}}\\
\textit{{\color{teal}Answer choices were randomly ordered.}}
\label{app:part2:q4}

\item Please rate how \textbf{helpful} the following \textbf{information} was for \textbf{deciding how to react} to this email?\\
\textit{{\color{teal}Answer choice per item: Not at all helpful (1) -- Extremely helpful (5).}}\\
\myspecial{Affected account name (i.e., email address)}
\myspecial{Location}
\myspecial{Date}
\myspecial{Device}
\label{app:part2:q5}
\end{enumerate}

\noindent \textbf{Time \& Location}
\begin{enumerate}[leftmargin=3em, label=\textbf{MQ\arabic*}]
\setcounter{enumi}{5}
\item \textbf{When} did you read the email?\\
\mychoice{I never read it}
\mychoice{Immediately after I noticed it}\\
\mychoice{Less than 1 hour after I noticed it}
\mychoice{A few hours after I noticed it}\\
\mychoice{One day after I noticed it}
\mychoice{More than one day after I noticed it}\\
\mychoice{I do not remember}
\label{app:part2:q6}
\end{enumerate}

\noindent \textit{{\color{teal}If participant has not selected ``Never'' in \ref{app:part2:q6}: }}

\begin{enumerate}[leftmargin=3em, label=\textbf{MQ\arabic*}]
\setcounter{enumi}{6}
\item \textbf{In which US state} have you been when you \textbf{read} the \textbf{email}?\\
\textit{Dropdown with all 50 US states + District of Columbia}.\\If somewhere outside the USA: \rule{1.5cm}{.1pt}
\label{app:part2:q7}

\item \textbf{Where} did you read the email?\\
\mychoice{At home}
\mychoice{At work}
\mychoice{On the go}
\mychoice{Somewhere else: \rule{1.5cm}{.1pt}}\\
\mychoice{I do not remember}
\label{app:part2:q8}
\end{enumerate}

\begin{enumerate}[leftmargin=3em, label=\textbf{MAC2},nolistsep]
\item Please select `Agree' as the \textbf{answer} to this question. \\
\mychoice{Strongly disagree}
\mychoice{Disagree}
\mychoice{Neither agree or disagree}
\mychoice{Agree} 
\mychoice{Strongly agree}
\label{app:part2:ac2}
\end{enumerate}

\begin{enumerate}[leftmargin=3em, label=\textbf{MQ\arabic*}]
\setcounter{enumi}{8}
\item In case you received the email at a \textbf{different location or different time}, would your \textbf{reaction} to it been any \textbf{different}?\\
\mychoice{Yes}
\mychoice{No}
\mychoice{Do not know}
\label{app:part2:q9}
\end{enumerate}

\noindent \textit{{\color{teal}If participant selected ``Yes'' in \ref{app:part2:q8}: }}

\begin{enumerate}[leftmargin=3em, label=\textbf{MQ\arabic*}]
\setcounter{enumi}{9}
\item \textbf{What} would you have \textbf{done differently}, if you had received the email at a \textbf{different location or different time}?\\
Answer: \rule{3.5cm}{.1pt}
\label{app:part2:q10}
\end{enumerate}

\noindent\textbf{Comprehension}
\begin{enumerate}[leftmargin=3em, label=\textbf{MQ\arabic*}]
\setcounter{enumi}{10}
\item In your opinion, \textbf{why} have you \textbf{received} this email?\\
Answer: \rule{3.5cm}{.1pt}
\label{app:part2:q11}
\end{enumerate}

\noindent \textbf{Expectation}
\begin{enumerate}[leftmargin=3em, label=\textbf{MQ\arabic*}]
\setcounter{enumi}{11}

\item In your opinion, \textbf{when} should real companies \textbf{send emails} like this one? (Select all that apply)\\
\mymultchoice{Never}\\
\mymultchoice{After every detected sign-in which suggests that something is suspicious or wrong} \\
\mymultchoice{After every detected sign-in when I have not signed in for a while}\\
\mymultchoice{After every detected sign-in from a new device}\\
\mymultchoice{After every detected sign-in at an unusual time of the day (e.g., in the middle of the night)}\\
\mymultchoice{After every detected sign-in from a new location}\\
\mymultchoice{After every detected sign-in}\\
\mymultchoice{Other: \rule{1.5cm}{.1pt}}
\label{app:part2:q12}
\end{enumerate}

\noindent \textit{{\color{teal}If participant selected ``Never'' in \ref{app:part2:q12}: }}

\begin{enumerate}[leftmargin=3em, label=\textbf{MQ\arabic*}]
\setcounter{enumi}{12}
\item In your opinion, \textbf{why} do you think real companies should \textbf{never send emails} like this one?\\
Answer: \rule{3.5cm}{.1pt}
\label{app:part2:q13}
\end{enumerate}

\noindent \textbf{Prior Experience}
\begin{enumerate}[leftmargin=3em, label=\textbf{MQ\arabic*}]
\setcounter{enumi}{13}
\item Have you had any \textbf{negative experiences} with a \textbf{security or privacy} incident within the \textbf{last two years} (e.g., data breach, identity theft)?\\
\mychoice{Yes}
\mychoice{No}
\label{app:part2:q14}

\item \textbf{Regularly changing} my \textbf{password} (e.g., every 90 days) \textbf{increases} the \textbf{security} of my account.\\
\mychoice{Strongly disagree}
\mychoice{Disagree}
\mychoice{Neither agree or disagree}
\mychoice{Agree} 
\mychoice{Strongly agree}
\label{app:part2:q15}

\item \textbf{Changing} my \textbf{password} after it \textbf{has been breached} \textbf{increases} the \textbf{security} of my account.\\
\mychoice{Strongly disagree}
\mychoice{Disagree}
\mychoice{Neither agree or disagree}
\mychoice{Agree} 
\mychoice{Strongly agree}
\label{app:part2:q16}

\end{enumerate}
}

\noindent{\textbf{One More Thing}}\\
Please indicate if you've honestly participated in this survey and followed instructions completely. You will not be penalized/rejected for indicating `No' but your data may not be included in the analysis: \\
\mychoice{Yes}
\mychoice{No}

\noindent\textbf{Thank you for taking the survey!}\\
We will send you the compensation for completing the final stage of this study shortly. You can now close this window.
\begin{itemize}
    \item[] \textit{{\color{teal}Only for participants in the malicious group:}}
    \item[] \textit{Note, as part of this research, we have sent you an email about a new sign-in. 
    \item[] This sign-in did not take place; \textbf{at no time was your account at risk}.}
\end{itemize}
If you want to learn more about sign-in emails, feel free to visit: \{link\} There we have created some information material for you. The info website will stay online even after the end of this study, so feel free to save the link or share it.
\clearpage

\onecolumn
\section{Real-World Notifications: Email Metadata}\label{app:real-world:emailmetadata}
\setcounter{table}{1}
\begin{table*}[!htb]
    \centering
    \caption{Sender, email address, and subject of the notifications sent by real-world services.}
    \label{tab:notificationsSenderSubject}
    \Description[Information about services of analyzed real-world notifications]{Table shows information about the sender, email address, and subject of the notifications sent by real-world services. Column 1 (Rank) represents the rank of the service on the Tranco list, Column 2 (Display Name) is the name of the service. Column 3 (Email Address) shows the mail address the notification was sent from, and Column 4 (Subject) contains the subject of each notification.}
    \scriptsize
    \begin{tabular}{r l | l l | l }
    \toprule
    \textbf{Rank} & \textbf{Domain} & \textbf{Display Name} & \textbf{Email Address} & \textbf{Subject} \\   
    \midrule
    1 & google.com & Google & no-reply@accounts.google.com & Security alert \\
    \rowcolor[gray]{.9}
    & workspace.google.com & Google Workspace Alerts & google-workspace-alerts-noreply@google.com & Alert: Suspicious login \\
    2 & facebook.com & noreply & noreply@facebookmail.com & Did you use Facebook from somewhere new? \\
    \rowcolor[gray]{.9}
    6 & microsoft.com & Microsoft account team & account-security-noreply@ \dots microsoft.com & Microsoft account unusual sign-in activity \\
    7 & twitter.com & Twitter & verify@twitter.com & New login to Twitter from \{browser\} on \{OS\} \\
    \rowcolor[gray]{.9}
    9 & instagram.com & Instagram & security@mail.instagram.com & New login to Instagram from \{browser\} on \{OS\} \\
    10 & cloudflare.com & Cloudflare & noreply@notify.cloudflare.com & Your Cloudflare account has been accessed from a new IP Address \\
    \rowcolor[gray]{.9}
    13 & apple.com & Apple & noreply@email.apple.com & Your Apple ID was used to sign in to iCloud on a \{device\} \\
    14 & linkedin.com & LinkedIn & security-noreply@linkedin.com & \{Name\}, please verify your new device \\
    \rowcolor[gray]{.9}
    15 & netflix.com & Netflix & info@mailer.netflix.com & A new device is using your account \\ 
    17 & wikipedia.org & Wikipedia & wiki@wikimedia.org & Login to Wikipedia as \{account name\} from a device you have not recently used \\
    \rowcolor[gray]{.9}
    20 & amazon.com & amazon.com & account-update@amazon.com & amazon.com, action needed: Sign-in \\
    25 & yahoo.com & Yahoo & no-reply@cc.yahoo-inc.com & Unexpected sign-in attempt \\
    \rowcolor[gray]{.9}
    32 & github.com & GitHub & noreply@github.com & [GitHub] Please review this sign in \\
    36 & pinterest.com & Pinterest & noreply@account.pinterest.com & New login on your Pinterest account \\
    \rowcolor[gray]{.9}
    63 & vk.com & VK & admin@notify.vk.com & Someone has accessed your account from \{OS\} through \{browser\}, \{country\} \\
    65 & tiktok.com & TikTok & noreply@account.tiktok.com & New device login detected \\
    \rowcolor[gray]{.9}
    72 & mozilla.org & Firefox Accounts & accounts@firefox.com & New sign-in to Firefox \\
    80 & spotify.com & Spotify & no-reply@spotify.com & New login to Spotify \\ 
    \rowcolor[gray]{.9}
    82 & tumblr.com & Tumblr & no-reply@tumblr.com & Your account has been logged into. \\
    83 & paypal.com & service@paypal.com & service@paypal.com & Login from a new device \\
    \rowcolor[gray]{.9}
    97 & ebay.com & eBay & ebay@ebay.com & A new device is using your account \\
    99 & dropbox.com & Dropbox & no-reply@dropbox.com & We noticed a new sign in to your Dropbox \\
    \rowcolor[gray]{.9}
    103 & csdn.net & CSDN & service@register.csdn.net & [CSDN] Notification of remote login \\
    104 & imdb.com & imdb.com & account-update@imdb.com & imdb.com, action needed: Sign-in \\
    \rowcolor[gray]{.9}
    125 & soundcloud.com & SoundCloud Login & no-reply@login.soundcloud.com & SoundCloud sign-in detected from a new device \\
    155 & twitch.tv & Twitch & no-reply@twitch.tv & Your Twitch Account - Successful Log-in \\
    \rowcolor[gray]{.9}
    157 & etsy.com & Etsy & noreply@mail.etsy.com & \{Name\}, did you recently sign into Etsy? \\
    164 & booking.com & - & noreply@booking.com & New sign in to your account \\
    \rowcolor[gray]{.9}
    171 & sourceforge.net & SourceForge & noreply@sourceforge.net & Foreign login to your SourceForge.net account \\
    179 & researchgate.net & ResearchGate & no-reply@researchgatemail.net & New login from \{browser\} on \{OS\} \\
    \rowcolor[gray]{.9}
    180 & oracle.com & Oracle & no-reply@oracle.com & New Device Login Detected with Your Account \\
    186 & slack.com & Slack & feedback@slack.com & Slack account sign in from a new device \\
    \rowcolor[gray]{.9}
    206 & weebly.com & - & noreply@messaging.squareup.com & New login from \{browser\} on \{OS\} \\
    236 & samsung.com & Samsung Account & sa.noreply@samsung-mail.com & New sign in to your Samsung account \\
    \rowcolor[gray]{.9}
    322 & grammarly.com & Grammarly & hello@info.grammarly.com & New Login to Grammarly \\
    328 & fiverr.com & Fiverr & noreply@e.fiverr.com & New login on your Fiverr account \\
    \rowcolor[gray]{.9}
    344 & snapchat.com & Team Snapchat & no\_reply@snapchat.com & New Snapchat Login \\
    381 & yelp.com & Yelp & no-reply@yelp.com & New login to your Yelp account (\{account name\}) \\
    \rowcolor[gray]{.9}
    392 & binance.com & Binance & do-not-reply@ses.binance.com & [Binance] Login Attempted from New IP address \{IP\} - \{time\}(\{timezone\}) \\
    524 & netease.com & NetEase Account Center & passport@service.netease.com & NetEase mailbox account abnormal login reminder \\
    \rowcolor[gray]{.9}
    541 & gitlab.com & GitLab & gitlab@mg.gitlab.com & gitlab.com sign-in from new location \\
    545 & atlassian.com & Atlassian & noreply@am.atlassian.com & Unusual login attempts on your Atlassian account \\
    \rowcolor[gray]{.9}
    563 & uber.com & Uber & noreply@uber.com & New device sign-in \\
    753 & airbnb.com & Airbnb & automated@airbnb.com & Account activity: New login from \{browser\} \\
    \rowcolor[gray]{.9}
    885 & nintendo.com & - & no-reply@accounts.nintendo.com & [Nintendo Account] New sign-in \\
    924 & xing.com & XING & mailrobot@mail.xing.com & New login on XING: \{browser\} \{OS\} \\
    \rowcolor[gray]{.9}
    1205 & wayfair.com & Wayfair & noreply@wayfair.com & New device sign-in \\
    1327 & deezer.com & Deezer Security Team & securityteam@deezer.com & Login troubles? \\
    \rowcolor[gray]{.9}
    1387 & lyft.com & Lyft & noreply@lyftmail.com & New Login \\
    1413 & battle.net & Blizzard Entertainment & noreply@blizzard.com & Help us keep your Blizzard Account safe with a security check \\
    \rowcolor[gray]{.9}
    1576 & agoda.com & Agoda & no-reply@security.agoda.com & New Login to Your Agoda-Account \\
    2476 & 1password.com & 1Password & hello@1password.com & New 1Password sign-in from \{browser\} \\
    \rowcolor[gray]{.9}
    2645 & porkbun.com & Porkbun Support & support@porkbun.com & porkbun.com $\vert$ account security notice - successful login \\
    2705 & synology.com & Synology Account & noreply@synologynotification.com & Synology Account - Security alert \\
    \rowcolor[gray]{.9}
    3179 & faceit.com & FACEIT & no-reply@faceit.com & Login from a new IP \\
    3210 & bitwarden.com & Bitwarden & no-reply@bitwarden.com & New Device Logged In From \{browser\} \\
    \rowcolor[gray]{.9}
    3605 & plex.tv & Plex & noreply@plex.tv & New sign-in to your Plex account \\
    4189 & dhl.de & - & noreply.kundenkonto@dhl.de & Successful login to your DHL account with a new device or browser \\
    \rowcolor[gray]{.9}
    4250 & dashlane.com & Dashlane & no-reply@dashlane.com & New device added to Dashlane \\
    5383 & logmein.com & LogMeIn.com Auto-Mailer & do-not-reply@logmein.com & LogMeIn Audit Notification - Login from an unfamiliar location \\
    \rowcolor[gray]{.9}
    8544 & maxmind.com & - & support@maxmind.com & MaxMind Notification: Unrecognized Device Login \\
    10625 & check24.com & CHECK24 Accounts & customeraccount@check24.com & New Login to Your Customer Account \\
    \rowcolor[gray]{.9}
    16460 & myunidays.com & UNiDAYS & no-reply@myunidays.com & Important: UNiDAYS Log-in Notification \\
    16993 & n26.com & N26 & noreply@n26.com & Action needed: Unusual login to your N26 account \\
    \rowcolor[gray]{.9}
    19535 & neteller.com & NETELLER & no-reply@emails.neteller.com & New device has been detected \\
    25667 & splitwise.com & Splitwise & hello@splitwise.com & New sign-in to your Splitwise account \\
    \rowcolor[gray]{.9}
    27539 & decathlon.com & DECATHLON Service & noreply@services.decathlon.com & DECATHLON: New login to your account \\
    31988 & netatmo.com & Legrand - Netatmo - Bticino & do-not-reply@netatmo.com & Someone has logged into your account \\
    \rowcolor[gray]{.9}
    40161 & stacksocial.com & StackSocial & shop@email.stackcommerce.com & Account Activity Notification \\
    46969 & kinguin.net & Kinguin & help@kinguin.net & New browser login detected \\
    \rowcolor[gray]{.9}
    48031 & traderepublic.com & Trade Republic & service@traderepublic.com & Registration from a new device \\
    \bottomrule
    \end{tabular}
\end{table*}
\clearpage

\section{Real-World Notifications: Features}\label{app:real-world:features}
\setcounter{table}{2}
\begin{table*}[!htb]
    \centering
    \caption{Information contained in notifications sent by real-world services.}
    \label{tab:notificationsGrantedAccess}
    \Description[Information contained in real-world notifications]{Column 1 (Rank) represents the rank of the service on the Tranco list, Column 2 (Domain) contains the services' domain. Columns 3-11 show whether the notification contained the respective information. This is indicated with filled circles (notification contained information) or empty circles (notification did not contain information). Columns 12 and 13 represent whether the notification explicitly contained instructions for the legitimate and the malicious cases.}
    \scriptsize
    \begin{tabular}{r l | c c c c c c c c c | c c }
    \toprule
    \textbf{Rank} & \textbf{Domain} & \textbf{Account Name} & \textbf{Browser} & \textbf{Country} & \textbf{State} & \textbf{City} & \textbf{IP} & \textbf{OS} & \textbf{Time} & \textbf{Time Zone} & \textbf{Instructions Legitimate} & \textbf{Instructions Malicious} \\
    \midrule
    1 & google.com & \nnnewmoon & \nnfullmoon & \nnfullmoon & \nnfullmoon & \nnfullmoon & \nnfullmoon & \nnnewmoon & \nnfullmoon & \nnfullmoon & \nnnewmoon & \nnnewmoon \\
    \rowcolor[gray]{.9}
    & workspace.google.com & \nnnewmoon & \nnfullmoon & \nnfullmoon & \nnfullmoon & \nnfullmoon & \nnnewmoon & \nnfullmoon & \nnfullmoon & \nnfullmoon & \nnfullmoon & \nnfullmoon \\
    2 & facebook.com & \nnnewmoon & \nnnewmoon & \nnnewmoon & \nnfullmoon & \nnnewmoon & \nnfullmoon & \nnnewmoon & \nnnewmoon & \nnnewmoon & \nnnewmoon & \nnnewmoon \\
    \rowcolor[gray]{.9}
    6 & microsoft.com & \nnnewmoon & \nnnewmoon & \nnnewmoon & \nnfullmoon & \nnfullmoon & \nnnewmoon & \nnnewmoon & \nnnewmoon & \nnnewmoon & \nnnewmoon & \nnnewmoon \\
    7 & twitter.com & \nnnewmoon & \nnnewmoon & \nnnewmoon & \nnnewmoon & \nnnewmoon & \nnfullmoon & \nnnewmoon & \nnfullmoon & \nnfullmoon & \nnnewmoon & \nnnewmoon \\
    \rowcolor[gray]{.9}
    9 & instagram.com & \nnnewmoon & \nnnewmoon & \nnnewmoon & \nnfullmoon & \nnnewmoon & \nnfullmoon & \nnnewmoon & \nnnewmoon & \nnnewmoon & \nnnewmoon & \nnnewmoon \\
    10 & cloudflare.com & \nnnewmoon & \nnnewmoon & \nnfullmoon & \nnfullmoon & \nnfullmoon & \nnnewmoon & \nnnewmoon & \nnnewmoon & \nnnewmoon & \nnnewmoon & \nnnewmoon \\
    \rowcolor[gray]{.9}
    13 & apple.com & \nnnewmoon & \nnfullmoon & \nnfullmoon & \nnfullmoon & \nnfullmoon & \nnfullmoon & \nnnewmoon & \nnnewmoon & \nnnewmoon & \nnnewmoon & \nnnewmoon \\
    14 & linkedin.com & \nnnewmoon & \nnnewmoon & \nnnewmoon & \nnnewmoon & \nnnewmoon & \nnfullmoon & \nnnewmoon & \nnnewmoon & \nnnewmoon & \nnfullmoon & \nnnewmoon \\
    \rowcolor[gray]{.9}
    15 & netflix.com & \nnnewmoon & \nnfullmoon & \nnnewmoon & \nnnewmoon & \nnfullmoon & \nnfullmoon & \nnfullmoon & \nnnewmoon & \nnnewmoon & \nnnewmoon & \nnnewmoon \\
    17 & wikipedia.org & \nnfullmoon & \nnfullmoon & \nnfullmoon & \nnfullmoon & \nnfullmoon & \nnfullmoon & \nnfullmoon & \nnfullmoon & \nnnewmoon & \nnnewmoon & \nnnewmoon \\
    \rowcolor[gray]{.9}
    20 & amazon.com & \nnfullmoon & \nnnewmoon & \nnnewmoon & \nnnewmoon & \nnfullmoon & \nnfullmoon & \nnnewmoon & \nnnewmoon & \nnnewmoon & \nnnewmoon & \nnnewmoon \\
    25 & yahoo.com & \nnnewmoon & \nnnewmoon & \nnnewmoon & \nnfullmoon & \nnfullmoon & \nnnewmoon & \nnnewmoon & \nnnewmoon & \nnfullmoon & \nnnewmoon & \nnnewmoon \\
    \rowcolor[gray]{.9}
    32 & github.com & \nnnewmoon & \nnfullmoon & \nnfullmoon & \nnfullmoon & \nnfullmoon & \nnfullmoon & \nnfullmoon & \nnfullmoon & \nnfullmoon & \nnnewmoon & \nnnewmoon \\
    36 & pinterest.com & \nnfullmoon & \nnnewmoon & \nnfullmoon & \nnfullmoon & \nnfullmoon & \nnfullmoon & \nnnewmoon & \nnfullmoon & \nnnewmoon & \nnnewmoon & \nnnewmoon \\ 
    \rowcolor[gray]{.9}
    63 & vk.com & \nnfullmoon & \nnnewmoon & \nnfullmoon & \nnfullmoon & \nnfullmoon & \nnfullmoon & \nnfullmoon & \nnnewmoon & \nnfullmoon & \nnfullmoon & \nnnewmoon \\
    65 & tiktok.com & \nnnewmoon & \nnfullmoon & \nnfullmoon & \nnnewmoon & \nnfullmoon & \nnfullmoon & \nnfullmoon & \nnnewmoon & \nnnewmoon & \nnnewmoon & \nnnewmoon \\
    \rowcolor[gray]{.9}
    72 & mozilla.org & \nnfullmoon & \nnfullmoon & \nnnewmoon & \nnnewmoon & \nnnewmoon & \nnnewmoon & \nnfullmoon & \nnnewmoon & \nnnewmoon & \nnfullmoon & \nnnewmoon \\
    80 & spotify.com & \nnfullmoon & \nnfullmoon & \nnnewmoon & \nnfullmoon & \nnfullmoon & \nnfullmoon & \nnfullmoon & \nnnewmoon & \nnnewmoon & \nnnewmoon & \nnnewmoon \\
    \rowcolor[gray]{.9}
    82 & tumblr.com & \nnfullmoon & \nnnewmoon & \nnnewmoon & \nnnewmoon & \nnfullmoon & \nnnewmoon & \nnnewmoon & \nnnewmoon & \nnfullmoon & \nnnewmoon & \nnnewmoon \\
    83 & paypal.com & \nnnewmoon & \nnnewmoon & \nnnewmoon & \nnnewmoon & \nnnewmoon & \nnfullmoon & \nnnewmoon & \nnnewmoon & \nnnewmoon & \nnnewmoon & \nnnewmoon \\
    \rowcolor[gray]{.9}
    97 & ebay.com & \nnnewmoon & \nnnewmoon & \nnnewmoon & \nnnewmoon & \nnnewmoon & \nnfullmoon & \nnnewmoon & \nnnewmoon & \nnnewmoon & \nnnewmoon & \nnnewmoon \\
    99 & dropbox.com & \nnnewmoon & \nnnewmoon & \nnfullmoon & \nnfullmoon & \nnfullmoon & \nnfullmoon & \nnnewmoon & \nnnewmoon & \nnnewmoon & \nnnewmoon & \nnnewmoon \\
    \rowcolor[gray]{.9}
    103 & csdn.net & \nnnewmoon & \nnfullmoon & \nnnewmoon & \nnfullmoon & \nnfullmoon & \nnnewmoon & \nnfullmoon & \nnnewmoon & \nnfullmoon & \nnfullmoon & \nnnewmoon \\
    104 & imdb.com & \nnnewmoon & \nnnewmoon & \nnnewmoon & \nnnewmoon & \nnfullmoon & \nnfullmoon & \nnnewmoon & \nnnewmoon & \nnnewmoon & \nnnewmoon & \nnnewmoon \\
    \rowcolor[gray]{.9}
    125 & soundcloud.com & \nnnewmoon & \nnnewmoon & \nnnewmoon & \nnfullmoon & \nnnewmoon & \nnnewmoon & \nnnewmoon & \nnnewmoon & \nnnewmoon & \nnnewmoon & \nnnewmoon \\
    155 & twitch.tv & \nnnewmoon & \nnnewmoon & \nnnewmoon & \nnnewmoon & \nnnewmoon & \nnnewmoon & \nnnewmoon & \nnnewmoon & \nnnewmoon & \nnnewmoon & \nnnewmoon \\
    \rowcolor[gray]{.9}
    157 & etsy.com & \nnnewmoon & \nnnewmoon & \nnnewmoon & \nnnewmoon & \nnnewmoon & \nnnewmoon & \nnnewmoon & \nnfullmoon & \nnnewmoon & \nnnewmoon & \nnnewmoon \\
    164 & booking.com & \nnfullmoon & \nnnewmoon & \nnnewmoon & \nnfullmoon & \nnnewmoon & \nnfullmoon & \nnnewmoon & \nnnewmoon & \nnnewmoon & \nnnewmoon & \nnnewmoon \\
    \rowcolor[gray]{.9}
    171 & sourceforge.net & \nnnewmoon & \nnnewmoon & \nnnewmoon & \nnfullmoon & \nnfullmoon & \nnfullmoon & \nnnewmoon & \nnnewmoon & \nnnewmoon & \nnfullmoon & \nnnewmoon \\
    179 & researchgate.net & \nnnewmoon & \nnnewmoon & \nnnewmoon & \nnfullmoon & \nnfullmoon & \nnfullmoon & \nnnewmoon & \nnnewmoon & \nnfullmoon & \nnfullmoon & \nnnewmoon \\
    \rowcolor[gray]{.9}
    180 & oracle.com & \nnnewmoon & \nnnewmoon & \nnnewmoon & \nnnewmoon & \nnnewmoon & \nnnewmoon & \nnnewmoon & \nnnewmoon & \nnnewmoon & \nnnewmoon & \nnnewmoon \\
    186 & slack.com & \nnfullmoon & \nnnewmoon & \nnnewmoon & \nnfullmoon & \nnfullmoon & \nnnewmoon & \nnfullmoon & \nnnewmoon & \nnfullmoon & \nnnewmoon & \nnnewmoon \\
    \rowcolor[gray]{.9}
    206 & weebly.com & \nnnewmoon & \nnnewmoon & \nnnewmoon & \nnnewmoon & \nnnewmoon & \nnfullmoon & \nnnewmoon & \nnnewmoon & \nnnewmoon & \nnnewmoon & \nnnewmoon \\
    236 & samsung.com & \nnnewmoon & \nnfullmoon & \nnnewmoon & \nnfullmoon & \nnfullmoon & \nnfullmoon & \nnfullmoon & \nnnewmoon & \nnnewmoon & \nnfullmoon & \nnnewmoon \\
    \rowcolor[gray]{.9}
    322 & grammarly.com & \nnnewmoon & \nnnewmoon & \nnnewmoon & \nnfullmoon & \nnnewmoon & \nnnewmoon & \nnnewmoon & \nnnewmoon & \nnnewmoon & \nnfullmoon & \nnnewmoon \\
    328 & fiverr.com & \nnnewmoon & \nnnewmoon & \nnnewmoon & \nnfullmoon & \nnnewmoon & \nnfullmoon & \nnnewmoon & \nnnewmoon & \nnnewmoon & \nnfullmoon & \nnnewmoon \\
    \rowcolor[gray]{.9}
    344 & snapchat.com & \nnnewmoon & \nnfullmoon & \nnnewmoon & \nnfullmoon & \nnnewmoon & \nnnewmoon & \nnfullmoon & \nnnewmoon & \nnnewmoon & \nnnewmoon & \nnnewmoon \\
    381 & yelp.com & \nnnewmoon & \nnnewmoon & \nnfullmoon & \nnfullmoon & \nnfullmoon & \nnfullmoon & \nnnewmoon & \nnnewmoon & \nnnewmoon & \nnnewmoon & \nnnewmoon \\
    \rowcolor[gray]{.9}
    392 & binance.com & \nnnewmoon & \nnfullmoon & \nnfullmoon & \nnfullmoon & \nnfullmoon & \nnnewmoon & \nnfullmoon & \nnnewmoon & \nnnewmoon & \nnfullmoon & \nnnewmoon \\
    524 & netease.com & \nnnewmoon & \nnfullmoon & \nnnewmoon & \nnfullmoon & \nnfullmoon & \nnnewmoon & \nnfullmoon & \nnnewmoon & \nnfullmoon & \nnfullmoon & \nnnewmoon \\
    \rowcolor[gray]{.9}
    541 & gitlab.com & \nnnewmoon & \nnfullmoon & \nnfullmoon & \nnfullmoon & \nnfullmoon & \nnnewmoon & \nnfullmoon & \nnnewmoon & \nnnewmoon & \nnnewmoon & \nnnewmoon \\
    545 & atlassian.com & \nnnewmoon & \nnnewmoon & \nnnewmoon & \nnnewmoon & \nnnewmoon & \nnnewmoon & \nnnewmoon & \nnnewmoon & \nnnewmoon & \nnnewmoon & \nnnewmoon \\
    \rowcolor[gray]{.9}
    563 & uber.com & \nnfullmoon & \nnnewmoon & \nnnewmoon & \nnfullmoon & \nnnewmoon & \nnnewmoon & \nnnewmoon & \nnnewmoon & \nnnewmoon & \nnfullmoon & \nnnewmoon \\
    753 & airbnb.com & \nnfullmoon & \nnnewmoon & \nnnewmoon & \nnnewmoon & \nnfullmoon & \nnfullmoon & \nnnewmoon & \nnnewmoon & \nnnewmoon & \nnnewmoon & \nnnewmoon \\
    \rowcolor[gray]{.9}
    885 & nintendo.com & \nnnewmoon & \nnnewmoon & \nnnewmoon & \nnfullmoon & \nnfullmoon & \nnfullmoon & \nnfullmoon & \nnnewmoon & \nnfullmoon & \nnfullmoon & \nnnewmoon \\
    924 & xing.com & \nnnewmoon & \nnnewmoon & \nnnewmoon & \nnfullmoon & \nnfullmoon & \nnfullmoon & \nnnewmoon & \nnnewmoon & \nnfullmoon & \nnnewmoon & \nnnewmoon \\
    \rowcolor[gray]{.9}
    1205 & wayfair.com & \nnnewmoon & \nnfullmoon & \nnfullmoon & \nnfullmoon & \nnfullmoon & \nnfullmoon & \nnnewmoon & \nnnewmoon & \nnnewmoon & \nnnewmoon & \nnnewmoon \\
    1327 & deezer.com & \nnfullmoon & \nnnewmoon & \nnnewmoon & \nnfullmoon & \nnnewmoon & \nnfullmoon & \nnfullmoon & \nnnewmoon & \nnnewmoon & \nnnewmoon & \nnnewmoon \\
    \rowcolor[gray]{.9}
    1387 & lyft.com & \nnnewmoon & \nnnewmoon & \nnnewmoon & \nnnewmoon & \nnnewmoon & \nnfullmoon & \nnfullmoon & \nnnewmoon & \nnfullmoon & \nnnewmoon & \nnnewmoon \\
    1413 & battle.net & \nnnewmoon & \nnfullmoon & \nnfullmoon & \nnfullmoon & \nnfullmoon & \nnfullmoon & \nnfullmoon & \nnfullmoon & \nnnewmoon & \nnfullmoon & \nnnewmoon \\
    \rowcolor[gray]{.9}
    1573 & agoda.com & \nnnewmoon & \nnnewmoon & \nnnewmoon & \nnfullmoon & \nnfullmoon & \nnfullmoon & \nnnewmoon & \nnnewmoon & \nnnewmoon & \nnnewmoon & \nnnewmoon \\
    2476 & 1password.com & \nnfullmoon & \nnnewmoon & \nnnewmoon & \nnnewmoon & \nnfullmoon & \nnnewmoon & \nnfullmoon & \nnnewmoon & \nnnewmoon & \nnfullmoon & \nnfullmoon \\
    \rowcolor[gray]{.9}
    2645 & porkbun.com & \nnnewmoon & \nnfullmoon & \nnfullmoon & \nnfullmoon & \nnfullmoon & \nnnewmoon & \nnfullmoon & \nnfullmoon & \nnnewmoon & \nnfullmoon & \nnnewmoon \\
    2705 & synology.com & \nnnewmoon & \nnnewmoon & \nnnewmoon & \nnfullmoon & \nnfullmoon & \nnnewmoon & \nnnewmoon & \nnnewmoon & \nnnewmoon & \nnnewmoon & \nnnewmoon \\
    \rowcolor[gray]{.9}
    3210 & bitwarden.com & \nnfullmoon & \nnnewmoon & \nnfullmoon & \nnfullmoon & \nnfullmoon & \nnnewmoon & \nnfullmoon & \nnnewmoon & \nnnewmoon & \nnfullmoon & \nnfullmoon \\
    3179 & faceit.com & \nnfullmoon & \nnfullmoon & \nnnewmoon & \nnfullmoon & \nnfullmoon & \nnnewmoon & \nnfullmoon & \nnfullmoon & \nnfullmoon & \nnnewmoon & \nnnewmoon \\
    \rowcolor[gray]{.9}
    3605 & plex.tv & \nnnewmoon & \nnfullmoon & \nnnewmoon & \nnnewmoon & \nnnewmoon & \nnnewmoon & \nnfullmoon & \nnfullmoon & \nnnewmoon & \nnfullmoon & \nnnewmoon \\
    4189 & dhl.de & \nnfullmoon & \nnfullmoon & \nnfullmoon & \nnfullmoon & \nnfullmoon & \nnfullmoon & \nnfullmoon & \nnfullmoon & \nnnewmoon & \nnfullmoon & \nnnewmoon \\
    \rowcolor[gray]{.9}
    4250 & dashlane.com & \nnfullmoon & \nnnewmoon & \nnnewmoon & \nnfullmoon & \nnfullmoon & \nnnewmoon & \nnnewmoon & \nnfullmoon & \nnnewmoon & \nnfullmoon & \nnnewmoon \\
    5383 & logmein.com & \nnnewmoon & \nnnewmoon & \nnnewmoon & \nnnewmoon & \nnnewmoon & \nnnewmoon & \nnnewmoon & \nnnewmoon & \nnfullmoon & \nnfullmoon & \nnnewmoon \\
    \rowcolor[gray]{.9}
    8544 & maxmind.com & \nnfullmoon & \nnfullmoon & \nnnewmoon & \nnnewmoon & \nnnewmoon & \nnnewmoon & \nnfullmoon & \nnnewmoon & \nnnewmoon & \nnnewmoon & \nnnewmoon \\
    10625 & check24.com & \nnnewmoon & \nnnewmoon & \nnnewmoon & \nnfullmoon & \nnfullmoon & \nnfullmoon & \nnnewmoon & \nnnewmoon & \nnnewmoon & \nnfullmoon & \nnnewmoon \\
    \rowcolor[gray]{.9}
    16460 & myunidays.com & \nnnewmoon & \nnfullmoon & \nnfullmoon & \nnfullmoon & \nnfullmoon & \nnnewmoon & \nnfullmoon & \nnnewmoon & \nnnewmoon & \nnfullmoon & \nnnewmoon \\
    16993 & n26.com & \nnnewmoon & \nnnewmoon & \nnnewmoon & \nnfullmoon & \nnnewmoon & \nnnewmoon & \nnnewmoon & \nnnewmoon & \nnnewmoon & \nnnewmoon & \nnnewmoon \\
    \rowcolor[gray]{.9}
    19535 & neteller.com & \nnnewmoon & \nnnewmoon & \nnfullmoon & \nnfullmoon & \nnfullmoon & \nnfullmoon & \nnnewmoon & \nnnewmoon & \nnnewmoon & \nnnewmoon & \nnnewmoon \\
    25667 & splitwise.com & \nnnewmoon & \nnnewmoon & \nnfullmoon & \nnfullmoon & \nnfullmoon & \nnfullmoon & \nnnewmoon & \nnfullmoon & \nnnewmoon & \nnnewmoon & \nnnewmoon \\
    \rowcolor[gray]{.9}
    27539 & decathlon.com & \nnfullmoon & \nnnewmoon & \nnnewmoon & \nnfullmoon & \nnfullmoon & \nnfullmoon & \nnnewmoon & \nnnewmoon & \nnnewmoon & \nnnewmoon & \nnnewmoon \\
    31988 & netatmo.com & \nnnewmoon & \nnfullmoon & \nnnewmoon & \nnfullmoon & \nnnewmoon & \nnnewmoon & \nnnewmoon & \nnnewmoon & \nnfullmoon & \nnnewmoon & \nnnewmoon \\
    \rowcolor[gray]{.9}
    40161 & stacksocial.com & \nnnewmoon & \nnfullmoon & \nnfullmoon & \nnfullmoon & \nnfullmoon & \nnfullmoon & \nnfullmoon & \nnnewmoon & \nnnewmoon & \nnnewmoon & \nnnewmoon \\
    46969 & kinguin.net & \nnnewmoon & \nnnewmoon & \nnfullmoon & \nnfullmoon & \nnfullmoon & \nnnewmoon & \nnnewmoon & \nnnewmoon & \nnnewmoon & \nnfullmoon & \nnnewmoon \\
    \rowcolor[gray]{.9}
    48031 & traderepublic.com & \nnnewmoon & \nnfullmoon & \nnfullmoon & \nnfullmoon & \nnfullmoon & \nnfullmoon & \nnnewmoon & \nnfullmoon & \nnfullmoon & \nnfullmoon & \nnnewmoon \\
    \bottomrule
    \end{tabular}
\end{table*}
\clearpage

\twocolumn
\onecolumn
\section{Codebook}\label{sec:codebook}
\setcounter{table}{3}
\begin{table}[H]
\footnotesize
\centering
\caption{Codebook for \ref{app:part2:q2a}, \ref{app:part2:q2b}, \ref{app:part2:q3a}, \ref{app:part2:q3b}, and \ref{app:part2:q11} used in Section~\ref{sec:study2-comprehension-and-reaction} 
{RQ1: Reaction \& Comprehension}.}
\label{tab:codebook:study2-rq1}
\Description[Codebook for questions MQ2, MQ3, and MQ11]{Codebook for questions MQ2, MQ3, and MQ11 answering research question 1. Column 1 shows the code name, column 2 how often the code emerged. Column 3 describes the meaning of the code and column 4 provides an example quote.}
\begin{tabular}{p{2.4cm}p{0.45cm}p{6.5cm}p{7cm}}
\toprule
\textbf{Code} & \textbf{Freq.} & \textbf{Description} & \multicolumn{1}{l}{\textbf{Example}} \\
\midrule
\multicolumn{4}{c}{\ref{app:part2:q2a}: In reaction to this email, you decided to change your password. Please describe any other actions you took. } \\
\multicolumn{4}{c}{\ref{app:part2:q2b}: What did you do in reaction to it? } \\
\midrule

Nothing & \multicolumn{1}{r}{121} & Participant did nothing. & \emph{``After I read it, I didn't do anything as it was me who signed in.''} (L17-N) \\

Change PW & \multicolumn{1}{r}{26} & Participant changed the password. & \emph{``I took no other actions than to change my password as directed because I had not signed in.''} (M71-C) \\

Check Details & \multicolumn{1}{r}{10} & Participant checked the login details in the notification. & \emph{``I just made sure it was my device, and on the day I accessed''} (L69-N) \\

Reaction Unclear & \multicolumn{1}{r}{10} & Participant did not know how to react. & \emph{``I was confused and decided to wait and see.''} (M93-N) \\

Archive Email & \multicolumn{1}{r}{6} & Participant archived the notification. & \emph{``save it in my personal files in gmail''} (M8-N) \\

Mark as Spam & \multicolumn{1}{r}{4} & Participant marked the notification as spam. & \emph{``Put it in my spam folder''} (M25-N) \\

Understand & \multicolumn{1}{r}{4} & Participant tried to understand the notification. & \emph{``I thought about it for a couple of minutes and then deleted it.''} (M105-N) \\

\midrule
\multicolumn{4}{c}{\ref{app:part2:q3a}: Why did you react this way, i.e., change your password and take the other actions you described. } \\
\multicolumn{4}{c}{\ref{app:part2:q3b}: Why did you react this way? } \\
\midrule

Was Me & \multicolumn{1}{r}{71} & Participant described the own login being the reason. & \emph{``because it was me that logged in''} (L10-N) \\

Spontaneous & \multicolumn{1}{r}{27} & Participant reacted spontaneously. & \emph{``I just didn't think much of it''} (M51-N) \\

Not Me & \multicolumn{1}{r}{26} & Participant was not the one signing in. & \emph{``Because the wrong state especially the opposite coast is a huge red flag.''} (M77-C) \\

Suspicious & \multicolumn{1}{r}{18} & Participant questioned the legitimacy the notification. & \emph{``I hadn't logged in and the location was California so I was afraid it was a phishing attempt.''} (M107-N) \\

Don't Understand & \multicolumn{1}{r}{14} & Participant did not understand the notification. & \emph{``Wasn't sure what it was for''} (L30-N) \\

Fatigue & \multicolumn{1}{r}{9} & Participant felt fatigued by seeing the notification. & \emph{``it's good for security but I get these all the time.''} (M74-N) \\

Low Value & \multicolumn{1}{r}{7} & Account has a low value for the participant. & \emph{``Why should I care if someone accesses my SRS survey?''} (M70-N) \\

Unsure & \multicolumn{1}{r}{6} & Participant did not know how to react. & \emph{``I unsure it was me why I received it''} (L40-N) \\

Feel Protected & \multicolumn{1}{r}{3} & Participant felt protected by receiving the notification. & \emph{``I was glad that they sent me this in case there was anything out of the ordinary going on.''} (L17-N) \\

\midrule
\multicolumn{4}{c}{\ref{app:part2:q11}: In your opinion, why have you received this email? } \\
\midrule
Inform About Login & \multicolumn{1}{r}{64} & Notification informed about a new login. & \emph{``Because your system recognized that a device signed into my account.''} (L47-N) \\

Check Login & \multicolumn{1}{r}{41} & Notification was a prompt to check the login that just happened. & \emph{``To make sure that it was in fact you who had signed in to the account.''} (M104-N) \\

(Potential) Compromise & \multicolumn{1}{r}{46} & Notification informed about an actual or a potential compromise. & \emph{``My reaction was that someone from California somehow got into my account.''} (M116-C) \\

Don't Know & \multicolumn{1}{r}{39} & Participant did not know why the notification was sent. & \emph{``I had no idea, which is why I deleted it.''} (M93-N) \\

Unusual Login & \multicolumn{1}{r}{28} & Notification informed about a login that was somehow unusual. & \emph{``It sounded like someone other than my typical device had logged into my account.''} (M45-N) \\

Security & \multicolumn{1}{r}{8} & Notification was sent for security reasons. & \emph{``Security purposes.''} (L9-N) \\

Phishing & \multicolumn{1}{r}{3} & Notification was phishing. & \emph{``I thought it was phishing''} (L30-N) \\

\bottomrule
\end{tabular}
\end{table}

\setcounter{table}{4}
\begin{table}[H]
\footnotesize
\centering
\caption{Codebook for \ref{app:part2:q10} used in Section~\ref{sec:study2-decision-and-execution} {RQ2: Decision-Making \& Execution}.}
\label{tab:codebook:study2-rq2}
\Description[Codebook for questions MQ10]{Codebook for question MQ10 answering research question 2. Column 1 shows the code name, column 2 how often the code emerged. Column 3 describes the meaning of the code and column 4 provides an example quote.}
\begin{tabular}{p{2.4cm}p{0.45cm}p{6.5cm}p{7cm}}
\toprule
\textbf{Code} & \textbf{Freq.} & \textbf{Description} & \multicolumn{1}{l}{\textbf{Example}} \\
\midrule
\multicolumn{4}{c}{\ref{app:part2:q10}: What would you have done differently, if you had received the email at a different location or different time? } \\
\midrule

Pay More Attention & \multicolumn{1}{r}{12} & Participant would have payed more attention to the email. & \emph{``I might have taken a closer look at it.''} (L56-N) \\

Change PW & \multicolumn{1}{r}{6} & Participant would have changed the password. & \emph{``I would have done as the email said and changed my password ''} (L104-N) \\

Contact Support & \multicolumn{1}{r}{4} & Participant would have contacted the support. & \emph{``Read it very carefully. If anything didn't look right I’d have contacted your organization''} (L19-N) \\

Panic & \multicolumn{1}{r}{2} & Participant would have panicked because then someone else would have been signing in.  & \emph{``If i was outside I might panic a bit more, or if the email came at a weird or random time''} (L69-N) \\

\bottomrule
\end{tabular}
\end{table}

%
%
%
%
\setcounter{table}{5}
\begin{table}[!htbp]
\footnotesize
\centering
\caption{Codebook for \ref{app:part2:q13} used in Section~\ref{sec:study2-perception-and-expectation} {RQ3: Perception \& Expectation}.}
\label{tab:codebook:study2-rq3}
\Description[Codebook for questions MQ13]{Codebook for question MQ13 answering research question 3. Column 1 shows the code name, column 2 how often the code emerged. Column 3 describes the meaning of the code and column 4 provides an example quote.}
\begin{tabular}{p{2.4cm}p{0.45cm}p{6.5cm}p{7cm}}
\toprule
\textbf{Code} & \textbf{Freq.} & \textbf{Description} & \multicolumn{1}{l}{\textbf{Example}} \\
\midrule
\multicolumn{4}{c}{\ref{app:part2:q13}: In your opinion, why do you think real companies should never send emails like this one? } \\
\midrule

Feels Like Scam & \multicolumn{1}{r}{4} & Email notification in the current form feels like scam. & \emph{``I got very concern, since include a link in the email instead of suggesting go to the website.''} (M10-N) \\

Annoying & \multicolumn{1}{r}{2} & Receiving the email notifications is annoying. & \emph{``They take too much time''} (M107-N) \\

\bottomrule
\end{tabular}
\end{table}

\end{document}